\documentclass[11pt]{article}
\usepackage{amsmath,amsthm,latexsym,amssymb,amsfonts,epsfig}

\makeatletter
\@addtoreset{equation}{section}
\makeatother

\newtheorem{Theorem}{Theorem}[section]
\newtheorem{Definition}{Definition}[section]

\newtheorem{Corollary}{Corollary}[section]

\def\be{\begin{equation}}
\def\ee{\end{equation}}
\def\ba{\begin{eqnarray}}
\def\ea{\end{eqnarray}}
\def\a{{\cal A}}
\def\ab{\overline{\a}}

\def\Nl{{\mathchoice
{\setbox0=\hbox{$\displaystyle\rm N$}\hbox{\hbox to0pt
{\kern0.4\wd0\vrule height0.9\ht0\hss}\box0}}
{\setbox0=\hbox{$\textstyle\rm N$}\hbox{\hbox to0pt
{\kern0.4\wd0\vrule height0.9\ht0\hss}\box0}}
{\setbox0=\hbox{$\scriptstyle\rm N$}\hbox{\hbox to0pt
{\kern0.4\wd0\vrule height0.9\ht0\hss}\box0}}
{\setbox0=\hbox{$\scriptscriptstyle\rm N$}\hbox{\hbox to0pt
{\kern0.4\wd0\vrule height0.9\ht0\hss}\box0}}}}
\def\Zl{{\mathchoice
{\setbox0=\hbox{$\displaystyle\rm Z$}\hbox{\hbox to0pt
{\kern0.4\wd0\vrule height0.9\ht0\hss}\box0}}
{\setbox0=\hbox{$\textstyle\rm Z$}\hbox{\hbox to0pt
{\kern0.4\wd0\vrule height0.9\ht0\hss}\box0}}
{\setbox0=\hbox{$\scriptstyle\rm Z$}\hbox{\hbox to0pt
{\kern0.4\wd0\vrule height0.9\ht0\hss}\box0}}
{\setbox0=\hbox{$\scriptscriptstyle\rm Z$}\hbox{\hbox to0pt
{\kern0.4\wd0\vrule height0.9\ht0\hss}\box0}}}}
\def\Ql{{\mathchoice
{\setbox0=\hbox{$\displaystyle\rm Q$}\hbox{\hbox to0pt
{\kern0.4\wd0\vrule height0.9\ht0\hss}\box0}}
{\setbox0=\hbox{$\textstyle\rm Q$}\hbox{\hbox to0pt
{\kern0.4\wd0\vrule height0.9\ht0\hss}\box0}}
{\setbox0=\hbox{$\scriptstyle\rm Q$}\hbox{\hbox to0pt
{\kern0.4\wd0\vrule height0.9\ht0\hss}\box0}}
{\setbox0=\hbox{$\scriptscriptstyle\rm Q$}\hbox{\hbox to0pt
{\kern0.4\wd0\vrule height0.9\ht0\hss}\box0}}}}
\def\Rl{{\mathchoice
{\setbox0=\hbox{$\displaystyle\rm R$}\hbox{\hbox to0pt
{\kern0.4\wd0\vrule height0.9\ht0\hss}\box0}}
{\setbox0=\hbox{$\textstyle\rm R$}\hbox{\hbox to0pt
{\kern0.4\wd0\vrule height0.9\ht0\hss}\box0}}
{\setbox0=\hbox{$\scriptstyle\rm R$}\hbox{\hbox to0pt
{\kern0.4\wd0\vrule height0.9\ht0\hss}\box0}}
{\setbox0=\hbox{$\scriptscriptstyle\rm R$}\hbox{\hbox to0pt
{\kern0.4\wd0\vrule height0.9\ht0\hss}\box0}}}}
\def\Cl{{\mathchoice
{\setbox0=\hbox{$\displaystyle\rm C$}\hbox{\hbox to0pt
{\kern0.4\wd0\vrule height0.9\ht0\hss}\box0}}
{\setbox0=\hbox{$\textstyle\rm C$}\hbox{\hbox to0pt
{\kern0.4\wd0\vrule height0.9\ht0\hss}\box0}}
{\setbox0=\hbox{$\scriptstyle\rm C$}\hbox{\hbox to0pt
{\kern0.4\wd0\vrule height0.9\ht0\hss}\box0}}
{\setbox0=\hbox{$\scriptscriptstyle\rm C$}\hbox{\hbox to0pt
{\kern0.4\wd0\vrule height0.9\ht0\hss}\box0}}}}
\def\Hl{{\mathchoice
{\setbox0=\hbox{$\displaystyle\rm H$}\hbox{\hbox to0pt
{\kern0.4\wd0\vrule height0.9\ht0\hss}\box0}}
{\setbox0=\hbox{$\textstyle\rm H$}\hbox{\hbox to0pt
{\kern0.4\wd0\vrule height0.9\ht0\hss}\box0}}
{\setbox0=\hbox{$\scriptstyle\rm H$}\hbox{\hbox to0pt
{\kern0.4\wd0\vrule height0.9\ht0\hss}\box0}}
{\setbox0=\hbox{$\scriptscriptstyle\rm H$}\hbox{\hbox to0pt
{\kern0.4\wd0\vrule height0.9\ht0\hss}\box0}}}}
\def\Ol{{\mathchoice
{\setbox0=\hbox{$\displaystyle\rm O$}\hbox{\hbox to0pt
{\kern0.4\wd0\vrule height0.9\ht0\hss}\box0}}
{\setbox0=\hbox{$\textstyle\rm O$}\hbox{\hbox to0pt
{\kern0.4\wd0\vrule height0.9\ht0\hss}\box0}}
{\setbox0=\hbox{$\scriptstyle\rm O$}\hbox{\hbox to0pt
{\kern0.4\wd0\vrule height0.9\ht0\hss}\box0}}
{\setbox0=\hbox{$\scriptscriptstyle\rm O$}\hbox{\hbox to0pt
{\kern0.4\wd0\vrule height0.9\ht0\hss}\box0}}}}

\DeclareMathOperator{\MC}{\boldsymbol{\mathsf{M}}}
\DeclareMathOperator{\MCO}{\boldsymbol{\widehat{\mathsf{M}}}}

\DeclareMathOperator{\MCOW}{\boldsymbol{\mathsf{Master\;\;Constraint\;\;
Operator}}}

\title{The LQG -- String:\\
Loop Quantum Gravity Quantization\\ of\\ String Theory\\ ~\\
I. Flat Target Space}
\author{T. Thiemann\thanks{tthiemann@perimeterinstitute.ca} \\
\\
Perimeter Institute for Theoretical Physics\\
35 King St. N., Waterloo, ON N2J 2W9, Canada\\
\\
and\\
\\
University of Waterloo}
\date{{\small 
PI-2004-001}}

\begin{document}

\maketitle

\begin{abstract}
We combine I. background independent {\bf Loop Quantum Gravity (LQG)}
quantization techniques, II.  the mathematically rigorous framework of 
{\bf Algebraic Quantum Field 
Theory (AQFT)} and III. the theory of integrable systems resulting in the 
invariant {\bf Pohlmeyer Charges} in order to set up the general 
representation theory (superselection theory) for the closed bosonic 
quantum string on flat target space.

While we do not solve the, expectedly, rich representation theory 
completely, we present a, to the best of our knowledge new, non -- trivial 
solution to the representation problem. This solution exists 
1. for any 
target space dimension, 2. for Minkowski signature of the target space, 3.
without tachyons, 4. manifestly ghost -- free (no negative norm states), 
5. without fixing a worldsheet or target space gauge, 6. without 
(Virasoro) anomalies (zero central charge), 7. while preserving manifest
target space Poincar\'e invariance and 8. without picking up UV 
divergences.

The existence of this stable solution is, on the one hand, exciting 
because it 
raises the hope that among all the solutions to the representation problem 
(including fermionic degrees of freedom) we 
find stable, phenomenologically acceptable ones in lower dimensional 
target spaces, possibly without supersymmetry, that are much simpler 
than the solutions that arise via compactification of the standard 
Fock representation of the string. On the other hand, if such 
solutions are found, then this would prove that neither a critical 
dimension (D=10,11,26) nor supersymmetry is a prediction of string theory.
Rather, these would be features of the particular
Fock representation of current string theory and hence would not be 
generic. 

The solution presented in this paper exploits the flatness of the target 
space in several important ways. In a companion paper we treat the  
more complicated case of curved target spaces.   
\end{abstract}

\newpage

\tableofcontents

\section{Introduction}
\label{s1}

String Theory (ST) \cite{1} and Loop Quantum Gravity (LQG) \cite{2} 
(see \cite{2a} for recent reviews) are 
currently the two major approaches towards a quantum theory of gravity. 
They are complementary in many senses. For example, in ST a 
central idea is the {\it Unification of all Forces (UF)} while in LQG the 
unification of the {\it Background Independence (BI) Principle} with the 
principles of Quantum Theory is considered as the most important 
guideline.
Hence, in ST the BI is currently not implemented and vice versa LQG 
presently does not put any constraints on the matter content of the world.
It is not clear if any of these principles or both should be realized 
in quantum gravity at all, however, historically there is evidence for the 
success of both. On the one hand, the non-renormalizable Fermi model of 
the weak 
interaction was replaced by the renormalizable electroweak theory which 
unifies the weak and electromagnetic interaction. On the other hand the 
puzzles of non-relativistic quantum mechanics (e.g. negative energy 
particles) were resolved by unifying special relativity and quantum 
mechanics in QFT (``second quantization''). 

In the absence of experimental input (so far) it is 
therefore worthwhile to keep our minds open and push complementary ideas 
to their frontiers and to learn from the advantages and disadvantages of 
competing programmes. It is the purpose of this paper to make a small
contribution to that extent. Namely, we ask the question: 

\newpage

~~~~\\
{\bf Can the BI methods of LQG be employed in order to provide an 
alternative quantization of ST? If yes, what are the differences?}\\
\\
By ST we mean here old-fashioned perturbative string theory and not its
(yet to be defined) M -- Theory generalization. 

In this paper we precisely define the general quantization problem for the 
closed, bosonic string on flat (Minkowski) target space. Namely, we set
up the representation theory for the closed, bosonic string. We can 
fruitfully combine three different frameworks:
\begin{itemize}
\item{I.} {\it Background Independent LQG}\\
The string can be viewed as a worldsheet diffeomorphism invariant,
two -- dimensional QFT. Hence it precisely falls into class of theories 
that can be quantized by LQG methods. Moreover, LQG provides a general 
framework for how to implement quantum constraints without gauge fixing.
\item{II.} {\it Algebraic Quantum Field Theory (AQFT)}\\
The Haag -- Kastler 
approach to QFT \cite{3} provides a clean distinction 
between the physical object to be quantized, namely the algebra of 
physical observables, and the corresponding set of representations 
(Hilbert spaces) thereof. The latter can be viewed as different phases of 
the the theory which may be or may be not realized in nature. AQFT 
provides very powerful tools in order to solve the classification 
problem of the corresponding representations, see e.g. \cite{4} for a 
recent review. For infinite dimensional 
systems such as the string there is no Stone -- von Neumann uniqueness 
theorem \cite{5} and the corresponding representation theory is usually 
very complex.
\item{III.} {\it Integrable Systems (Pohlmeyer Charges)}\\
In constrained dynamical systems it is highly non -- trivial
to identify the gauge invariant observables of the theory. It is even 
harder to find faithful representations of the corresponding, usually
highly non -- linear, commutation relations. Fortunately, for the closed 
bosonic string on flat target space the complete set of gauge invariant 
observables, the {\bf Pohlmeyer Charges}, has been found \cite{6} and 
their classical Poisson algebra 
is under complete control \cite{16,7,17}. This will come to the surprise 
of 
most string theorists who are 
used to work in the so -- called conformal
worldsheet and/or lightcone target space gauge. In contrast, these 
observables are manifestly gauge invariant and manifestly Lorentz 
covariant. 
\end{itemize}
Hence, what we are looking for is a background independent, 
that is, gauge invariant (in the sense of {\bf LQG}) and Lorentz 
invariant, representation (in the 
sense of {\bf AQFT}) of the {\bf Pohlmeyer Algebra}. Notice that by 
definition
these representations do not allow us to pick either a worldsheet gauge
(such as the conformal gauge) nor a target space gauge (such as the 
lightcone gauge). Consequently, the problem can be set up directly for the
Nambu -- Goto String rather than the Polyakov -- String. Therefore, 
Conformal Field Theory Methods never play 
even the slightest role because there is never any need to introduce and 
eventually fix any worldsheet metric. Notice also that by definition we 
only consider representations without Virasoro and Lorentz anomalies, the 
central charge is zero by definition. Finally, in the modern framework of 
QFT \cite{3} there is no need for mathematically ill -- defined objects 
such as negative 
norm states (ghosts) in the Gupta -- Bleuler quantization procedure so 
that all our representations will be ghost -- free by definition.

The above mentioned three frameworks can now be combined as follows: 
The {\bf Pohlmeyer Charges} provide the algebra to be quantized,
{\bf AQFT} provides us with techniques to construct representations 
and finally {\bf LQG} equips us with methods to obtain representations
of the algebra of invariants from diffeomorphism invariant representations 
of kinematical (not gauge invariant) observables. 

In this paper we set up 
the general framework for the representation theory of the closed bosonic 
string on flat target space and present a non -- trivial solution thereof.   
This solution exists in any target space dimension and has no tachyons.
While fermionic worldsheet degrees of freedom are certainly needed for
phenomenological reasons, our solution shows that supersymmetry is not 
required in all representations of quantum string theory (the degrees of 
freedom do not necessarily form a supersymmetry multiplett).

One may ask how these celebrated predictions of string theory, 1. a 
critical dimension of $D=26$ for the bosonic string and $D=10$ for the 
superstring are circumvented. The answer is very simple: In the sense of 
our defintion, the representation used in ordinary string theory is rather
unnatural both from the point of view of AQFT and and LQG: From the 
point of view of AQFT, the usual Fock representation of string theory does 
not manifestly define a positive linear functional on the corresponding 
$^\ast-$algebra. It is therefore not surprising that the ``no -- ghost
theorem'' only holds in critical dimensions and that 
our solution to a manifestly ghost -- free problem works in any dimension. 
From the LQG point of view on the other hand  
the implementation of the constraints in ordinary string theory is rather
unnatural because a) one works in a particular worldsheet gauge thus 
breaking worldsheet diffeomorphism invarinance down to the conformal 
symmtries of the flat worldsheet metric and b) the contraints are 
implemented only weakly rather than strongly so that one does not perform 
an honest Dirac quantization. It is therefore not surprising that one 
usually finds a central charge and that in contrast in our framework 
CFT methods never play any role as we never have to fix a worldsheet 
metric. Finally the tachyon in ordinary bosonic string theory is a
direct consequence of an ultraviolet divergence in one of the Virasoro 
generators which is explicitly avoided in LQG. It is therefore not too 
surprising that we do not find a tachyon.

One can read the results of the present paper partly positively and partly
negatively. On the negative side one should notice that, from the purely 
mathematical point of view, our solution 
demonstrates that neither a critical dimension nor supersymmetry is a 
prediction of string theory. Rather, these notions are features of the 
particular Fock representation of current string theory which is just one 
solution among possibly zillions of others of the representation problem. 
Of course, one must show that there are solutions which are physically
acceptable from the pheomenological point of view. On the positive side 
one should 
notice that the existence of our solution is very encouraging in view of 
of the fact that solutions with tachyons are unstable and supersymmetric 
ones without tachyons have yet to be shown to be consistent with 
phenomenology. There are possibly 
an infinite number of other stable solutions, including fermionic degrees 
of freedom, some of which might be closer 
to the usual Fock representation of string theory than the one we will 
give in this paper
but also much simpler, especially in lower dimensions, which has obious 
advantages for model building.
Indeed, we encourge string 
theorists and algebraic quantum field theorists to look at the string more 
abstractly from the algebraic point of view and to systematically 
develop its representation theory.\\
\\
The present paper is organized as follows:\\ Sections two through five 
merely summarize background material on the classical Nambu -- Goto 
string, its theory of invariants, AQFT and LQG respectively. We have 
included this material for the benefit of readers from various 
backgrounds in order to make the paper accessible to a wide audience.
These sections can be safely skipped by the experts. The main results of 
this paper are contained in section six which we summarize once more in 
section seven. More in detail, here is what we will do:\\ 
\\

In section two we recall the canonical formulation of the Nambu -- Goto 
string, the actual geometrical object under consideration. The Polyakov 
string usually employed in string theory introduces an auxiliary 
worldsheet metric which is locally pure gauge 
due to the worldsheet diffeomorphism invariance and an additional Weyl 
invariance which is absent in the Nambu -- Goto formulation. 
We thus emphasize that there is never any worldsheet metric to be 
discussed and that Weyl invariance never appears in our formulation.
The only local symmetry group is the diffeomorphism 
(or reparameterization) group Diff$(M)$ of the 
two -- dimensional worldsheet $M$. We never gauge fix that symmetry
in contrast to usual string theory in the Polyakov formulation where one 
usually uses the conformal gauge which allows to fix the worldsheet 
metric $g$ to be locally flat $\eta$. This gauge fixes the Weyl invariance 
of the Polyakov string, however, it only partially fixes worldsheet 
diffeomorphism invariance since conformal symmetries 
$\varphi\in$Conf$_\eta(M)\subset Diff(M)$ (with respect to the
flat worldsheet metric $\eta$) are still allowed. This residual symmetry 
is the reason for the importance of conformal QFT techniques in usual
string theory, however, in our manifestly Diff$(M)$ -- invariant 
formulation such techniques never play any role.

After having analyzed the Nambu -- Goto string as a constrained dynamical 
system \'a la Dirac,
in section three we recall the theory of the {\bf Pohlmeyer Charges}. The 
string on flat target spaces turns out to be a completely integrable 
two -- dimensional system and the {\bf Pohlmeyer Charges} are nothing else 
than the 
invariants constructed from the corresponding monodromies via Lax pair 
methods.
It is these charges that we want to study interesting representations of.

In section four we recall elements from Algebraic Quantum Field Theory
(AQFT). In particular, we describe how cyclic representations of a 
given $^\ast-$algebra $\mathfrak{A}$ arise via the Gel'fand -- Naimark -- 
Segal (GNS) construction once a positive linear functional (a state) 
$\omega$ is given. This is the same construction that underlies 
the Wightman reconstruction theorem \cite{8} (reconstructing a Hilbert 
space from a set of $n-$point functions subject to the positivity 
requirement). Moreover, if a symmetry group acts on $\mathfrak{A}$ as a 
group
of automorphisms and if $\omega$ is invariant then the symmetry group can 
be implemented as a group of unitary operators on the GNS Hilbert space
{\it without anomalies}.    

In section five we recall basics from Loop Quantum Gravity (LQG). In 
particular, we review the background independent canonical quantization 
programme which aims at constructing the physical Hilbert space, on 
which the quantum constraints are identically satisfied, from a given 
kinematical representation of an algebra of non -- gauge invariant 
operators. The reason for not considering the algebra of invariants right 
away is that the complete set of invariants is rarely known explicitly 
for a sufficiently complicated theory (e.g. General Relativity). However,
if the kinematical algebra separates the points of the full phase space 
then the invariant algebra is contained (possibly as a limit) in the 
kinematical algebra and hence the kinematical representation is a 
representation of the invariants as well, generically with quantum 
corrections. The kinematical representation is admissable, however, only
if the physical representation induced from it still carries a 
representation of the invariants. It is in this step that 
non -- trivial regularization techniques come into play.

In section six we collect the results of sections three, four and five 
to formulate the general representation problem of the closed bosonic 
string on flat target space. We then present a non -- trivial solution to 
it. Basically, we found a worldsheet diffeomorphism invariant and target 
space Poincar\'e invariant state for a kinematical Weyl algebra 
$\mathfrak{A}$ which contains the {\bf Pohlmeyer Charges} as a limit.
We see here how LQG, AQFT and the theory of integrable systems click 
together: LQG provides a suitable Weyl algebra on which the worldsheet
symmetries act as automorphisms, AQFT provides tools to construct 
representations thereof and finally the theory of integrable systems 
provides us with the invariant {\bf Pohlmeyer Charges} which can be 
defined in our representation as well -- defined operators and whose
vacuum expectation values 
play the role of {\it gauge invariant Wightman 
functions}. In this representation Weyl invariance never arises, 
worldsheet diffeomorphism invariance and target space Poincar\'e 
invariance are exact symmetries without anomalies (central charges),
ghosts (negative norm states) never arise  
and all physical states turn out to be of non -- negative mass so that 
there is no tachyon.  

In section seven we conclude, compare our results with ordinary 
string theory and repeat their consequences.

It should be emphasized that in this paper we heavily exploited that the 
target space is flat and hence we only need a minimal amount of the 
techniques of LQG. The full power of LQG techniques however comes into 
play when we discuss curved target spaces and higher p -- brane theories
such as the (super --)membrane \cite{10} which, in contrast to string 
theory, is an interacting theory even on flat target spaces. These issues 
will be discussed in our companion paper \cite{11} which overlaps in  
part with the pioneering work \cite{11a} but, as the present paper, 
departs 
from it in most aspects. See also \cite{11b} for a different new approach 
to string theory using a modification of the Lorentz group.

\section{The Nambu -- Goto String}
\label{s2}

In introductory texts to string theory \cite{1} the Nambu -- Goto string 
is barely mentioned, one almost immediately switches to the Polyakov 
string whose corresponding action has the same set of classical solutions 
as the Nambu -- Goto action. The advantage of the Polyakov action is that 
it is bilinear in the string degrees of freedom on flat target space, 
however, this comes at the price of introducing an additional Weyl
invariance and an auxiliary worldsheet metric. From the point of a 
geometer this classical reformulation of the geometerical object, the 
string, as a theory of scalars interacting with topological gravity is 
rather unnatural. We thus review below the canonical formulation of 
the Nambu -- Goto action, especially for the benefit of readers 
without much knowledge of string theory. Experts can safely skip this 
section. See \cite{12,13} for more details.\\
\\
Let us in fact consider the bosonic $p-$brane with $p\ge 2$. Its action is 
defined in 
terms of an embedding $X:\;M\to T;\;y\mapsto X^\mu(y)$ from a 
$p-$dimensional worldsheet $M$ to a $D-$dimensional target space $T$
and a target space metric tensor field $X\mapsto \eta_{\mu\nu}(X)$
of Minkowskian signature $(-1,1,..,1)$ (the case of Euclidean signature is 
treated in 
the companion paper \cite{11}). Here we take the worldsheet coordinates 
$y^\alpha,\;\alpha=0,..,p-1$ to be dimension -- free while the target
space coordinates $X^\mu,\mu=0,..,D-1$ have dimension of length. The 
bosonic $p-$brane action is nothing else than the volume of $M$ as 
measured by $\eta$, that is,
\be \label{2.1}
S[X]:=-\frac{1}{\alpha'}\;\int_M\; d^py\; \sqrt{-\det([X^\ast \eta](y))}
\ee 
Here $\alpha'$ is called the $p-$brane tension and its has dimensions such 
that $\hbar\alpha'=:\ell_s^p$ has dimensions of cm$^p$. 
In order that $X$ be an embedding, the vector fields $\partial/\partial 
y^\alpha$ must be everywhere linearly independent on $M$. Furthermore, the 
pull-back metric on $M$ given by $X^\ast \eta$ is supposed to have 
everywhere Minkowskian signature.

We proceed to the canonical analysis. We assume that $M=\Rl\times \sigma$
where $\sigma$ is a $(p-1)-$dimensional manifold of fixed but arbitrary 
topology. The momentum canoniclly conjugate to $X^\mu$ is given by
\be \label{2.2}
\pi_\mu(y):=\alpha'\frac{\delta S}{\delta \dot{X}^\mu(y)}=-
\sqrt{-\det(X^\ast\eta)(y)} 
([(X^\ast \eta)(y)]^{-1})^{t\alpha}\eta_{\mu\nu}(X(y)) X^\nu_{,\alpha}(y)
\ee
Introducing temporal and spatial coordinates $y=(t,x)$ 
we have the elementary Poisson brackets
\be \label{2.2a}
\{X^\mu(t,x),X^\nu(t,x')\}=\{\pi_\mu(t,x),\pi_\nu(t,x')\}=0,\;\;
\{\pi_\mu(t,x),X^\nu(t,x')\}=\alpha' \delta_\mu^\nu \delta(x,x')
\ee
In the sense of Dirac's analysis of constrained systems the Legendre 
transform (\ref{2.2}) is not onto, the Langrangean in (\ref{2.1}) is 
singular and we arrive at the $p$ primary constraints 
\ba \label{2.3}
D_a &:=& \pi_\mu X^\mu_{,a}=0 
\nonumber\\
C &:=& \frac{1}{2}[\eta^{\mu\nu} \pi_\mu\pi_\nu + \det(q)]=0
\ea
where 
\be \label{2.4}
q_{ab}(y)=[X^\ast \eta]_{ab}(y)
\ee
and $a,b,..=1,..,p-1$. Notice that $\eta$ maybe an arbitrary curved 
metric. The constraint $D_a$ is often called spatial diffeomorphism 
constraint in the LQG literature because its Hamiltonian vector field 
generates spatial diffeomorphisms of $\sigma$. Likewise, the constraint
$C$ is called the Hamiltonian constraint because, on the solutions to the 
equations of motion, it generates canonically temporal 
reparameterizations. See our companion paper \cite{11} for a more detailed
elaboration on this point.  

These $p$ constraints are in fact first class so that there are no 
secondary constraints. Indeed we find the {\bf Hypersurface Deformation 
Algebra $\mathfrak{H}$} 
\ba \label{2.5}
\{D(\vec{N}),D(\vec{N}')\} &=& \alpha'
D({\cal L}_{\vec{N}}\vec{N}')
\nonumber\\
\{D(\vec{N}),C(N')\} &=& \alpha'
C({\cal L}_{\vec{N}}N')
\nonumber\\
\{C(N),C(N')\} &=& \alpha'\int_\sigma\; d^{p-1}x \;(N_{,a} N'-N N_{,a})\;
[\det(q) q^{ab}] D_b
\ea 
where $\cal L$ denotes the Lie derivative and 
where we have smeared the constraints with smooth test functions of rapid
decrease, that is, $C(N)=\int_\sigma d^{p-1}x N C$ and  
$D(\vec{N})=\int_M d^{p-1}x N^a D_a$.

Notice that precisely when $p=2$ we have that $\det(q) q^{ab}=1$ is a 
constant. Thus, among all $p-$brane actions the string is singled out by 
the fact that its constraints form an honest Lie Poisson algebra. For 
$p>2$ we get nontrivial structure functions just like for General 
Relativity. We will strongly exploit this fact for the rest of the paper
for which we restrict our attention to the case $p=2$ and $\eta$ is the 
target space Minkowski metric $\eta=$diag$(-1,1,..,1)$. See our companion 
paper \cite{11} for the more general case.

Notice that throughout our analysis we have not introduced a worldsheet
metric and there is no Weyl symmetry at all. The only local 
worldsheet symmetry is 
Diff$(M)$. Also we work manifestly gauge free, that is, we never
(partially) fix Diff$(M)$, there is no conformal gauge or anything 
like that. Finally we want to construct a true Dirac quantization of the 
system, hence we do not introduce any gauge such as the lightcone gauge 
in order to solve the constraints classically. This will have the 
advantage to preserve manifest Lorentz invariance in all steps of our 
construction.

\section{{\bf Pohlmeyer Charges}}
\label{s3}

In the rest of the paper we focus on the closed bosonic string, i.e. 
$p=2$ and $\sigma=S^1$ on 
flat Minkowski space. We take our spatial coordinate $x$ to be in the 
range $[0,2\pi)$. All functions in what follows are periodic with period
$2\pi$ unless specified otherwise.\\
\\
From the classical point of view we must find the (strong) Dirac 
observables of the 
system, that is, functions on the phase space coordinatized by $X,\pi$
which have (strongly) vanishing Poisson brackets with the constraints. 
While the string is a parameterized system so that the Hamiltonian defined 
by the Legendre transform vanishes identically, the initial value 
constraints do generate arbitrary time reparameterizations and in this 
sense the problem of finding the Dirac observables is closely related to 
finding the integrals of motion of a dynamical system \cite{14}. This 
leads us to the theory of integrable systems \cite{14a} and the very 
powerful techniques that have been developed for those, especially in two 
dimensions.
Pohlmeyer et. al. have shown in an impressive series of papers 
\cite{6,16,7,17}
that the string is completely integrable. Since to the best of our 
knowledge these important works are basically unknown even among string 
theorists, in what follows we will summarize
the basics of these developments, following the beautiful thesis 
\cite{15}.

\subsection{Automorphisms of Gauge -- and Symmetry Transformations}
\label{s3.1}

The first step is to introduce an equivalent set of constraints called the 
two Virasoro constraints
\be \label{3.1}
V_\pm(u):=\pm\frac{1}{2\alpha'}\int_{S^1}\;dx\; u (C\pm D)\equiv
\pm\frac{1}{2\alpha'}\int_{S^1}\;dx\; \xi  \eta_{\mu\nu} Y^\mu_\pm 
Y^\nu_\pm
\ee
where $u$ is a smearing function and 
\be \label{3.2}
Y^\mu_\pm:=\eta^{\mu\nu}\pi_\nu \pm X^{\mu\prime}
\ee
Here and in what follows a prime denotes derivation with respect to $x$ 
while a dot denotes derivation with respect to $t$. The advantage of 
(\ref{3.1}) over (\ref{2.3}) is that the 
constraint algebra simplifies to
\be \label{3.3}
\{V_\pm(u),V_\mp(v)\}=0,\;\;
\{V_\pm(u),V_\pm(v)\}=V_\pm(s(u,v))
\ee
where $s(u,v)=u' v- u v'$. Thus the constraint algebra can be displayed as 
the direct sum of two algebras each of which is isomorphic to the Lie 
algebra of the diffeomorphism group Diff$(S^1)$ of the circle. These two
algebras are, however, not the same as yet a third copy of diff$(S^1)$
generated by $D=V_+ + V_-$ which generates diffeomorphisms of the 
circle for all phase space functions while $V_\pm$ do that only 
for functions of $Y_\pm$.
Hence, there are three different diffeomorphism groups 
of the circle at play which have to be cleanly distinguished.

The important functions $Y_\pm(f):=\int_{S^1}\;dx\; f_\mu\; Y^\mu_\pm$ 
themselves obey the following algebra
\be \label{3.4}
\{Y_\pm(f),Y_\mp(g)\}=0,\;\;
\{Y_\pm(f),Y_\pm(g)\}=\pm \alpha'
\eta^{\mu\nu}\int_{S^1}\;dx\;
(f_\mu' g_\nu- f_\mu g_\nu')
\ee
From the geometrical point of view the $Y_\pm$ are one forms on $S^1$
and the $f$ are scalars while the $u,v$ are vector fields. In one 
dimension, $p-$times covariant and $q-$times contravariant tensors are the 
same thing as scalar densities of weight $p-q$, hence all integrals are
over scalar densities of weight one which are spatially diffeomorphism 
invariant. However, the Hamiltonian vector fields corresponding to the 
$V_\pm(\xi)$ only act on the $Y_\pm$ not on the $u,v,f,f'$. Specifically 
we have 
\be \label{3.5}
\{V_\pm(u),Y_\mp(f)\}=0,\;\; 
\{V_\pm(u),Y_\pm(f)\}=Y_\pm(u f')
\ee
Hence the Virasoro generators $V_\pm(u)$ act on $Y_\pm$ by infinitesimal 
diffeomorphisms while they leave $Y_\mp$ invariant. Consider the 
Hamiltonian flow of $V_\pm(u)$ defined by 
\be \label{3.6}
[\alpha^\pm_u(t)](F):=\sum_{n=0}^\infty \frac{t^n}{n!} 
\{V_\pm(u),F\}_{(n)}
\ee
where $F$ is any smooth function on phase space and the repeated Poisson 
bracket 
is inductively defined by $\{G,F\}_{(0)}=F,\;
\{G,F\}_{(n+1)}=\{G,\{G,F\}_{(n)}\}$. It is easy to check that
\be \label{3.7}
[\alpha^\pm_u(t)](Y_\mp(x))=Y_\mp(x),\;\;
[\alpha^\pm_u(t)](Y_\pm(x))=([\varphi^u_t]^\ast Y_\pm)(x)
\ee
where $t\mapsto \varphi^u_t$ is the one parameter group of diffeomorphisms 
of $S^1$ defined by the integral curves $t\mapsto c^u_x(t)$ through $x$ of 
the vector field $u$ on $S^1$, that is, $\varphi^u_t(x)=c^u_x(t)$.
Here $\varphi^\ast$ denotes the pull -- back of $p-$forms. 

Since the Hamiltonian flow of a smooth phase space function defines an 
automorphism of the Poisson algebra of smooth functions on phase space,
we obtain for any smooth function of the functional form 
$F=F[Y_+,Y_-]$ that 
\be \label{3.8}
[\alpha^+_u(t)](F)=F[(\varphi^u_t)^\ast Y_+,Y_-],\;\;
[\alpha^-_u(t)](F)=F[Y_+,(\varphi^u_t)^\ast Y_-]
\ee
Writing $\alpha^\pm_{\varphi^u_t}:=\alpha^\pm_u(t)$ we may extend 
(\ref{3.8}) to all elements $\varphi$ of Diff$(S^1)$
\be \label{3.9}
[\alpha^+_\varphi](F)=F[\varphi^\ast Y_+,Y_-],\;\;
[\alpha^-_\varphi](F)=F[Y_+,\varphi^\ast Y_-]
\ee
Besides this local gauge freedom of the string we also have global 
Poincar\'e symmetry. The generators of infinitesimal translations and 
Lorentz transformations respectively are given by 
\ba \label{3.10}
p_\mu &=& \frac{1}{\alpha'}\int_{S^1}\; dx\; \pi_\mu 
\nonumber\\
J^{\mu\nu} &=& \frac{1}{\alpha'}\int_{S^1}\; dx\; 
[X^\nu \eta^{\mu\rho} \pi_\rho-X^\mu \eta^{\nu\rho} \pi_\rho]
\ea
It is straightforward to check that $p_\mu,J^{\mu\nu}$ have vanishing 
Poisson brackets with the $V_\pm(u)$ and thus are strong Dirac 
observables. Moreover, we have 
\be \label{3.11}
\{p_\mu,Y^\nu_\pm(x)\}=0,\;\;
\{J^{\mu\nu},Y^\rho_\pm(x)\}=(
\eta^{\mu\rho} Y^\nu_\pm-\eta^{\nu\rho} Y^\mu_\pm)(x)
\ee
Hence for the corresponding Hamiltonian flows we obtain
\ba \label{3.12}
\alpha_a(Y_\pm)&:=&\sum_{n=0}^\infty\; \frac{1}{n!}\;
\{a^\mu p_\mu,Y_\pm\}_{(n)}=Y_\pm
\nonumber\\
\alpha_\Lambda(Y_\pm)&:=&\sum_{n=0}^\infty\; \frac{1}{n!}\;
\{\Lambda_{\mu\nu} J^{\mu\nu},Y_\pm\}_{(n)}=
\exp(\Lambda_{\mu\nu} \tau^{\mu\nu}) \cdot Y_\pm
\ea 
where $\Lambda$ is an antisymmetric matrix and $\tau^{\mu\nu}$ are 
appropriate basis elements of $so(1,D-1)$. Denoting 
$L=\exp(\Lambda_{\mu\nu} \tau^{\mu\nu})$ the automorphisms extend to 
smooth functions of $Y_\pm$ as 
\be \label{3.13}
\alpha_a(F)=F,\;\;\alpha_L(F)=F[L\cdot Y_+,L\cdot Y_-]
\ee

\subsection{Algebra of Invariants}
\label{s3.2}

This finishes our analysis of the actions of the constraints and symmetry 
generators on the $Y_\pm$. We will now construct Dirac 
observables $F[Y_+,Y_-]$ by using the theory of integrable systems.
The idea is to use the method of Lax pairs, that is, one reformulates the 
equations of motion for $Y_\pm$ as a matrix equation of the form
$M_{,\alpha}=[A_\alpha,M]$ where $M(t,x)$ is an $N\times N$ matrix and 
$A_\alpha$ is a ``connection'', that is, a one form on $\Rl\times S^1$ 
with values in $GL(N,\Cl)$. The integrability condition for the Lax pair
$L,A$ is the zero curvature equation 
$F_{\alpha\beta}=
\partial_\alpha A_\beta
-\partial_\beta A_\alpha+[A_\alpha,A_\beta]=0$ and one looks for $L,A$
in such a way that $F_{\alpha\beta}=0$ is equivalent to the equations of 
motion. Given such a setup, it follows that the functions Tr$(L^n),\;
n=1,..,N$ are constants of the motion ($n>N$ leads to algebraically 
dependent invariants due to the theorem of Hamilton -- Caley). 

We proceed to the details. The ``Hamiltonian'' of the string is
given by 
\be \label{3.14}
H(u,v)=\frac{1}{\alpha'}(C(u)+D(v))=
V_+(v+u)+V_-(v-u)
\ee
where $u,v$ are arbitrary test functions.
``Time'' evolution is defined by
\be \label{3.15}
\dot{Y}_\pm:=\{H(u,v),Y_\pm\}=[(v\pm u) Y_\pm]'
\ee
Notice that the ``time'' evolution depends on $u,v$ and it will be our 
task 
to show that our final invariants constructed by the Lax pair method do 
not depend on $u,v$.

Let $\tau_I$ be a basis of $GL(n,\Rl)$ and $T_\mu^I$ some complex valued 
matrices. We define the following connection 
\be \label{3.16}
A^\pm_x:=Y^\mu_\pm T_\mu^I \tau_I=:Y^\mu_\pm T_\mu,\;\;
A^\pm_t:=(v\pm u) A^\pm_x
\ee
The zero curvature condition reproduces the equations of motion 
\be \label{3.17}
F^\pm_{tx}=\partial_t A^\pm_x-\partial_x A^\pm_t+[A^\pm_t,A^\pm_x]=
\partial_t A^\pm_x-\partial_x A^\pm_t=
\partial_t A^\pm_x-\{H(u,v),A^\pm_x\}=0
\ee
Given a curve $c$ on $M=\Rl\times S^1$ we define its holonomy as the usual
path ordered product
\be \label{3.18}
h_c(A^\pm)={\cal P}\exp(\int_c\; dx \; A^\pm)
\ee
where $\cal P$ denotes path ordering with the lowest parameter value to 
the outmost left. Given a parameterization of the path 
$[0,b]\to c;\;s\mapsto c(s)$ the holonomy is the unique solution to the 
ordinary differential equation
\be \label{3.18a}
\frac{d}{ds} h_{c_s}(A^\pm)=h_{c_s}(A^\pm) A^\pm_\alpha(c(s)) 
\dot{c}^\alpha(s),\;\; h_{c_0}(A^\pm):=1_N,\;\;h_{c_b}(A^\pm)\equiv 
h_c(A^\pm)
\ee
where $c_s=c([0,s])$.

Supposing that the curvature of $A$ vanishes, that is, 
that $A$ is flat, it is clear that for loops $c$ on $\Rl\times S^1$ only
those which are not contractible lead to a non -- trivial value of 
$h_c(A)$. Among the non contractible loops of winding number one all are 
homotopic to the loop $c_{t,x}=\{t\}\times [x,x+2\pi)$ in the 
worldsheet time slice $t=$const. winding once around the cylinder with 
starting point $x$. In the theory of integrable systems, the holonomy 
of that loop is called the monodromy matrix
\be \label{3.19}
h_{t,x}(A^\pm):=h_{c_{t,x}}(A^\pm)
\ee
Notice that (\ref{3.19}) is not a periodic function of $x$ even if the 
connection $A$ is flat. 

Consider an arbitrary interior point $x_0$ of $c_{t,x}$. This point 
subdivides the loop into two edges $e_{t,x}=\{t\}\times [x_0,x]$ and 
$e'_{t,x}=\{t\}\times [x_0,x+2\pi]$, that is, 
$c_{t,x}=e_{t,x}^{-1}\circ e'_{t,x}$.  
Hence we have by basic properties of the holonomy 
\be \label{3.20}
h_{t,x}(A^\pm):=(h_{e_{t,x}}(A^\pm))^{-1} h_{e'_{t,x}}(A^\pm)
\ee
By the very definition of the holonomy of a connection along a path 
(\ref{3.18a}) we find with the parameterization $c_{t,x}(s)=x+s,\;s\in 
[0,2\pi)$
 \be \label{3.21}
\partial_x h_{t,x}(A^\pm)=[h_{t,x}(A^\pm),A_x(t,x)]
\ee
On the other hand we have 
\ba \label{3.22}
\partial_t h_{t,x}(A^\pm) 
&=& 
\lim_{\epsilon\to 0} \frac{1}{\epsilon}
(h_{t+\epsilon,x}(A^\pm)-h_{t,x}(A^\pm))
\nonumber\\
&=& 
\lim_{\epsilon\to 0} \frac{1}{\epsilon}
(h_{e_{\epsilon,t,x}}(A^\pm)^{-1}[h_{e_{\epsilon,t,x}}(A^\pm)
h_{t+\epsilon,x}(A^\pm)]-h_{t,x}(A^\pm))
\nonumber\\
&=& 
\lim_{\epsilon\to 0} \frac{1}{\epsilon}
(h_{e_{\epsilon,t,x}}(A^\pm)^{-1}[h_{t,x}(A^\pm) h_{e_{\epsilon,t,x}}(A^\pm)]
-h_{t,x}(A^\pm))
\nonumber\\
&=& [h_{t,x},A^\pm_t(t,x)]
\ea
where $e_{\epsilon,t,x}(s)=(t+s,x),\;s\in[0,\epsilon]$. Here we have used
that the loop $c_{t,x}\circ e_{\epsilon,t,x} \circ c_{t+\epsilon,x}^{-1}
\circ e_{\epsilon,t,x}^{-1}$ is contractible and the zero curvature 
condition in the third step. 

Conversely, postulating equations (\ref{3.21}) and (\ref{3.22}) we 
discover, using the Jacobi identity
\be \label{3.23}
2\partial_{[\alpha} \partial_{\beta]} h_{t,x}(A^\pm)
=[h_{t,x}(A^\pm),F^\pm_{\alpha,\beta}]
\ee
where $F^\pm$ is the curvature of $A^\pm$. Hence the zero curvature 
condition is the integrability condition for the equations 
$\partial_\alpha h_{t,x}(A^\pm)=[h_{t,x}(A^\pm),A^\pm_\alpha(t,x)]$.

We now claim that for any $n\le N$ the functions 
\be \label{3.24}
T^n_{t,x}(A^\pm):=\mbox{Tr}_N([h_{t,x}(A^\pm)]^n)
\ee
are independent of both $x,t$. This follows immediately from (\ref{3.21})
and (\ref{3.22}) as well as the cyclicity of the trace. What is, however,
even more remarkable is that (\ref{3.24}) has vanishing Poisson brackets 
with $H(u,v)$ {\it for all $u,v$} even though 
$A^\pm_t=(v\pm u) A^\pm_x$ depends explicitly
on $u,v$. The reason for this is that $A_x^\pm$ is actually independent of 
$u,v$ and that the dependence of the time evolution of $h_{t,x}(A^\pm)$ 
consists just in a prefactor 
\be \label{3.25}
\partial_t h_{t,x}(A^\pm)=(v\pm u)\partial_x h_{t,x}(A^\pm)
\ee
and the second factor in (\ref{3.25}) is independent of $u,v$. More 
precisely we have 
\ba \label{3.26}
\{H(u,v),h_{t,x}(A^\pm)\} &=& 
\int_{S^1}\; dy\;\{H(u,v),Y^\mu_\pm(t,y)\}\;
\frac{\delta h_{t,x}(A^\pm)}{\delta Y^\mu_\pm(t,y)}
\nonumber\\
&=:&
\int_{S^1}\; dy\;[\partial_t Y^\mu_\pm(t,y)]\;
\frac{\delta h_{t,x}(A^\pm)}{\delta Y^\mu_\pm(t,y)}
\nonumber\\
&=& \partial_t h_{t,x}(A^\pm)
\ea
One can of course verify by direct computation from the expression 
(\ref{3.18}) that the $T^n(A^\pm)$ are Dirac observables for the string.

Since we can trade the power $n\le N$ for considering increasing rank
$N$ of the matrices $\tau_I$ the only interesting invariant is the 
generating functional
\be \label{3.27}
Z^T_\pm:=T^1(A^\pm)=N+\sum_{n=1}^\infty\; Z_\pm^{\mu_1..\mu_N}\;
\frac{\mbox{Tr}_N(T_{\mu_1}..T_{\mu_N})}{N}
\ee
where $T^\mu=T^\mu_I \tau^I$ and we have used the expansion (\ref{3.18}). 
It is understood that $N$ can be arbitrarily large and that the 
complex matrices $T^\mu\in GL(N,\Cl)$ are freely specificable while 
$Z^T_\pm$ is an invariant. Thus we conclude that the expansion 
coefficients are themselves invariants. These are the {\bf Pohlmeyer
Charges}
\ba \label{3.28}
Z_\pm^{\mu_1..\mu_N} &=& 
[R^{\mu_1..\mu_N}(x)+R^{\mu_2..\mu_N \mu_1}(x)+..
+R^{\mu_N \mu_1..\mu_{N-1}}(x)]
\nonumber\\
R_\pm^{\mu_1..\mu_N}(x) &=& \int_x^{x+2\pi}\; dx_1\;\int_{x_1}^{x+2\pi}\; 
dx_2\;..\int_{x_{N-1}}^{2+2\pi}\; dx_N\; Y^{\mu_1}_\pm (x_1)\;..
\;Y^{\mu_N}_\pm(x_N)
\nonumber\\
&=:& \int_{x\le x_1 \le ..\le x_N \le x+2\pi}\; d^Nx \;
Y^{\mu_1}_\pm (x_1)\;..\;Y^{\mu_N}_\pm(x_N)
\ea
The functionals $R^{(N)}_\pm(x)$ depend explicitly on $x$, they are not 
invariants. It is only after cyclic summation $C_N\cdot R^{(N)}_\pm=
Z^{(N)}_\pm$ 
that 
they become invariants.

The {\bf Pohlmeyer Charges} exepectedly form a complicated, closed 
subalgebra of the 
full Poisson algebra of the string. We just quote the result 
and refer the reader to the literature \cite{6,16,7,17,15}. One finds 
after
a lot of algebra
\ba \label{3.29}
&& \{Z_\pm^{\mu_1..\mu_M},Z_\mp^{\nu_1..\nu_N}\}=0
\\
&& \{Z_\pm^{\mu_1..\mu_M},Z_\pm^{\nu_1..\nu_N}\}=
\mp 2\alpha'\sum_{m=1}^M \;\sum_{n=1}^N\;\eta^{\mu_m\nu_n}
\times \nonumber\\
&&\times 
[S_{\mu_{m+2}..\mu_{m-1};\nu_{n+1}..\nu_{n-2}} \cdot
Z_\pm^{\mu_{m+1}..\mu_{m-1} \nu_{n+1}..\nu_{n-1}} 
-
S_{\mu_{m+1}..\mu_{m-2};\nu_{n+2}..\nu_{n-1}} \cdot
Z_\pm^{\nu_{n+1}..\nu_{n-1} \mu_{m+1}..\mu_{m-1}} ]
\nonumber
\ea
Here the symbol $S_{a_1.. a_m;b_1..b_n}$ denotes the shuffle operator on 
multi -- indexes, that is, it imposes summation over all permutations 
of the $a_1,..,a_m,b_1,..,b_n$ such that 
$a_k$ is always to the left of $a_l$ for $1\le k\le l \le m$ and such that  
$b_k$ is always to the left of $b_l$ for $1\le k\le l \le n$, ie. the 
sequence of the indices $a_1,..,a_m$ and $b_1,..,b_n$ remains unaltered.
They even form a $^\ast-$algebra, namely
\ba \label{3.30}
&& (Z_\pm^{\mu_1..\mu_M})^\ast:=\overline{Z_\pm^{\mu_1..\mu_M}}=
Z_\pm^{\mu_1..\mu_M}
\nonumber\\
&& Z_\pm^{\mu_1..\mu_M}\;\; Z_\pm^{\mu_{M+1}..\mu_{M+N}}
= C_N\cdot[S_{\mu_1..\mu_M;\mu_{M+1}..\mu_{M+N-1}}\cdot 
Z_\pm^{\mu_1..\mu_{M+N}}]
\ea
where the cyclic summation acts on the $\mu_1,..,\mu_N$ only.
Notice the relations $p^\mu=Z^\mu_\pm$ and
\be \label{3.31}
\{p^\mu,Z_\pm^{\mu_1..\mu_N}\}=\{Z^\mu_\mp,Z_\pm^{\mu_1..\mu_N}\}=0
\ee

As one can show, together with $J^{\mu\nu}$ the invariants $Z_\pm$
provide a {\bf complete} system of invariants for the string in the sense 
that one can reconstruct $X^\mu(t,x)$ up to gauge transformations 
(parameterizations) and up
to translations in the direction of $p^\mu$.   
Unfortunately, the invariants $Z_\pm$ are not algebraically 
independent, that is, they are overcomplete because there are polynomial 
relations between them.  
However, it is possible to construct so -- called {\it standard 
invariants}
\cite{6,16} of which all the $Z_\pm$ are polynomials. Although we do not 
need them for what follows we will briefly sketch their definition in 
order to summarize the state of the art of Pohlmeyer's algebraic approach 
to string theory.

The first step is to consider the logarithm of the monodromy matrix
by means of the identity
\be \label{3.32}
h_{t,x}(A^\pm)=\exp(\ln(h_{t,x}(A^\pm)))=\sum_{k=0}^\infty 
\frac{1}{k!}[\ln(h_{t,x}(A^\pm))]^k
\ee
Close to $T_\mu=0$ or $h_{t,x}(A^\pm)=1_N$ we can expand the logarithm as 
for $|x-1|<1$
\be \label{3.33}
\ln(x)=\ln(1-(1-x))=-\sum_{n=1}^\infty \frac{1}{n} (1-x)^n 
=\sum_{n=1}^\infty \frac{(-1)^{n+1}}{n} (x-1)^n 
\ee
so that 
\be \label{3.34}
\frac{1}{k!}[\ln(x)]^k=\sum_{n=k}^\infty c_{kn} (x-1)^n
\ee
for certain coefficients $c_{kn}$ satisfying $c_{kn}=0$ for $k>n$. Noting 
that 
\be \label{3.35}
h_{t,x}(A^\pm)-1_N=\sum_{M=1}^\infty R_\pm^{\mu_1..\mu_M} 
T_{\mu_1}..T_{\mu_M}
\ee
we see that the coefficients $R_{\pm,k}$ of 
\be \label{3.36}
\frac{1}{k!}[\ln(h_{t,x}(A^\pm))]^k=\sum_{n=k}^\infty 
R_{\pm,k}^{\mu_1..\mu_n} T_{\mu_1}..T_{\mu_n}
\ee
are given by polynomomials of the tensors $R_\pm$ of tensor rank $1\le 
M\le n$ whose tensor ranks add up to $n$. Specifically they are given
by the multi -- shuffle -- sums
\be \label{3.36a}
R_{\pm,k}^{\mu_1..\mu_n}=\sum_{m=k}^n\; c_{km}\;
\sum_{0<a_1<..<a_{m-1}<n} 
R^{\mu_1..\mu_{a_1}}_\pm\;
R^{\mu_{a_1+1}..\mu_{a_2}}_\pm\;..\;
R^{\mu_{a_{m-1}+1}..\mu_n}_\pm\;
\ee
They are therefore called the 
{\it homogeneous} tensors. The {\it truncated} tensors are simply
\be \label{3.37}
R_{\pm,t}^{\mu_1..\mu_n}:=R_{\pm,k=1}^{\mu_1..\mu_n}
\ee
Their importance lies in the fact that one can write the $R_\pm$ in terms 
of the $R_{\pm,t}$ namely
\ba \label{3.38}
R_\pm^{\mu_1..\mu_n} &=& \sum_{k=1}^n R_{\pm,k}^{\mu_1..\mu_n}
\nonumber\\
R_{\pm,k}^{\mu_1..\mu_n} &=& \frac{1}{k!} \sum_{0<a_1<..<a_{k-1}<n}\;
R_{\pm,t}^{\mu_1..\mu_{a_1}}\; 
R_{\pm,t}^{\mu_{a_1+1}..\mu_{a_2}}\;..\;
R_{\pm,t}^{\mu_{a_{k-1}+1}..\mu_n}\; 
\ea 
The first relation in (\ref{3.38}) has a direct analog for the 
invariants themselves 
\be \label{3.39}
Z_\pm^{\mu_1..\mu_n} = \sum_{k=1}^n Z_{\pm,k}^{\mu_1..\mu_n}
\mbox{ where } Z_{\pm,k}^{\mu_1..\mu_n}=C_n\cdot 
R_{\pm,k}^{\mu_1..\mu_n}
\ee
so that all invariants are polynomials in the truncated tensors. 

We notice that the the homogeneous invariants $Z^{(n)}_{\pm,k}$ carry two 
gradings, the tensor rank $n\ge 0$ and the homogeneity degree $k\ge 0$.
Under Poisson brackets these two gradings behave as follows
\be \label{3.40}
\{Z^{(n)}_{\pm,k},Z^{(n')}_{\pm,k'}\}=Z^{(n+n'-2)}_{\pm,k+k'-1}
\ee
We can define a new grading degree $L:=n-k-1,\;n\ge k$, called the 
{\it layer}, which behaves additively under Poisson brackets
\be \label{3.41}
\{Z_\pm(L),Z_\pm(L')\}=Z_\pm(L+L')
\ee
where $Z_\pm(L)$ is any linear combination of homogeneous invariants
$Z^{(n)}_{\pm,k}$ such that $n-k-1=L$. Notice that the vector space 
$V_L$ of the $Z_\pm(L)$ is infinite dimensional for each finite $L$.

Consider now the layer $V_L$. It contains the following {\it standard 
invariants} 
\be \label{3.42}
Z^{0,..,0,a,\mu_2,..,\mu_{l+1},b}_{\pm,L+1}
\ee
where $a,b=1,..D-1$ are spacelike tensor components and the first $L$
tensor components are zero (timelike). One can show \cite{15} that 
these standard invariants, together with $p_\mu$, define an {\it 
algebraic basis} for the string, that is, they are polynomially 
independent and all other invariants are polynomials of those.

Remarkably, almost all standard invariants of the layers $L\ge 2$ can be 
obtained by multiple Poisson brackets of the layers $L=0,1$ (the layer
$L=-1$ are the momentum components $p_\mu$ which are the central elements
of the invariant algebra). In each layer with $L$ odd one finds so -- 
called {\it exceptional elements}
\be \label{3.43}
Z^{0\mu 0..0 \nu}_{\pm,2}
\ee
with $L$ zeroes in between $\mu,\nu$ which cannot be generated via Poisson 
brackets from the $L=0,1$ standard invariants. In an incredible 
computational effort it has been verified that up to layer $L=7$ in $D=3$ 
that the exceptional elements are the only such invariants which cannot 
be generated. It is assumed, but has not been proved yet, that the 
{\it quadratic generation hypothesis} holds, namely that the standard 
invariants of the first two layers $L=0,1$ together with the 
infinite number of exceptional 
elements from the odd layers generate the full algebra of invariants via 
multiple Poisson brackets (the name ``quadratic'' is due to the fact that 
the exceptional elements (\ref{3.43}) have homogeneity degree $k=2$).
 
Let us assume that the quadratic generation hypothesis holds and denote 
by $\mathfrak{G}$ the vector space of generators of the full algebra 
$\mathfrak{Z}$ of invariants. A given invariant in $\mathfrak{Z}$ can, 
however, be written in many ways as a linear combination of 
multiple Poisson brackets between elements from $\mathfrak{G}$, for 
instance, due to the Jacobi identity for the Poisson bracket, but there 
are even more relations as one can see by computing the number of 
standard invariants in each layer $L$ and the number of possible Poisson 
brackets between generators with range in that layer. In other 
words, the vector space $\mathfrak{G}$ does not generate $\mathfrak{Z}$
freely. Denote by $S(\mathfrak{G})$ the symmetric (i.e. commuting under
the tensor product) 
envelopping Lie algebra
of $\mathfrak{G}$ and by $\mathfrak{J}\subset S(\mathfrak{G})$ the ideal
generated by these polynomial relations between multiple Poisson brackets.
Then $\mathfrak{Z}=S(\mathfrak{G})/\mathfrak{J}$.    

Pohlmeyer's programme to quantize the string algebraically now consists
in the construction of the universal (i.e. non -- commutating under the 
tensor product) envelopping algebra $\hat{U}(\mathfrak{G})$ of 
$\mathfrak{G}$ quotiented by a quantum ideal $\widehat{\mathfrak{J}}$.
The quantum ideal roughly arises from the classical one by replacing 
Poisson brackets by commutators divided by $i\hbar$ with symmetric 
ordering, whereby higher order (in $\hbar$) quantum corrections are 
allowed. These corrections have to be chosen {\it self -- consistently} 
in such a way that the number of quantum relations is the same as the 
number of classical relations in order that both classical and quantum 
theory have the same number of degrees of freedom. Due to the non -- 
commutativity of quantum theory it might happen that commutators between 
quantum relations and quantum generators produce 
{\it anomalies}, that is, new relations without classical counterpart.
Again, in an enormous compuational effort it has been shown in $D=4$
that there are suitable non -- anomalous choices for the quantum relations 
up to layer five.

It appears hopelessly difficult to complete this programme to all layers. 
However, recently \cite{16} there has been made important progress in this 
respect:
It is possible to identify $\mathfrak{Z}$ as the kernel of a larger 
algebra $S(\mathfrak{R})$ without relations under a suitable derivation 
$\partial_x:\;S(\mathfrak{R})\to \Rl^D\times S(\mathfrak{R})$ where 
$\mathfrak{R}$ stands for a set of generators (constructed from 
the truncated tensors) larger than $\mathfrak{G}$. One then 
quantizes, like in BRST quantization, a quantum derivation $\delta$ on
the universal envelopping algebra $\hat{U}(\mathfrak{R})$ of $\mathfrak{R}$ 
and defines the quantum algebra as the kernel of $\delta$. This takes, 
self-consistently, care of all quantum relations in all layers in 
{\it a single stroke}.\\
\\
\\ 
We hope to have given a useful and fair introduction to the 
{\bf Pohlmeyer String}. The outstanding problems in this programme are 
1. the proof of the quadratic generation hypothesis and 2. the proof that 
the final quantum algebra $\widehat{\mathfrak{Z}}$ of invariants has 
interesting representations. Notice that the approach is completely 
intrinsic, it is a {\it quantization after solving the constraints} 
because one works on the reduced phase space of the unsconstrained phase 
space and seeks a quantization of the corresponding Dirac observables.

For the LQG string we proceed differently, namely we adopt Dirac's 
programme of {\it solving the constraints after quantizing}. This means,
in particular, that we have to introduce a kinematical, that is, 
unphysical algebra $\mathfrak{A}$ of observables which contains the {\bf 
Pohlmeyer Charges} as certain limits. We define the representation theory 
of $\mathfrak{A}$, taking care of the quantum constraints by methods of 
LQG and AQFT and then define physically interesting representations 
as those which support the {\bf Quantum Pohlmeyer Charges}. 
This presents an alternative route to quantizing $\mathfrak{Z}$.\\
\\
Let us clarify this difference between the two programmes: In Pohlmeyer's 
programme one starts directly
from the classical Poisson algebra of invariant charges. One then seeks 
for representations of the corresponding quantum algebra which may involve 
$\hbar$ corrections. In this intrinsic approach it is important that the 
quantum corrections are again polynomials of the charge operators because 
it is only those that one quantizes. Moreover, one must make sure that 
there are no anomalies in the corresponding quantum ideal. This is a very 
difficult programme 
because the charge algebra is very complicated and it is not a priori 
clear that the infinitely many consistency conditions among the quantum 
corrections are satisfied. However, modulo establishing the quadratic 
generation hypothesis, appropriate quantum corrections can be found and 
one is left with the problem of finding irreducible rpresentations.  
This is even more difficult given the expectedly difficult 
structure of the quantum corrections. In this paper we will take a much 
less ambitious approach which bears a lot of similarity with the proposal 
of \cite{16}: Namely, we start with a fundamental 
algebra $\mathfrak{A}$ which is much simpler but also much larger than 
the extended algebra considered in \cite{16} and seek representations of 
that. The quantum 
algebra $\widehat{\mathfrak{Z}}$ is then a derived object. The advantage 
is that 
the quantum corrections for $\mathfrak{Z}$ that we necessarily get are
not supposed to be generated by elements of $\mathfrak{Z}$ again, rather 
they may be generated from elements of $\mathfrak{A}$ which is clearly the 
case. Moreover, the quantum ideal corresponding to $\mathfrak{A}$ is much 
simpler, in fact it is rather standard in AQFT. This not only 
simplifies the algebraic problem tremendously but also the representation 
theory. The only thing that one has to make sure is that the quantum 
corrections really vanish in the classical limit.

\section{Operator Algebras and Algebraic Quantum Field Theory}
\label{s4}

We include this section only as background material for the unfamiliar 
reader, see e.g. \cite{18} for further information. Everything that we 
summarize here is standard knowledge in  
mathematical physics and can be safely skipped by the experts.
\begin{itemize}
\item[I.] {\it Operator Algebras}\\
An algebra $\mathfrak{A}$ is simply a vector space over $\Cl$ in which 
there is defined an associative and distributive multiplication.
It is unital if there is a unit {\bf 1} which satisfies 
$a\mbox{{\bf 1}}=\mbox{{\bf 1}}a=a$ for all $a\in \mathfrak{A}$. It is a 
$^\ast-$algebra if there is defined an involution satisfying 
$(ab)^\ast=b^\ast a^\ast,\;(a^\ast)^\ast=a$ which reduces to complex 
conjugation on the scalars $z\in \Cl$. 

A Banach algebra is an algebra with 
norm $a\mapsto ||a||\in \Rl^+$ which satisfies the usual axioms 
$||a+b||\le ||a||+||b||,\;||ab||\le ||a||\;||b||,\;||za||=|z|\;||a||,\;
||a||=0\;\; \Leftrightarrow \;\;a=0$ and with respect to which it is 
complete.

A $C^\ast-$algebra is a Banach $^\ast-$algebra whose norm satisfies the 
$C^\ast-$property $||a^\ast a||=||a||^2$ for all $a\in \mathfrak{A}$.
Physicists are most familiar with the $C^\ast-$algebra 
${\cal B}({\cal H})$ of bounded operators on a Hilbert space $\cal H$. 
\item[II.] {\it Representations}\\
A representation of a $^\ast-$algebra $\mathfrak{A}$ is a pair $({\cal 
H},\pi)$ consisting of a Hilbert space $\cal H$ and a morphism
$\pi:\;\mathfrak{A}\to {\cal L}({\cal H})$ into the algebra of linear
(not necessarily bounded) operators on $\cal H$ with common and invariant 
dense domain. This means that 
$\pi(za+z'a')=z\pi(a)+z'\pi(a'),\; \pi(ab)=\pi(a)\pi(b),\;
\pi(a^\ast)=[\pi(a)]^\dagger$ where $^\dagger$ denotes the adjoint in 
$\cal H$. 

The representation is said to be faithful if Ker$\pi=\{0\}$
and non -- degenerate if $\pi(a)\psi=0$ for all $a\in \mathfrak{A}$ 
implies $\psi=0$. 

A representation  
is said to be cyclic if there exists a normed vector $\Omega\in {\cal H}$
in the common domain of all the $a\in \mathfrak{A}$ such that 
$\pi(\mathfrak{A})\Omega$ is dense in $\cal H$. Notice that the existence 
of a cyclic vector implies that the states $\pi(b)\Omega,\;b\in 
\mathfrak{A}$ lie in the common dense and invariant domain for all 
$\pi(a),\; a\in 
\mathfrak{A}$. A representation is said to be irreducible if every vector 
in a common dense and invariant (for $\mathfrak{A}$) domain is cyclic.
\item[III.] {\it States}\\
A state on a $^\ast-$algebra is a linear functional 
$\omega:\;\mathfrak{A}\to \Cl$ which is positive, that is,
$\omega(a^\ast a)\ge 0$ for all $a\in \mathfrak{A}$. 
If $\mathfrak{A}$ is unital we require that $\omega(\mbox{{\bf 1}})=1$.
The states that 
physicists are most familiar with are vector states, that is, if we are 
given a representation $({\cal H},\pi)$ and an element $\psi$ in the 
common domain of all the $a\in \mathfrak{A}$ then $a\mapsto 
<\psi,\pi(a)\psi>_{{\cal H}}$ evidently defines a state. These are 
examples of pure states, i.e. those which cannot be written as convex 
linear combinations of other states. However, the concept of states is 
much more general and includes what physicists would call mixed (or 
temperature) states.
\item[IV.] {\it Automorphisms}\\
An automorphism of a $^\ast-$algebra is an isomorphism of $\mathfrak{A}$
which is compatible with the algebraic structure. If $G$ is a group then 
$G$ is said to be represented on $\mathfrak{A}$ by a group of 
automorphisms $\alpha:\;G\to \mbox{Aut}(\mathfrak{A});\;g\mapsto \alpha_g$
provided that $\alpha_g\circ \alpha_{g'}=\alpha_{gg'}$ for all 
$g,g'\in G$. A state $\omega$ on $\mathfrak{A}$ is said to be invariant 
for an automorphism $\alpha$ provided that $\omega\circ \alpha=\omega$.
It is said to be invariant for $G$ if it is invariant for all 
$\alpha_g,\;g\in G$.
\end{itemize}
The following two structural theorems combine the notions introduced above 
and are of fundamental importance for the construction and analysis of 
representations.
\begin{Theorem}[GNS Construction] \label{th4.1} ~\\
Let $\omega$ be a state on a unital $^\ast-$algebra $\mathfrak{A}$. Then 
there 
are GNS data $({\cal H}_\omega,\pi_\omega,\Omega_\omega)$ consisting of a 
Hilbert space ${\cal H}_\omega$, a cyclic representation $\pi_\omega$ of 
$\mathfrak{A}$ on ${\cal H}_\omega$ and a normed, cyclic vector 
$\Omega_\omega\in {\cal H}_\omega$ such that 
\be \label{4.1}
\omega(a)=<\Omega_\omega,\pi_\omega(a)\Omega_\omega>_{{\cal H}_\omega}
\ee
Moreover, the GNS data are determined by (\ref{4.1}) uniquely up to 
unitary equivalence.
\end{Theorem}
The name GNS stands for Gel'fand -- Naimark -- Segal. The idea is very 
simple. The algebra $\mathfrak{A}$ is in particular a vector space and we 
can equip it with a sesqui -- linear form $<a,b>:=\omega(a^\ast b)$. 
This form is not necessarily positive definite. However, by exploiting the 
Cauchy -- Schwarz inequality $|\omega(a^\ast b)|^2\le 
\omega(a^\ast a)\omega(b^\ast b)$ one convinces oneself that the set
$\mathfrak{I}_\omega$ consisting of the elements of $\mathfrak{A}$ 
satisfying $\omega(a^\ast a)=0$ defines a left ideal. We can thus pass to 
the equivalence classes $[a]=\{a+b;\;b\in \mathfrak{I}_\omega\}$ and
define a positive definite scalar product by 
$<[a],[b]>:=\omega(a^\ast b)$ for which one checks independence of the 
representative. Since $\mathfrak{I}_\omega$ is a left ideal one checks 
that $[a]+[b]:=[a+b],\;z[a]:=[za],\;[a][b]:=[ab]$ are well defined 
operations. Then ${\cal H}_\omega$ is simply the Cauchy completion of the 
vectors $[a]$, the representation is simply $\pi_\omega(a)[b]:=[ab]$ and
the cyclic vector is just given by $\Omega_\omega:=[\mbox{{\bf 1}}]$.
Finally, if $({\cal H}'_\omega,\pi'_\omega,\Omega'_\omega)$ are other 
GNS data then the operator $U:\;{\cal H}_\omega\to {\cal H}'_\omega$ 
defined densely by 
$U\pi_\omega(a)\Omega_\omega:=\pi'_\omega(a)\Omega'_\omega$ is unitary.
\begin{Theorem} \label{th4.2} ~~\\
Let $\omega$ be a state over a unital $^\ast-$algebra $\mathfrak{A}$ which 
is invariant for an element $\alpha\in\mbox{Aut}(\mathfrak{A})$. Then 
there exists a uniquely determined unitary operator $U_\omega$ on the 
GNS Hilbert space ${\cal H}_\omega$ such that
\be \label{4.2}
U_\omega\;\pi_\omega(a)\;\Omega_\omega=\pi_\omega(\alpha(a))\Omega_\omega
\ee
\end{Theorem}
The proof follows from the uniqueness part of theorem \ref{th4.1} applied 
to the alternative data 
$({\cal H}_\omega,\pi_\omega\circ \alpha,\Omega_\omega)$. 
\begin{Corollary} \label{col4.1} ~~\\
Let $\omega$ be a $G-$invariant state on a unital $^\ast-$algebra. Then 
there is a unitary representation $g\mapsto U_\omega(g)$ of $G$ on the 
GNS Hilbert space ${\cal H}_\omega$ defined by 
\be \label{4.3}
U_\omega(g)\;\pi_\omega(a)\;\Omega_\omega:=
\pi_\omega(\alpha_g(a))\;\Omega_\omega
\ee
where $g\mapsto\alpha_g$ is the corresponding automorphism group.
\end{Corollary}
Notice that this means that the group $G$ is represented {\it without 
anomalies}, that is, there are e.g. no central extensions with non -- 
vanishing obstruction cocycle. 

An important concept in connection with a state $\omega$ is its 
{\it folium}. This is defined as the set of states $\omega_\rho$ 
on $\mathfrak{A}$ defined by 
\be \label{4.4}
\omega_\rho(a):=\frac{
\mbox{Tr}_{{\cal H}_\omega}(\rho \pi_\omega(a))}{\mbox{Tr}_{{\cal H}_\omega}
(\rho)}
\ee
where $\rho$ is a positive trace class operator on the GNS Hilbert 
space ${\cal H}_\omega$. 

If $\mathfrak{A}$ is not only a unital $^\ast-$algebra but in fact a 
$C^\ast-$algebra then there are many more structural theorems available.
For instance one can show, using the Hahn -- Banach theorem that 
representations always exist, that every non -- degenerate representation 
is a direct sum of cyclic representations and that every state is 
continuous so that the GNS representations are always by bounded 
operators. While the $C^\ast-$norm implies this huge amount of extra 
structure, a reasonable $C^\ast-$norm on a $^\ast-$algebra is usually very 
hard to guess unless one actually constructs a representation by bounded 
operators. We have thus chosen to keep with the more 
general concept of $^\ast-$algebras. \\
\\
\\
In AQFT \cite{3} one uses the mathematical framework of operator 
algebras, the basics of which we just sketched and combines it with 
the physical concept of locality of nets of local algebras 
${\cal O}\mapsto\mathfrak{A}({\cal O})$. That is, given a background 
spacetime $(M,\eta)$ consisting of a differentiable $D-$manifold and 
a background metric $\eta$, for each open region $\cal O$ one assigns 
a $C^\ast-$algebra $\mathfrak{A}({\cal O})$. These are mutually 
(anti)commuting for spacelike separated (with respect to $\eta$) regions.
This is the statement of the 
most important one of the famous Haag -- Kastler axioms. The
framework is 
ideally suited to formulate and prove all of the structural theorems of 
QFT on Minkowski space and even to a large extent on curved spaces 
\cite{19}. In AQFT one cleanly separates two steps of quantizing a field 
theory, namely first to define a suitable algebra $\mathfrak{A}$ and then 
to study its representations in a second step.

In what follows we apply this framework to the bosonic string. However,
since the string is a worldsheet background independent theory, we must 
use a background independent quantization scheme similar to the programme 
sketched in section \ref{s3}. In particular, we must extend the framework 
just presented as to deal with constraints.

\section{Elements from Loop Quantum Gravity}
\label{s5}

LQG is a canonical approach towards quantum gravity based on Dirac's 
quantization programme for field theories's with constraints. The 
canonical approach is ideally suited to constructing background 
metric independent representations of the canonical commutation relations 
as is needed e.g. in quantum gravity. We will describe this 
programme in some detail below, again experts can safely skip this 
section. As we will see, in its modern form Dirac's programme uses some 
elements of the theory of operator algebras and AQFT but what is different 
from AQFT is that the canonical approach is, by definition, a quantum
theory of the initial data, that is, operator valued distributions are 
smeared with test functions supported in $(D-1)-$dimensional slices rather 
than $D-$dimensional regions. This is usually believed to be a bad 
starting point in AQFT because of the singular behaviour of the 
$n-$point Wightman distributions of interacting scalar fields in 
perturbation theory when smeared with ``test functions'' supported in 
lower dimensional submanifolds. The way out of this ``no-go theorem''
is twofold: 1. In usual perturbation theory one uses very specific 
(Weyl) algebras and corresponding Fock representations to formulate the 
canonical commutation relations but the singular 
behaviour might be different for different algebras and their associated 
representations. 2. In a reparameterization invariant theory such as 
General Relativity or string theory all 
observables are by definition time independent, see e.g. the {\bf 
Pohlmeyer Charges} constructed above, so that smearing in the 
additional time direction just multiplies the observables with a 
constant and thus does not change their singularity structure.

As a pay -- off, in contrast to AQFT the framework of LQG does not need a 
background spacetime as we will see below.\\
\\
\\
We assume to be given a (possibly infinite dimensional) phase 
space $\cal M$ with a Poisson bracket $\{.,.\}$ (technically a strong 
symplectic structure). Furthermore, we assume to be given a set of
first class constraints $C_I$ on $\cal M$ which we take to be real valued
w.l.g. (pass to real and imaginary part if necessary). Here the index set 
${\cal I}$ in which the indices $I$ take values can be taken to be 
discrete. This seems not to be the most general situation because e.g. in the 
case of the string we have the continuous label $x\in S^1$, e.g. for the 
constraints $C(x)$, however, the constraints always must be smeared 
with test functions. In fact, let $b_I$ be an orthonormal basis of smooth
functions of  
$L_2(S^1,dx/(2\pi))$ (e.g. $b_I(x)=\exp(iIx),\;I\in\Zl$) and define 
$C_I:=<b_I,C>$. Then $C_I=0\;\forall I$ is equivalent with $C(x)=0$
for a.a. $x\in S^1$ and since $C(x)$ is classically a continuous function 
it follows that $C(x)=0$ for all $x\in S^1$. That the constraints be first 
class means that there are 
structure functions $f_{IJ}\;^K$ on $\cal M$ (not necessarily independent
of $\cal M$) such that $\{C_I,C_J\}=f_{IJ}\;^K C_K$. We may also have a 
Hamiltonian $H$ (not a linear combination of the $C_I$) which is supposed 
to be gauge invariant, that is, $\{C_I,H\}=0$ for all $I$. 

Given this set -- up, the Canonical Quantization Programme consists, 
roughly, of the following steps:
\begin{itemize}
\item[I.] {\it Kinematical Poisson Algebra $\mathfrak{P}$ and 
Kinematical Algebra $\mathfrak{A}$}\\
The first choice to be made is the selection of a suitable 
Poisson$^\ast$ subalgebra of $C^\infty({\cal M})$. This means that we 
must identify a subset of functions $f\in C^\infty({\cal M})$ which 
is closed under complex conjugation and Poisson brackets and which 
separates the points of $\cal M$. This choice is guided by gauge 
invariance, that is, the functions $f$ should have a simple behaviour 
under the gauge transformations generated by the Hamiltonian vector fields 
of the $C_I$ on $\cal M$. At this point it is not important that 
$\mathfrak{P}$ consists of bounded functions. However, when promoting
$\mathfrak{P}$ to an operator algebra, it will be  
convenient to choose bounded functions of the elements $p$ of 
$\mathfrak{P}$, say the usual Weyl elements $W=\exp(itp),\; t\in \Rl$,
and to define the algebra of the $W$ to be given by formally imposing the 
canonical commutation relations among the $p$, namely that commutators are
given by $i\hbar$ times the Poisson bracket and that the operators 
corresponding to real valued $p$ are 
self-adjoint. This has the advantage of 
resulting in bounded operators $W$ which avoids domain questions later on.
We will denote the resulting $^\ast-$algebra generated by the operators
$W$ by $\mathfrak{A}$.  
\item[II.] {\it Representation Theory of $\mathfrak{A}$}\\
We study the representation theory of $\mathfrak{A}$, that is, 
all $^\ast-$algebra morphisms 
$\pi:\;\mathfrak{A}\to {\cal B}({\cal H}_{Kin})$ where 
${\cal B}({\cal H}_{Kin})$ denotes the $C^\ast-$algebra of bounded 
operators on a kinematical Hilbert space ${\cal H}_{Kin}$. In particular,
\be \label{5.1}
\pi(\bar{f})=[\pi(f)]^\dagger,\;\;[\pi(f),\pi(f')]=i\hbar\pi(\{f,f'\})
\ee
and the corresponding (Weyl) relations for the exponentiated elements.
Guiding principles here are again gauge invariance and (weak) continuity.
Moreover, the representation should be irreducible on physical 
grounds (otherwise we have superselection sectors implying that the 
physically relevant information is already captured in a closed subspace).
Operator algebra theoretic methods such as the GNS construction are of 
great importance here.
\item[III.] {\it Selection of a suitable kinematical representation}\\
Among all possible representations $\pi$ we are, of course, only 
interested in those which support the constraints $C_I$ as operators.
Since, by assumption, $\mathfrak{A}$ separates the points of $\cal M$
it is possible to write every $C_I$ as a function of the 
$f\in \mathfrak{A}$, however, that function is far from unique due to
operator ordering ambiguities and in field theory usually involves 
a limiting procedure (regularization and renormalization). We must make 
sure that the resulting limiting operators $\pi(C_I)$ are 
densely defined and closable (i.e. their adjoints are also 
densely defined) on a suitable domain of ${\cal H}_{Kin}$. This step
usually severely restricts the abundance of representations. 
Alternatively,
in rare cases it is possible to quantize the finite gauge transformations 
generated by the classical constraints provided they exponentiate to a 
group. This is actually what we will do in this paper.
\item[IV.] {\it Solving the Quantum Constraints}\\
There are essentially two different strategies for solving the 
constraints, the first one is called ``Group Averaging'' and the second 
one is called ``Direct Integral Decomposition''. The first method makes 
additional assumtions about the structure of the quantum constraint 
algebra while the second does not and is therefore of wider applicability. 

For ``Group Averaging'' \cite{19a}
the first assumption is that the $\pi(C_I)$ are 
actually self-adjoint
operators on ${\cal H}_{Kin}$ and that the structure functions 
$f_{IJ}\;^K$ are actually constants on $\cal M$. In this case, under 
suitable functional analytic assumptions we can define the unitary 
operators 
\be \label{5.2}
U(t):=\exp(i\sum_I t^I \pi(C_I))
\ee
where the parameters take range in a subset of $\Rl$ depending on the 
$\pi(C_I)$ in such a way that the $U(t)$ define a unitary representation 
of the Lie group $G$ determined by the Lie algebra generators $C_I$. 
In particular, this means that there is no anomaly, i.e. 
$[\pi(C_I),\pi(C_J)]=i\hbar f_{IJ}\;^K \pi(C_K)$. The 
second assumption is that $G$ has an invariant (not necessarily finite)
bi -- invariant Haar measure $\mu_H$. In this case we may define 
an anti-linear rigging map
\be \label{5.3}   
\rho:\;{\cal H}_{Kin}\to{\cal H}_{Phys};\; \psi\mapsto 
\int_G\; d\mu_H(t)\; <U(t)\psi,.>_{{\cal H}_{Kin}}
\ee
with physical inner product
\be \label{5.4}
<\rho(\psi),\rho(\psi')>_{{\cal H}_{Phys}}:=[\rho(\psi')](\psi)
\ee
Notice that $\rho(\psi)$ defines a distribution on (a dense subset of)
${\cal H}_{Kin}$ and solves the constraints in the sense that
$[\rho(\psi)](U(t)\psi')=[\rho(\psi)](\psi')$ for all $t\in G$. 
Moreover, given any kinematical algebra element $O\in \mathfrak{A}$ we may 
define a corresponding Dirac observable by
\be \label{5.5}
[O]:=\int_G\; d\mu_H(t)\;U(t) \; O \; U(t)^{-1}
\ee

Let us now come to the ``Direct Integral Method'' \cite{9a,13a}. Here we 
do not need to
assume that the $\pi(C_I)$ are self-adjoint. Also the structure functions 
$f_{IJ}\;^K$ may have non-trivial dependence on $\cal M$. In contrast to 
the string, this is actually the case in 4D General Relativity and 
for higher p -- brane theories which is 
why only this method is available there. Let us now consider an operator
valued positive definite matrix $\hat{Q}^{IJ}$ such that the $\MCOW$
\be \label{5.6}
\MCO:=\frac{1}{2}\sum_{I,J} [\pi(C_I)]^\dagger \hat{Q}_{IJ} [\pi(C_J)]
\ee
is densely defined. Obvious candidates for $\hat{Q}_{IJ}$ are 
quantizations $\pi(Q^{IJ})$ of postive definite ${\cal M}-$valued matrices 
with suitable decay behaviour in ${\cal I}-$space. Then, since $\MCO$
is positive by construction it has self-adjoint extensions (e.g. its
Friedrichs extension \cite{12}) and its spectrum is supported on the 
positive real line. Let $\lambda^{\MC}_0=\inf \sigma(\MCO)$ be the 
minimum of the spectrum of $\MCO$ and redefine $\MCO$ by 
$\MCO-\lambda^{\MC}_0\mbox{id}_{{\cal H}_{Kin}}$. Notice that 
$\lambda^{\MC}_0<\infty$ by assumption and proportional to $\hbar$
by construction. We now use the well-known fact that ${\cal H}_{Kin}$,
if separable, can be represented as a direct integral of Hilbert spaces 
\be \label{5.7}
{\cal H}_{Kin}\cong\int_{\Rl^+}^\oplus\;d\mu(\lambda) {\cal 
H}^\oplus_{Kin}(\lambda)
\ee
where $\MCO$ acts on ${\cal H}^\oplus_{Kin}(\lambda)$ by multiplication by 
$\lambda$. The measure $\mu$ and the scalar product on
${\cal H}^\oplus_{Kin}(\lambda)$ are induced by the scalar product on 
${\cal H}_{Kin}$. The physical Hilbert space is then simply 
${\cal H}_{Phys}:={\cal H}^\oplus_{Kin}(0)$ and 
Dirac observables are now constructed from bounded 
self-adjoint operators 
$O$ on ${\cal H}_{Kin}$ by the {\it ergodic mean}
\be \label{5.8}
[O]=\lim_{T\to\infty} \frac{1}{2T} \int_{-T}^T\;dt\; e^{it\MCO}\;
O\; e^{-it\MCO}
\ee
and they induce bounded self-adjoint operators on ${\cal H}_{Phys}$.

Notice that both methods can be combined. Indeed, it may happen that 
a subset of the constraints can be solved by group averaging methods while 
the remainder can only be solved by direct integral decomposition methods.
In this case, one will construct an intermdediate Hilbert space which 
the first set of constraints annihilates and which carries a 
representation of the second set of constraints. This is actually the 
procedure followed in LQG and it will be convenient to do adopt this 
``solution in two steps'' for the string as well in our companion paper 
\cite{11} for curved target spaces. For the purposes of this paper, group 
averaging will be sufficient. 
%
%
\item[V.] {\it Classical Limit}\\
Notice that our construction is entirely non-perturbative, there are 
no (at least not necessarily) Fock spaces and there is no perturbative 
expansion (Feynman diagrammes) even if the theory is interacting.
While this is attractive, the prize to pay is that the 
representation ${\cal H}_{Kin}$ to begin with and also the final physical
Hilbert space ${\cal H}_{Phys}$ will in general be far removed from any
physical intuition. Hence, we must make sure that what we have constructed 
is not just some mathematical object but has, at the very least, the 
classical theory as its classical limit. In particular,
if classical Dirac observables are known, then the quantum Dirac 
observables (\ref{5.8}) and (\ref{5.8}) should reduce to them in the 
classical limit. To address such questions one must develop
suitable semiclassical tools. 
\end{itemize}
We see that the construction of the quantum field theory in AQFT as well
as in LQG is nicely separated: First one constructs the algebra and then 
its representations. What is new in LQG is that it also provides a 
framework for dealing with constraints. Now LQG is more than just offering 
a framework, this framework has been carried out systematically with quite 
some success, so far until step IV, see e.g. \cite{2} for exhaustive 
reports. That is, not only did one pick a suitable kinematical algebra 
$\mathfrak{A}$ but also constructed a non-trivial representation thereof
by use of the GNS construction 
which moreover supports the quantum constraints. In what follows we will
apply this programme systematically to the closed, bosonic string on flat 
backgrounds. Due to the flatness of the backgorund and the fact that the 
constraints close with structure constants rather than structure 
functions we can carry out 
our programme rather effortlessly in a novel representation that is not 
available in LQG. In our companion paper \cite{11} we will stick closer to 
the kind of representations used in LQG.

\section{The Closed, Bosonic LQG -- String on Flat Target Spaces}
\label{s6}

We can now finally pick up the fruits of our efforts and combine sections 
three through five to construct the LQG -- String. We will do this by 
systematically following the canonical quantization programme.

\subsection{General Representation Problem}
\label{s6.1}

Before we go into details, let us state the general representation 
problem:\\
\\
In view of the the theory of invariants of section \ref{s3}
we take as our classical Poisson $^\ast-$subalgebra $\mathfrak{P}$ the 
algebra of the $Y_\pm(f)$ of section \ref{s3}. By varying the smooth
smearing functions $f_\mu$ we can extract $\pi_\mu(x)$ as well as
$X^{\mu\prime}(x)$ for all $x\in S^1$. Only a global constant $X^\mu_0$
remains undetermined. Furthermore, we include into $\mathfrak{P}$ the 
Lorentz generators $J_{\mu\nu}$ which allows us to extract $X^\mu_0$
up to the the translation freedom $X^\mu_0\to t\ell_s^2 p^\mu,\; t\in 
\Rl$. This is the only information which we cannot extract from 
$\mathfrak{P}$ which is admissable in view of the fact that the physical 
invariants of the string do not depend on it. In this sense, 
$\mathfrak{P}$ seperates the points of the unconstrained phase space
$\cal M$.

Next we construct from $\mathfrak{P}$ bounded functions on $\cal M$ 
which still separate the points and 
promote them to operators by asking that Poisson brackets and complex 
conjugation on $\mathfrak{P}$ be promoted to commutators divided by 
$i\hbar$ and the adjoint respectively. Denote the resulting 
$^\ast-$algebra by $\mathfrak{A}$. 

The automorphism groups $\alpha^\pm_\varphi,\;\varphi\in \mbox{Diff}(S^1)$
generated by the Virasoro constraints as well as the Poincar\'e 
automorphism group $\alpha_{a,L}$ extend naturally from 
$\mathfrak{P}$ to $\mathfrak{A}$ simply by 
$\alpha_.(W(Y_\pm))=W(\alpha_.(Y_\pm))$. A general representation of 
$\mathfrak{A}$ should now be such that the automorphism groups $\alpha_.$
are represented by inner automorphisms, that is, by conjugation by unitary
operators representating the corresponding group elements. Physically the
representation property amounts to an anomaly -- free implementation of 
both the local gauge group and the global symmetry group while unitarity
implies that expectation values of gauge invariant or Poincar\'e invariant
observables does not depend on the gauge or frame of the measuring state.
Finally, the representation should be irreducible or at least cyclic.   

As we have sketched in section \ref{s4}, a powerful tool to arrive at such
representations is via the GNS construction. Hence we will be looking at 
representations that arise from a positive linear functional $\omega$ on 
$\mathfrak{A}$ which is invariant under all the automorphisms $\alpha_.$. 
Since $\mathfrak{A}$ is not (naturally) a $C^\ast-$algebra, this is not 
the most general representation but it is important subclass thereof.

These representations are still kinematical, that is, the corresponding
GNS Hilbert space ${\cal H}_{Kin}:={\cal H}_\omega$ still represents 
$\mathfrak{A}$ and 
not only physical observables. To arrive at the physical Hilbert space 
${\cal H}_{Phys}$ we make use of the group averaging method outlined in 
section \ref{s5}:\\
Let $\Phi_{Kin}:=\Phi_\omega$ be the dense subspace of ${\cal H}_\omega$
defined by the $\pi_\omega(a)\Omega_\omega,\;a\in \mathfrak{A}$. 
Let $\Phi^\ast_\omega$ be the algebraic dual of $\Phi_\omega$ (linear
functionals without continuity requirements).
Suppose that $\mu$ is a translation bi -- invariant measure 
on Diff$(S^1)$ such that the following anti -- linear map
exists
\be \label{6.1}
\eta_\omega:\;\Phi_\omega\to\Phi_\omega^\ast;\;
\pi_\omega(a)\Omega_\omega\mapsto 
\int_{[\mbox{Diff}(S^1)]^2}\;
d\mu(\varphi^+)\; d\mu(\varphi^-)\; 
<U_\omega^+(\varphi^+)U_\omega^-(\varphi^-)
\pi_\omega(a)\Omega_\omega,.>_{{\cal H}_\omega}
\ee
where we have used that the $U^+_\omega$ commute with the $U^-_\omega$. 
Consider the sesqui -- linear form
\ba \label{6.2}
&&<\eta_\omega(\pi_\omega(a)\Omega_\omega), 
\eta_\omega(\pi_\omega(b)\Omega_\omega)>_{Phys} 
\nonumber\\
&:=& 
[\eta_\omega(\pi_\omega(b)\Omega_\omega)](\pi_\omega(a)\Omega_\omega)
\nonumber\\
&=& \int_{[\mbox{Diff}(S^1)]^2}\;
d\mu(\varphi^+)\; d\mu(\varphi^-)\; 
<U_\omega^+(\varphi^+)U_\omega^-(\varphi^-)
\pi_\omega(b)\Omega_\omega,\pi_\omega(a)\Omega_\omega>_{{\cal H}_\omega}
\ea
Fortunately, the steps to rigorously define the rigging map (\ref{6.1}) 
and the physical inner product (\ref{6.2}) have been completed for a 
general theory already in \cite{24} so that we can simply copy those 
results.

This concludes the definition of our general representation problem. We 
now pass to a specific choice for the kinematical algebra.

\subsection{Concrete Implementation}
\label{s6.2}

We now propose a particular choice of $\mathfrak{A}$ based on our 
experiences with background independent representations that proved useful 
in LQG \cite{2}. Then, rather than studying its general representation 
theory, we confine ourselves in this paper to constructing a specific 
positive linear functional
and show that the {\bf Pohlmeyer Charges} can be defined on the 
corresponding physical Hilbert space in the sense of (\ref{6.2}).  
This then proves that the programme outlined in section \ref{s6.1} 
has non -- trivial solutions.\\
\\
Let $I$ denote Borel subsets of $S^1$, that is, sets generated by 
countable 
intersections and unions from closed subsets of $[0,2\pi)$ (with periodic
identifications). We consider instead of general smearing functions
$f_\mu(x)$ the more specific ones 
\be \label{6.5}
f^{I,k}_\mu(x)=k_\mu\chi_I(x)
\ee
where $k_\mu$ are real valued numbers with dimension cm$^{-1}$ and 
$\chi_I$ denotes the characteristic function of $I$. The corresponding 
smeared $Y^k_\epsilon(I):=Y_\epsilon(f^{I,k}),\;\epsilon=\pm 1$ satisfy 
the Poisson algebra
\be \label{6.6}
\{Y^k_\epsilon(I),Y^{k'}_{\epsilon'}(I')\}=
\epsilon\alpha'\delta_{\epsilon,\epsilon'}\eta^{\mu\nu} k_\mu k'_\nu
\{[\chi_I]_{\partial I'}-[\chi_{I'}]_{\partial I}\}
=:\alpha'\alpha(\epsilon,k,I;\epsilon,k',I')
\ee
where the boundary points in $\partial I$ of $I$ are signed according to 
whether they are right or left boundaries of closed intervals.
The functionals $Y^k_\epsilon(I)$, together with the $J_{\mu\nu}$, 
evidently still separate the points of $\cal M$ and form a closed
Poisson $^\ast-$subalgebra. 

The kinematical quantum algebra will be generated from the 
{\it Weyl elements}
\be \label{6.7}
\hat{W}^k_\epsilon(I):=e^{i\hat{Y}^k_\epsilon(I)}
\ee
for which we require canonical commutation relations induced from
\be \label{6.8}
{[}\hat{Y}^k_\epsilon(I),\hat{Y}^{k'}_{\epsilon'}(I')]=
i\hbar\widehat{\{Y^k_\epsilon(I),Y^{k'}_{\epsilon'}(I')\}}
\ee
It follows from the Baker -- Campbell -- Hausdorff formula that
\be \label{6.9}
W^k_\epsilon(I)W^{k'}_{\epsilon'}(I')=
\exp(-i\ell_s^2\alpha(\epsilon,k,I;\epsilon,k',I')/2)
\; \exp(i[Y^k_\epsilon(I)+Y^{k'}_{\epsilon'}(I')])
\ee
in particular, Weyl elements with $\epsilon\not=\epsilon'$ are commuting.
Since 
\be \label{6.10}
Y^k_\pm(I)+Y^{k'}_\pm(I')=
Y^{k+k'}_\pm(I\cap I')+Y^{k}_\pm(I-I')+Y^{k'}_\pm(I'- I)
\ee
we see that a general element of $\mathfrak{A}$ can be written as a 
finite, complex linear combinations of elements of the form
\be \label{6.11}
W_+^{k_1..k_M}(I_1,..,I_M) W_-^{l_1..l_N}(J_1,..,J_N)
\ee
where
\be \label{6.12}
W_\pm^{k_1,..,k_M}(I_1,..,I_M)=\exp(i[\sum_{m=1}^M Y^{k_{m}}_\pm(I_m)])
\ee
where all the $I_m$ are non -- empty and mutually non -- overlapping 
(they intersect at most in boundary points). These considerations
motivate the following definition.
\begin{Definition} \label{def6.1}  ~~~\\
i)\\
A momentum network $s=(\gamma(s),k(s))$ is a pair consisting of a finite 
collection $\gamma(s)$ of non -- empty and non -- overlapping 
(up to boundary points) closed, 
oriented 
intervals $I$ of $S^1$, together with an assignment $k(s)$ of momenta
$k_\mu^I(s)$ for each interval. We will use the notation $I\in \gamma$
if $I$ is an element of the collection of intervals $\gamma$ to 
which we also refer as edges.\\
ii)\\
A momentum network operator with parity $\epsilon$ is given by
\be \label{6.13}
W_\epsilon(s)=\exp(i[\sum_{I\in\gamma(s)} Y^{k^I(s)}_\epsilon(I)])
\ee
\end{Definition}
This definition has a precise counterpart in LQG in terms of spin networks 
and spin network (multiplication) operators where the intervals are 
replaced by oriented edges of a graph which is the counterpart of the 
collection of intervals here. The spin quantum numbers of LQG are replaced 
by the momentum labels of the string. 

The general {\it Weyl Relations} are then given by
\ba \label{6.14}
W_+(s_1)\; W_-(s'_1) W_+(s_2) W_-(s_2') &=&
e^{-\frac{i}{2}\ell_s^2[\alpha(s_1,s_2)-\alpha(s'_1,s_2')]}\;
W_+(s_1+s_2)\; W_-(s_1'+s_2')
\nonumber\\
{[}W_+(s_1) \; W_-(s_2)]^\ast &=& W_+(-s_1) \; W_-(-s_2)
\ea
where
\be \label{6.15}
\alpha(s_1,s_2)=
\sum_{I_1\in \gamma(s_1)}\;\sum_{I_2\in \gamma(s_2)}\;
{[}\eta^{\mu\nu} k^{I_1}_\mu(s_1) k^{I_1}_\mu(s_1)]\alpha(I_1,I_2)
\ee
with $\alpha(I,J)=[\chi_I]_{|\partial J}-[\chi_J]_{|\partial J}$.
The notation $s_1+s_2$ means that we decompose all the 
$I_1\in \gamma(s_1)$ and $I_2\in \gamma(s_2)$ into their unique, maximal,
mutually non -- overlapping segments and assign the momentum 
$k^{I_1}(s_1)+K^{I_2}(s_2)$ to the segment $I_1\cap I_2$ and 
$k^{I_1}(s_1),\;k^{I_2}(s_2)$ respectively to the segments 
$I_1-\gamma(s_2)$ and $I_2-\gamma(s_1)$ respectively. Likewise 
$-s$ is the same as $s$ just that $I\in \gamma(s)$ is assigned $-k^I(s)$.
See (\ref{6.10}) for an example. The relations (\ref{6.14}) are the direct 
analog of the holonomy -- flux Weyl algebra underlying LQG \cite{21}.
Notice that the ``holonomies'' along intervals $I$
\be \label{6.15a}
h^k_I(X):=\exp(2ik_\mu \int_I dx X^{\mu\prime})=W^k_+(I) W^{-k}_-(I)
\ee
are mutually commuting due to the difference in the sign of the phase 
factor in (\ref{6.14}). The holonomies define a maximal Abelean subalgebra
in $\mathfrak{A}$. The same is true for the exponentiated ``fluxes''
\be \label{6.15b}
F^k_I(\pi):=\exp(2ik^\mu \int_I dx \pi_\mu)=W^k_+(I) W^{k}_-(I)
\ee
 
We can combine the local gauge group generated by the Virasoro constraints 
and the global Poincar\'e group to the total group
\be \label{6.16}
G:=\mbox{Diff}(S^1)\times \mbox{Diff}(S^1) \times {\cal P}
\ee
given by the direct product of two copies of the diffeomorphism group of 
the circle and the Poincar\'e group which itself is the semi -- direct
product of the Lorentz group with the translation group. It has the 
following action on our Weyl algebra $\mathfrak{A}$ by the automorphisms 
derived in section \ref{s3}
\be \label{6.17}
\alpha_{\varphi_+,\varphi_-,(L,a)}(W_+(s_1)\; W_-(s_2))
=\alpha_{\varphi_+,{\rm id},(L,0)}(W_+(s_1))\;
\alpha_{{\rm id},\varphi_-,(L,0)}(W_-(s_2))
=W_+([\varphi_+,L]\cdot s_1)\;W_-([\varphi_-,L]\cdot s_2))
\ee
where 
\ba \label{6.18}
(\varphi,L)\cdot s
&=& (\gamma((\varphi,L)\cdot s),\{k^I((\varphi,L)\cdot s)\}_{I\in 
\gamma((\varphi,L)\cdot s)})
\nonumber\\
\gamma((\varphi,L)\cdot s) &:=& \varphi(\gamma(s))
\nonumber\\
k^{\varphi(I)}((\varphi,L)\cdot s) &:=& L\cdot k^I(s),\;I\in \gamma(s)
\ea
In other words, the diffeomorphism maps the intervals to their 
diffeomorphic image but leaves the momenta unchanged while the Lorentz
transformation acts only on the momenta. The translations have no effect
on the Weyl elements because they only depend on $X^\mu$ through the 
derivatives $X^\prime$.

In order to find representations of $\mathfrak{A}$ via the GNS 
construction we notice that 
\be \label{6.19}
\mathfrak{A}=\mathfrak{A}_+\;\otimes \mathfrak{A}_-
\ee
is the tensor product of two Weyl algebras which are isomorphic up to the
sign difference in the phase in (\ref{6.14}). Now we can use the lemma
that if $\omega_\pm$ is a positive linear functional on the 
$^\ast-$algebra $\mathfrak{A}_\pm$ then $\omega:=\omega_+ \otimes 
\omega_-$ defined by $\omega(a_+\otimes a_-):=\omega_+(a_+)\omega_-(a_-)$
is a positive linear functional on $\mathfrak{A}_+\otimes \mathfrak{A}_-$,
see e.g. \cite{20} for a simple proof. Hence it suffices to specify
$\omega_\pm$ separately.

\subsection{A Specific Example}
\label{s6.3}

We should now enter the general representation theory of 
$\mathfrak{A}_\pm$ or at least the subclass of cyclic and $G-$invariant
representations that arise via the GNS construction. 
Since we have not made any continuity assumptions 
about the representation of the one parameter unitary groups $t\mapsto 
W_\pm(ts)$ (where 
$ts$ is the same as $s$ just that the momenta are rescaled by $t$)
whithout which even the famous Stone -- von Neumann uniqueness theorem
for the representation theory of the Weyl algebra underlying simple 
quantum mechanics fails \cite{5,23}, we expect this problem to be rather
complex and we leave it as an important project for further research.
In this paper we will content ourselves with giving just one non -- 
trivial example. Here it is:
\be \label{6.20}
\omega_\pm(W_\pm(s)):=\delta_{s,0}
\ee
where $s=0$ denotes the trivial momentum network consisting of no 
intervals and zero momenta. This functional is tailored after the positive 
linear functional underlying the Ashtekar -- Lewandowski representation
for LQG \cite{24aa} which only recently \cite{21} has been shown to be the 
unique (cyclic, spatially diffeomorphism invariant and in part weakly 
continuous) representation of the LQG Weyl algebra. Expression
(\ref{6.20}) looks rather 
trivial at first sight but the rest of the paper is devoted to 
demonstrating that it contains some interesting structure.

Let us first check that it is indeed a $G-$invariant positive linear 
functional. First of all
\be \label{6.21}
\omega_\pm(\alpha_g(W_\pm(s)))=\omega_\pm(W_\pm(g\cdot s))
=\delta_{g\cdot s,0}=\delta_{s,0}=\omega_\pm(W_\pm(s))
\ee
since the $G-$action on the momentum networks preserves the cases $s=0$ 
and $s\not=0$. It follows that we obtain a unitary representation of 
$G$ defined by
\be \label{6.21bb}
U_\omega(g)\pi_\omega(W_+(s) W_-(s))\Omega_\omega
=\pi_\omega(W_+(g\cdot s) W_-(g\cdot s))\Omega_\omega
\ee
Next consider a generic element of $\mathfrak{A}_\pm$
given by
\be \label{6.22}
a_\pm=\sum_{n=1}^N z_n W_\pm(s_n)
\ee
where the $s_n$ are mutually different. Then 
\ba \label{6.23}
\omega_\pm((a_\pm)^\ast a_\pm) 
&=& \sum_{m,n}\; \bar{z}_m\;z_n\; \omega_\pm([W_\pm(s_m)]^\ast\; 
W_\pm(s_n))
\nonumber\\
&=& \sum_{m,n}\; \bar{z}_m\;z_n\; \omega_\pm(W_\pm(-s_m)\; W_\pm(s_n))
\nonumber\\
&=& \sum_{m,n}\; \bar{z}_m\;z_n \;
e^{\pm\frac{i}{2}\ell_s^2\alpha(s_m,s_n)}\;
\omega_\pm(W_\pm(s_n-s_m))
\nonumber\\
&=& \sum_{m,n}\; \bar{z}_m\;z_n \;
e^{\pm\frac{i}{2}\ell_s^2\alpha(s_m,s_n)}\;
\delta_{s_n-s_m,0}
\nonumber\\
&=& \sum_{n=1}^N |z_n|^2
\ea
since $\alpha(s,s)=0$. Thus $\omega_\pm([a_\pm]^\ast \;a_\pm)\ge 0$ and 
equality occurs only for $a_\pm=0$. Hence the functional is definite and 
there is no null ideal to be divided out, see section \ref{s4}. Thus
the GNS Hilbert space ${\cal H}_{\omega_\pm}$ is simply the Cauchy 
completion of the algebra $\mathfrak{A}_\pm$ considered as a vector space 
with cyclic vector $\Omega_{\omega_\pm}=\mbox{{\bf 1}}=W_\pm(0)$ and 
representation
$\pi_{\omega_\pm}(a_\pm)=a_\pm$. The GNS Hilbert space for $\mathfrak{A}$
is of course the tensor product 
${\cal H}_\omega={\cal H}_{\omega_+}\otimes {\cal H}_{\omega_-}$.
 
One might wonder whether this representation is unitarily equivalent
to the Ashtekar -- Isham -- Lewandowski representation 
\cite{24aa} for the string, or rather its direct analog for scalar
fields \cite{24ab,11a,24ac} 
which we are 
going to consider in great detail in the companion paper \cite{11}. 
If that were the case then we could identify $\Omega_\omega$ with the 
constant state {\bf 1}, the operators 
$A(s):=\pi_\omega(W_+(s/2))\pi_\omega(W_-(s/2))^{-1}$ would be 
multiplication ``holonomy'' operators on ${\cal H}_\omega=L_2(\ab,d\mu_0)$ 
where 
$\ab$ is a certain space of distributional ``connections'' $X^{\mu\prime}$
and the operators $\exp(i\pi(s)):=
\pi_\omega(W_+(s/2))\pi_\omega(W_-(s/2))$ would be exponentials of 
derivative operators with $\exp(i\pi(s))\Omega_\omega=\Omega_\omega$.
But this is not the case since $\omega(\exp(i\pi(s))=\delta_{s,0}$ while 
clearly $\omega_{AIL}(\exp(i\pi(s)))=1$. In the Ashtekar -- 
Isham -- Lewandowski 
representation the derivative operators exist (the corresponding 
one parameter unitary Weyl 
groups are weakly continuous) while in our representation the 
derivative operators do not exist (the Weyl groups are discontinuous).
Thus the corresponding GNS representations are unitarily inequivalent.

\subsection{Mass Spectrum}
\label{s6.4}

According to (\ref{6.21bb}) all states are translation invariant if we 
define 
a unitary representation of $G$, as in section \ref{s4}, by
\be \label{6.21a}
U_\omega(g)\pi_\omega(b)\Omega_\omega=\pi_\omega(\alpha_g(b))\Omega_\omega
\ee
for all $b\in \mathfrak{A}$ 
and since, as we will see, all invariant charges are constructed 
from the $\pi_\omega(W_\pm(s))$ by taking certain limits, also states 
created by the charges will be translation invariant. Hence the
total momentum and hence the mass of our states vanishes identically in 
drastic contrast with usual string theory. We now generalize our 
construction as to include {\it arbitrary non -- 
negative mass spectrum (tachyon -- freeness)}.

We introduce a D -- parameter family of states $p\mapsto \omega_p$ on
$\mathfrak{A}$ which as far as $\mathfrak{A}$ is concerned are just 
copies of our $\omega:=\omega_0$ constructed in the previous section.
However, we generalize (\ref{6.21a}) to
\be \label{6.21b}
V_\mu(g)\pi_{\omega_p}(b)\Omega_{\omega_p}=e^{i a\cdot p}
\pi_{\omega_{L^T p}}(\alpha_g(b))\Omega_{\omega_{L^T p}}
\ee
where $L^T$ is the transpose of the Lorentz transformation datum in
$g:=(\varphi_+,\varphi_-,L,a)$ and $({\cal H}_{\omega_p},\pi_{\omega_p},
\Omega_{\omega_p})$ are the GNS data associated with $\omega_p$. Of 
course, all these GNS data are mutually unitarily equivalent. It is easy 
to check that (\ref{6.21b}) satisfies the representation property.

Let $\overline{V}_+$ be the closure of the interior of the future 
lightcone and let $\mu$ be a
quasi -- $\mathfrak{L}^+_0$ -- invariant probability measure on $\Rl^D$ 
with support 
in $\overline{V}_+$. Here
$\mathfrak{L}^+_0$ denotes the connected component of $SO(1,D-1)$ which 
preserves the sign of $p^0$ (proper orthochronous Lorentz group). Recall
that a measure $\mu$ on a measurable space $\mathfrak{X}$ is said to be  
quasi -- invariant for a group $G$ acting on $\mathfrak{X}$ if $\mu$ and 
all its translates have the same measure zero sets. 
In other words, if 
$E$ is a measurable set of $\mathfrak{X}$ and $g\cdot E$ is the 
translation 
of $E$ by $g\in G$ then $\mu(E)=0$ implies $\mu_g(E):=\mu(g\cdot E)=0$
for all $g\in G$. This condition implies that the translated measures 
are mutually absolutely continuous and that the Radon -- Nikodym 
derivative $d\mu_p/d\mu$ is a well -- defined, positive 
$L_1(\mathfrak{X},d\mu)$ function. The choice of $\mu$ is not very 
important for the theory of induced representations as sketched below in 
the sense that mutally absolutely continuous measures lead to unitarily
equivalent representations. See e.g. \cite{24a} for all the 
details.

We now construct the direct integral of the Hilbert spaces 
${\cal H}_{\omega_p}$, that is,
\be \label{6.21c}
{\cal H}_\mu:=\int_{\Rl^D}^\oplus\;d\mu(p)\; {\cal H}_{\omega_p}
\ee
The definition of (\ref{6.21c}) is as follows: 
Let $\mathfrak{X}$ be a locally compact space, $\mu$ a 
measure on $\mathfrak{X}$ and $x\mapsto {\cal H}_x$ an assignment 
of Hilbert spaces such the function $x\mapsto n_x$, where $n_x$ 
is the dimension of of ${\cal H}_x$, is measurable. 
It follows that the sets $\mathfrak{X}_n=\{x\in \mathfrak{X};\;n_x=n\}$,
where $n$ denotes any cardinality, are measurable. Since Hilbert spaces
whose dimensions have the same cardinality are unitarily equivalent we may
identify all the ${\cal H}_x,\;n_x=n$ with a single ${\cal H}_n$. We now 
consider maps
\be \label{6.21d}
\xi:\; \mathfrak{X}\to \prod_{x\in \mathfrak{X}} {\cal H}_x;\;\;
x\mapsto (\xi(x))_{x\in \mathfrak{X}}
\ee
subject to the following two constraints: \\
1. The maps $x\mapsto <\xi,\xi(x)>_{{\cal H}_n}$ are measurable for
all $x\in \mathfrak{X}_n$ and all $\xi \in {\cal H}_n$. \\
2. If
\be \label{6.21e}
<\xi_1,\xi_2>:=\sum_n \int_{\mathfrak{X}_n}\; d\mu(x) 
<\xi_1(x),\xi_2(x)>_{{\cal H}_n}
\ee
then $<\xi,\xi> <\infty$.\\
The completion of the space of maps (\ref{6.21d}) in the inner product 
(\ref{6.21e}) is called the direct integral of the ${\cal H}_x$ with 
respect to $\mu$ and one writes
\be \label{6.21f}
{\cal H}_\mu=\int_{\mathfrak{X}}^\oplus\;d\mu(x)\; {\cal H}_x,\;\;\;
<\xi_1,\xi_2>=
\int_{\mathfrak{X}}\;d\mu(x)\; <\xi_1(x),\xi_2(x)>_{{\cal H}_x}
\ee
See \cite{18} for more details.

Since in our case the dimension of all the ${\cal H}_{\omega_p},
\; p\in \Rl^D=:\mathfrak{X}$ has the 
same cardinality, the measurability condition on the dimension function is 
trivially satisfied. In our case the $\xi(p), \; p\in \Rl^D$ are of the form 
\be \label{6.21g}
\xi(p)=\pi_{\omega_p}(b_p)\Omega_{\omega_p}
\ee
and provided the assignment $\Rl^D\to \mathfrak{A};\;p\mapsto b_p$ 
satisfies the covariance condition 
$\alpha_g(b_p)=\alpha_{\varphi_+,\varphi_-}(b_{L^T p})$ we obtain a 
{\it unitary} representation of $G$ on ${\cal H}_\mu$ by 
\be \label{6.21h}
[U_\mu(g)\xi](p):=\sqrt{\frac{d\mu_L(p)}{d\mu(p)}}
[V_\mu(g)\xi](p)
\ee
where $V_\mu$ was defined in (\ref{6.21b}). The construction (\ref{6.21h})
is, of course, standard in QFT and is related to the {\it induced 
representation} of $G$ by the little groups of the subgroup 
$H=$Diff$_+(S^1)\times$Diff$_-(S^1)\times \mathfrak{L}^+_0$. See 
\cite{24b} for more details.

A particularly nice set of states are the ``diagonal states'' which 
arise from the GNS data corresponding to the state 
\be \label{6.21i}
\omega_\mu:=\int_{\Rl^D}\; d\mu(p)\;\omega_p
\ee
which is just the convex linear combination of the $\omega_p$ so that 
\be \label{6.21j}
\Omega_{\omega_\mu}=(\Omega_{\omega_p})_{p\in \Rl^D},\;\;
\pi_{\omega_\mu}(b)\Omega_{\omega_\mu}=
(\pi_{\omega_p}(b)\Omega_{\omega_p})_{p\in \Rl^D}
\ee
hence $b_p=b$ does not depend on $p$ and the measurability condition is 
trivially satisfied. We clearly have $\omega_\mu(b)=\omega_0(b)$ so the 
diagonal states are normalizable if and only if $\omega_0(b^\ast 
b)<\infty$. More generally we consider ``almost diagonal'' states of the 
form
$(f(p)\pi_{\omega_p}(b)\Omega_{\omega_p})_{p\in \Rl^D})$ where $f$
is any $L_2(\Rl^D,d\mu)$ function.
They obviously define a closed invariant subspace of ${\cal H}_\mu$.
They do not satisfy the covariance condition because $\alpha_L$ does not 
act on $f$, however, since $\omega_p\circ\alpha_L=\omega_0$ for all $p$
this turns out to be sufficient to guarantee unitarity, see below.

We can now characterize an important subclass of quasi -- 
$\mathfrak{L}^+_0$ -- 
invariant probability measures with support in $\overline{V}_+$ a bit 
closer. 
Namely, let  
$\mu_0$ be an actually $\mathfrak{L}^+_0$ -- invariant measure and 
consider a 
quasi -- invariant measure of the form $\mu=|f|^2 \mu_0$ where $f$ is a
normalized element of $L_2(\Rl^D,d\mu_0)$ which is $\mu_0$ a.e. 
non -- vanishing. For instance, $f$ could be smooth and of rapid decrease.
It follows from the proof of the {\it K\"allen -- Lehmann} representation 
theorem for the two -- point function of an interacting Wightman scalar 
field that a polynomially bounded $\mu_0$ is necessarily of the form
\be \label{6.21k} 
d\mu_0(p)=c\delta(p)+d\rho(m)d\nu_m(\vec{p})
\ee
where $c\ge 0$, $d\nu_m(\vec{p})=d^{D-1}p/\sqrt{m^2+\vec{p}^2}$ is the 
standard $\mathfrak{L}^+_0$ -- invariant measure on the positive mass 
hyperboloid
$H_m=\{p\in \Rl^D;\; p\cdot p=-m^2\le 0,\;p^0\ge 0\}$ and $d\rho(m)$ is a 
polynomially bounded measure on $[0,\infty)$, sometimes called the 
{\it K\"allen -- Lehmann} spectral measure because it characterizes the 
mass spectrum of a Wightman field. See \cite{25} for an instructive proof.
Hence the freedom in $\mu_0$ boils down to $c,\rho$. 

The point of all these efforts is of course that the translation group 
of $G$, in contrast to the rest of $G$, is represented weakly continously 
on ${\cal H}_\mu$ because it acts trivially on $b\in 
\mathfrak{A}$. The 
momentum (generalized) eigenstates are precisely the $\xi(p)$ in 
(\ref{6.21g}), that is,
we can define a self -- adjoint operator $\pi_\mu(p_\nu)$ as the 
generator of the translation subgroupf of $G$ by
$[\pi_\mu(p_\nu)\xi](p)=p_\nu \xi(p)$. Notice that all the states of 
${\cal H}_{\omega_p}$ have the same mass $m^2=-p \cdot p\ge 0$, hence 
{\it there is no tachyon}.

There is a different way to look at the almost diagonal states: Define 
\be \label{6.21l}
\xi(p):=f(p) \pi_{\omega_p}(b)\Omega_{\omega_p},\;\;f\in L_2(\Rl^D,d\mu_0)
\ee
and $\mu:=\mu_0$ so that the Jacobean in (\ref{6.21h}) equals unity. Then
\ba \label{6.21m}
||U_{\mu_0}(g)\xi||^2 &=& \int_{\Rl^D}\; d\mu_0(p) \;
\omega_{L^T p}([f(p)\alpha_g(b)]^\ast\;[f(p)\alpha_g(b)])
\nonumber\\
&=& \int_{\Rl^D}\; d\mu_0(p) \; |f(p)|^2
\omega_{L^T p}(\alpha_g(b^\ast b))
\nonumber\\
&=& ||f||_2^2\; \omega_0(b^\ast b)=
||\xi||^2
\ea
reproving unitarity. In what follows we focus, for simplicity, on the 
closed subspace 
of ${\cal H}_\mu$ defined by the almost diagonal states (\ref{6.21l}) 
which in turn is completely characterized by
${\cal H}_\omega$ as far as $\mathfrak{A}$ is 
concerned but includes the additional twist with respect to the 
representation of $G$ displayed in this section which enables us to add 
massive states to the theory. For instance we could have 
\be \label{6.21n}
\rho(m)=\sum_{n=0}^\infty \;\delta(m,n/\ell_s)
\ee
{\it corresponding to the usual string theory mass spectrum with the 
tachyon removed}.  Keeping this in mind, it will suffice to consider 
${\cal H}_\omega$ in what follows (equivalent to setting $c=1,\rho=0$).

\subsection{Implementation of the {\bf Pohlmeyer Charges}}
\label{s6.5}

We saw at the end of section \ref{s6.3} that the one -- parameter unitary 
groups $t\mapsto \pi_\omega(W_\pm(ts))$
are not weakly weakly continuous on ${\cal H}_\omega$. Since the {\bf
Pohlmeyer Charges} $Z_\pm$ involve polynomials of the $Y_\pm$ rather 
than polynomials of the $W_\pm$ it seems that our representation does not 
support the {\bf Quantum Pohlmeyer Charges}. Indeed, one would need to
use derivatives with respect to $t$ at $t=0$ to ``bring down'' the 
$Y^k_\pm(I)$ from the exponent. In fact, notice the classical identity for 
the {\bf Pohlmeyer Charges}
\be \label{6.23a}
R^{\mu_1..\mu_n}_\pm(x)=\frac{1}{i^n} \int_{x\le x_1\le..x_n\le x+2\pi}\; 
d^nx \; [\frac{\delta^n}{\delta f_{\mu_1}(x_1)..
\delta f_{\mu_n}(x_n)}]_{|k_1=..=k_n=0} \; W_\pm(f_1)..W_\pm(f_n)
\ee
where $W_\pm(f)=\exp(iY_\pm(f))$. While it is indeed possible to extend
$\omega$ to the $W_\pm(f)$ by $\omega(W_\pm(f)):=\delta_{f,0}$ (notice 
that this is a Kronecker $\delta$, not a functional $\delta-$distribution)
the functional derivatives in (\ref{6.23}) are clearly ill -- defined 
when trying to extend $\omega$ to the invariants by
\ba\label{6.24}
\omega(R^{\mu_1..\mu_n}_\pm(x))
&:=& \frac{1}{i^n} \int_{x\le x_1\le..x_n\le 
x+2\pi}\; d^nx \; \frac{1}{n!}\sum_{\pi\in S_n}\;
\times \nonumber\\
&& \times 
[\frac{\delta^n}{\delta f_{\mu_1}(x_1)..
\delta f_{\mu_n}(x_n)}]_{|k_1=..=k_n=0} \; 
\omega(W_\pm(f_{\pi(1)})..W_\pm(f_{\pi(n)}))
\ea
where we have introduced a symmetric ordering of the the non -- 
commuting $W_\pm(f)$. Expression (\ref{6.24}) is the direct analog for how 
to define $n-$point Schwinger functions in Euclidean field theory from the 
generating (positive) functional of a probability measure. This works 
there because the Osterwalder -- Schrader axioms \cite{20} require the 
generating 
functional to be analytic in $f$. This is clearly not the case for our 
$\omega$.

In fact, it is quite hard to construct positive linear, $G-$invariant 
functionals $\omega$ which are weakly continuous or more regular than 
the one we have found above. The reason for this is the absence of any
(worldsheet) background metric in our worldsheet background independent 
theory. The analytic positive linear functionals for free ordinary Quantum 
Field Theories all make strong use of a spatially Euclidean background 
metric which we do not have at our diposal here.\\
Another difficulty is the Minkowski signature of the flat 
target space metric. The simplest more regular ansatz will try to invoke 
the positivity theorems of Gaussian measures from Euclidean QFT, however,
these theorems are not applicble due to the Minkowski signature.
If we would change to Euclidean target space signature then we could 
still construct {\bf Pohlmeyer Charges}, however, now they would be 
complex valued because they and the the Virasoro generators would be based 
on the complex objects $Y_\pm=\pi\pm i X'$. This would imply that the 
Weyl elements are no longer unitary, rather we would get something 
like $(W_\pm(f))^\ast=W_\mp(-f)$, and in 
particular the Weyl relations 
would pick up not only a phase but actually some unbounded positive factor 
which makes the construction of a positive linear functional even harder. 
In any case, it would be unclear how to relate this ``Euclidean string'' 
to the actual Minkowski string because our Weyl algebra is non -- 
commutative so that the usual Wick rotation is not well-defined.

In conclusion, while we certainly have not yet analyzed the issue 
systematically enough, (\ref{6.20}) is the only solution to our
representation problem that we could find so far and we must now try to
implement the quantum invariants by using a suitable regularization.
This means that we write the regulated invariants as polynomials in the 
$W_\pm(s)$ and then remove the regulator and see whether the result 
is well -- defined and meaningful.

The first step is to consider instead of the invariants 
$Z^{\mu_1..\mu_n}_\pm$ the functions
\be \label{6.25}
Z^{k_1..k_n}_\pm:=k_{\mu_1}\; ..\; k_{\mu_n}\; Z^{\mu_1..\mu_n}_\pm
\ee
from which the original invariants are regained by specializing the 
$k_1,..,k_n$.
Notice that the 
$Z^{k_1..k_n}$, in contrast to the $Z^{\mu_1..\mu_n}$, are dimensionless. 

Using the fact that classically 
\be \label{6.26}
\frac{W^{k}_\pm([a,b])-W^{-k}_\pm([a,b])}{2i[b-a]}=k_\mu 
Y^\mu_\pm(\frac{a+b}{2})+O((b-a)^2)
\ee
we can write $Z^{k_1..k_n}_\pm$ as the limit of a Riemann sum
\ba \label{6.27}
&&\pi_\omega(Z^{k_1..k_n}_\pm) = \lim_{{\cal P}\to S^1} 
\pi_\omega(Z^{k_1..k_n}_{\pm,{\cal P}})
\\
&&\pi_\omega(Z^{k_1..k_n}_{\pm,{\cal P}}) =
\pi_\omega(R^{k_1..k_n}_{\pm,{\cal P}}(0))+
\pi_\omega(R^{k_2..k_n k_1}_{\pm,{\cal P}}(0))+..
+\pi_\omega(R^{k_n k_1..k_{n-1}}_{\pm,{\cal P}}(0)) 
\nonumber\\
&&\pi_\omega(R^{k_1..k_n}_{\pm,{\cal P}}(0)) = \sum_{m_1=1}^M
\;\sum_{m_2=m_1}^M..\sum_{m_n=m_{n-1}}^M
\frac{1}{(2i)^n\;n!}\sum_{\pi\in S_n}
\times\nonumber\\
&& \times
[\pi_\omega(W^{k_{\pi(1)}}_\pm(I_{m_{\pi(1)}}))-
\pi_\omega(W^{-k_{\pi(1)}}_\pm(I_{m_{\pi(1)}}))]\; ..
[\pi_\omega(W^{k_{\pi(n)}}_\pm(I_{m_{\pi(n)}}))-
\pi(W^{-k_{\pi(n)}}_\pm(I_{m_{\pi(n)}}))]
\nonumber
\ea
where ${\cal P}=\{I_m;\;m=1,..,M:=|{\cal P}|\}$ is any partition of 
$[0,2\pi]$ into consecutive intervals, e.g. $I_m=[(m-1)2\pi/n,m\pi/n]$ in 
some coordinate system, 
and we have again introduced a symmetric 
ordering in order that the corresponding operator be at least symmetric.
Notice that the expression for $R_{\pm,{\cal P}}(0)$ had to arbitrarily 
choose a ``starting interval'' $I_1$ of $\cal P$ but the cyclic 
symmetrization in $Z_{\pm,{\cal P}}$ removes this arbitrariness again in 
analogy to the continuum expressions of section \ref{s3}.

Expression (\ref{6.27}) is an element of $\mathfrak{A}$ at finite $M$.
It has the expected transformation behaviour under Lorentz 
transformations, namely 
$\alpha_L(Z^{k_1..k_n}_{\pm,{\cal P}})= 
Z^{L\cdot k_1..L\cdot k_n}_{\pm,{\cal P}}$ and under Diff$(S^1)$ the 
partition
is changed to its diffeomorphic image, i.e 
$\alpha^\pm_\varphi(Z^{k_1..k_n}_{\pm,{\cal P}})= 
Z^{L\cdot k_1..L\cdot k_n}_{\pm,\varphi({\cal P})}$,
leaving $M$ unchanged. The 
invariance property of the charges is hence only to be recovered in the 
limit $M\to\infty$, that is, ${\cal P}\to\sigma$. Hence one would like to 
define
\be \label{6.28}
\pi_\omega(Z^{k_1..k_n}_\pm):=\lim_{|{\cal P}|\to S^1} 
\pi_\omega(Z^{k_1..k_n}_{\pm,{\cal P}})
\ee
in a suitable operator topology. 

It is easy to see that with this definition we have 
$\pi_\omega(Z^{k_1..k_n}_\pm)=0$ in the weak operator topology on ${\cal 
H}_\omega$
while $\pi_\omega(Z^{k_1..k_n}_\pm)=\infty$ in the strong operator 
topology on ${\cal 
H}_\omega$, the divergence being of the order of $M^{n/2}$. One can 
prove that this happens in general 
for diffeomorphism invariant operators in LQG \cite{2,24}
which are ``graph -- changing'', as the $\pi_\omega(Z_{\pm,{\cal P}})$ do
when defined as above, if one 
wants to define them on the kinematical Hilbert space, in this case
${\cal H}_\omega$.
This point will be explained in more detail in our companion paper 
\cite{11}. Suffice it to say, for the purposes of the present paper, that
there are two ways to get around this problem. The first one is to define 
the operators $Z_{\pm}$ on the physical Hilbert space (to be defined 
below), hence one uses the strong topology of the physical Hilbert space 
rather than the kinematical one which removes the above divergence which
is due to the ``infinite volume of the diffeomporphism group''. We will 
follow this approach in the companion paper \cite{11}. The second 
possibility is to define the operators $\pi_\omega(Z_{\pm})$ directly on 
the 
kinematical Hilbert space ${\cal H}_\omega$ such that they are 
diffeomorphism invariant but simultaneously {\it non -- graph changing}.
This is the possibility which we will explore below.  

The idea for defining a non -- graph changing operator comes from looking
at expression (\ref{6.27}):\\
We would like to define $\pi_\omega(Z_\pm)$ densely on ${\cal H}_\omega$,
hence it is sufficient to define it on the orthonormal basis given by
the $\pi_\omega(W_\pm(s))\Omega_\omega$. Now given $s$ we may assign
to it a ``distributional'' smearing function 
\be \label{6.28a}
f^s_\mu(x)=\sum_{I\in\gamma(s)}\;\chi_I(x)\;k^I_\mu(s)
\ee
such that $\exp(i Y_\pm(f^s))=W_\pm(s)$. The point is now that if we allow 
some of the momenta $k^I(s)$ to be zero then we can consider the graph 
$\gamma(s)$ as a {\it partition} of $S^1$. The only condition on the 
momenta $k^I(s)$ is that momenta assigned to neighbouring intervals $I$
are different from each other.

Now in the regularization 
$\pi_\omega(Z_{\pm,{\cal P}})$ of $\pi_\omega(Z_\pm)$ above we had to 
invoke a partition $\cal P$ as well, but any finite
partition does not render $Z_{\pm,{\cal P}}$ into a diffeomorphism 
invariant object, it is only diffeomorphism covariant. We now we bring 
these two things together:\\
We simply define for $n\ge 2$
\be \label{6.29}  
\pi_\omega(Z^{k_1..k_n}_\pm) \pi_\omega(W_\pm(s))\Omega_\omega
:=\pi_\omega(Z^{k_1..k_n}_{\pm,\gamma(s)}) 
\pi_\omega(W_\pm(s))\Omega_\omega
\ee
with $\pi_\omega(Z^{(n)}_{\pm,0}):=0$, the vacuum is annihilated by all 
charges. Also, if the decomposition of (\ref{6.29}) into states of the 
form $\pi_\omega(W_\pm(s'))\Omega_\omega$ contains states with 
$\gamma(s')$ strictly smaller than $\gamma(s)$ then we simply remove that 
state from the decomposition which ensures that (\ref{6.29}) becomes a 
symmetric operator as we will show in the next section.

Hence we have simply set ${\cal P}:=\gamma(s)$ in (\ref{6.27}). There is 
no limit ${\cal P}\to\infty$ to be taken, the partition is kept finite.
The idea is that the semiclassical limit of (\ref{6.29}) is reached only 
on states which have $|\gamma(s)|$ significantly large. This we will 
confirm below. We identify $\pi_\omega(Z^\mu_\pm)$ with the 
self -- adjoint generator of the translation subgroup of the Poincar\'e
group which is represented weakly continuously on ${\cal H}_\omega$.
In particular, it commutes with all charges $\pi_\omega(Z^{(n)}_\pm)$
since it commutes with all the $\pi_\omega(W_\pm(s))$.

Let us check that (\ref{6.29}) defines a diffeomorphism invariant 
operator. We must verify that
\ba \label{6.30}
&& U^\pm_\omega(\varphi)
\pi_\omega(Z^{k_1..k_n}_\pm) \pi_\omega(W_\pm(s))\Omega_\omega
=U^\pm_\omega(\varphi) \pi_\omega(Z^{k_1..k_n}_{\pm,\gamma(s)}) 
\pi_\omega(W_\pm(s))\Omega_\omega
\nonumber\\
&\equiv& 
\pi_\omega(Z^{k_1..k_n}_\pm) 
U^\pm_\omega(\varphi) \pi_\omega(W_\pm(s))\Omega_\omega
=\pi_\omega(Z^{k_1..k_n}_{\pm,\varphi(\gamma(s))}) 
\pi_\omega(\alpha^\pm_\varphi(W_\pm(s)))\Omega_\omega
\ea
Now by construction the right hand side in the first line of (\ref{6.30}) 
is a finite 
linear combination of vectors of the form 
$\pi_\omega(W_\pm(s'))\Omega_\omega$ with $\gamma(s')\subset \gamma(s)$
to which the unitary transformation $U^\pm_\omega(\varphi)$ is applied and 
hence these vectors are transformed into 
$\pi_\omega(W_\pm(\tilde{s}))\Omega_\omega$ with $\gamma(\tilde{s})\subset 
\varphi(\gamma(s))$. But this reproduces exactly the action of the 
operator in the second line of (\ref{6.30}) on the transformed states.

\subsection{Properties of the {\bf Quantum Pohlmeyer Charges}}
\label{s6.6}

In this section we wish to check that the algebra implied by (\ref{6.29}) 
defines a 
quantum deformation of the classical invariant algebra of section 
\ref{s3}. \\
\\
{\bf Adjointness Relations:}\\
\\
Notice that
the Hilbert space ${\cal H}^\pm_\omega$ can be written as an 
uncountable direct sum
\be \label{6.31}
{\cal H}^\pm_\omega=\overline{\oplus_\gamma \;\;
{\cal H}^\pm_{\omega,\gamma}}
\ee
where the overline denotes completion, the sum is over all partitions 
(graphs) of $S^1$ and  
${\cal H}^\pm_{\omega,\gamma}$ is the completion of the finite linear span 
of the vectors $\pi_\omega(W_\pm(s))$ with $\gamma(s)=\gamma$ with 
$s$ such that for neighbouring $I,J$ we have $k^I(s)\not=k^J(s)$. 

Let $P^\pm_\gamma$ be the orthogonal projection onto ${\cal 
H}^\pm_{\omega\gamma}$. Then it is easy to see that 
\be \label{6.32}
\pi_\omega(Z^{k_1..k_n}_\pm)=\oplus_\gamma\; P^\pm_\gamma\;\;
\pi_\omega(Z^{k_1..k_n}_{\pm,\gamma})\;\; P^\pm_\gamma\;\;
\ee
where the expression for $\pi_\omega(Z^{k_1..k_n}_{\pm,\gamma})$ is given 
in (\ref{6.29}). It is now manifest that (\ref{6.32}) defines 
a symmetric operator because $\pi_\omega(Z^{k_1..k_n}_{\pm,\gamma})$
is symmetric on ${\cal H}^\pm_{\omega,\gamma}$ provided it preserves 
$\gamma$ which, however, is ensured by the projections.
Since the expression (\ref{6.29}) is real 
valued (it maps basis elements into finite linear combinations of basis
elements with real valued components) it follows from von Neumann's
involution theorem \cite{25} that it has self-adjoint extensions. We do 
not need to worry about these extensions for what follows.

In order to define the $\pi_\omega(Z^{\mu_1..\mu_n}_\pm)$ themselves 
we use the trivial observation that classically 
$Z^{k_1..k_n}_\pm L^n=Z^{\mu_1..\mu_n}_\pm$ if we set $k_{j\mu}=
\delta^{\mu_j}_\mu/L$ where $L$ is an arbitrary but fixed parameter of 
dimension cm$^1$. Hence we define 
\be \label{6.32a}
\pi_\omega(Z^{\mu_1..\mu_n}_\pm):= L^n 
\pi_\omega(Z^{k_1..k_n}_\pm),\;\;k_{j\mu}:=\delta^{\mu_j}_\mu/L
\ee 
The parameter $L$ will enter the semiclassical analysis in section
\ref{s6.7}. We will see that (\ref{6.32a}) approximates the classical
expression the better the smaller the parameter $t:=(\ell_s/L)^2$ is.\\
\\
{\bf Algebraic Properties:}\\
\\
We can now study the algebra of our 
$\pi_\omega(Z_\pm)$ and check that up to quantum corrections 
the classical algebra of invariants is reproduced on each of the invariant 
${\cal H}^\pm_{\omega,\gamma}$ separately. To simplify the notation, let 
us introduce the shorthand 
\be \label{6.33}
\pi_\omega(Y^k_\pm(I)):=\frac{1}{2i}[
\pi_\omega(W^{k}_\pm(I))-\pi_\omega(W^{-k}_\pm(I))]
\ee
corresponding to (\ref{6.26}). In what follows we consider the algebra
of the $\pi_\omega(Z_\pm)$ restricted to a fixed ${\cal 
H}^\pm_{\omega,\gamma}$ with $|\gamma|=M$ and $\gamma=\{I_m\}_{m=1}^M$
where we have chosen an arbitrary starting interval $I_1$ and the 
intervals $I_m,\;I_{m+1}$ with $m\equiv m+M$ are next neighbours. We will
denote the corresponding restrictions by 
\ba \label{6.34}
\pi_\omega(Z^{k_1..k_n}_{\pm,M}) &=& 
\pi_\omega(R^{k_1..k_n}_{\pm,M})+
\pi_\omega(R^{k_2..k_n k_1}_{\pm,M})+\;..\;+
\pi_\omega(R^{k_n k_1..k_{n-1}}_{\pm,M})
\nonumber\\
\pi_\omega(R^{k_1..k_n}_{\pm,M}) &=&
\sum_{1\le m_1\le ..\le m_n\le M}\; \frac{1}{n!}\sum_{\pi\in S_n}
\pi_\omega(Y^{k_{\pi(1)}}_\pm(I_{m_{\pi(1)}}))\;..\;
\pi_\omega(Y^{k_{\pi(n)}}_\pm(I_{m_{\pi(n)}}))
\ea
We now notice that (\ref{6.34}) is the quantization of a Riemann sum 
approximation for the classical continuum integrals of section \ref{s3}
(of course, the approximation is classically good only for large $M$).
Since algebraic properties of iterated path ordered integrals have a 
precise analog
for iterated path ordered Riemann sums, it is clear that all 
algebraic relations of 
section \ref{s3} which only rely on manipulations of integrals are exactly
mirrored by expressions (\ref{6.34}) up to quantum corrections which come 
from 1. commuting products of the $\pi_\omega(Y^k_\pm(I))$, 2. ommission 
of states due to the projections in (\ref{6.32}) and 3. 
``finite 
size effects'' which come from the fact that we are dealing really with 
discrete objects (sums and intervals) rather than continuum ones 
(integrals and points). All of 
these corrections are suppressed in the semiclassical limit which is 
reached for states with large $M$ (they are, in a precise sense, of 
measure zero) and in the limit that $\ell_s\to 0$
as we will demonstrate below. Instead of going through a tedious 
bookkeeping exercise which would merely reproduce the results of 
\cite{6,16,7,15} in a discrete language while keeping track of the 
operator ordering, let us 
give a typical example which illustrates these effects.

Notice that 
\ba \label{6.35}
&& {[}\pi_\omega(Y^k_\pm(I_m)),\pi_\omega(Y^l_\pm(I_{m'}))]=
\frac{-i}{2}\sin(\mp \ell_s^2 [k\cdot l]\alpha(I_m,I_{m'})/2)
\times \nonumber\\
&& \times
\{\pi_\omega(W_\pm((k,I_m)+(l,I_{m'}))
+\pi_\omega([W_\pm((k,I_m)+(l,I_{m'})]^{-1})
\nonumber\\ && 
+
\{\pi_\omega(W_\pm((k,I_m)+(-l,I_{m'}))
+\pi_\omega([W_\pm((k,I_m)+(-l,I_{m'})]^{-1})\}
\ea
where 
\be \label{6.36}
\alpha(I_m,I_{m'})=(\chi_{I_m})_{|\partial 
I_{m'}}-(\chi_{I_{m'}})_{|\partial I_m}
=\left\{ 
\begin{array}{cc}
0 & m=m' \mbox{ or } |m-m'|>1 \\
-1 & m=m'+1\\
1 & m'=m+1
\end{array} 
\right.
\ee
where we have used the convention that $\chi_I(x)=1$ for $x\in I-\partial 
I$, $\chi_I(x)=1/2$ for $x\in \partial I$ and $\chi_I(x)=0$ for 
$I\not\in I$. This convention coincides $dx-$ a.e. with the usual 
convention but in our case does make a difference due to the singular 
support of our fields. It is the unique convention which ensures that for 
closed intervals $I,J$ with $f(I)=b(J)$ we have 
$\chi_{I\cup J}=\chi_I+\chi_J$. Here $b(I),f(I)$ denote beginning point 
and final point of $I$ respectively.

Hence 
\ba \label{6.37}
&& {[}\pi_\omega(Y^k_\pm(I_m)),\pi_\omega(Y^l_\pm(I_{m'}))] =\mp
\frac{i}{2}\sin(\ell_s^2 [k\cdot l]/2)
(\delta_{m,m'+1}-\delta_{m,m'-1})
\times\nonumber\\ &&\times 
\{\pi_\omega(W_\pm((k,I_m)+(l,I_{m'}))
+\pi_\omega([W_\pm((k,I_m)+(l,I_{m'})]^{-1})
\nonumber\\&&
+\pi_\omega(W_\pm((k,I_m)+(-l,I_{m'}))
+\pi_\omega([W_\pm((k,I_m)+(-l,I_{m'})]^{-1})\}
\nonumber\\
&=:&
\mp 2i\sin(\ell_s^2 [k\cdot l]/2)
[\delta_{m,m'+1}-\delta_{m,m'-1}]
\pi_\omega(a((m,k),(m',l))
\ea
Semiclassicaly the expression $\pi_\omega(a((m,k),(m',l))$ will tend to 
the constant $1$ so that (\ref{6.37}) is a specific quantum deformation of 
the classical Poisson bracket.

As an example we choose (notice that the first charge algebraically 
indpendent of $p_\mu$  
involves three indices because $Z^{\mu_1\mu_2}=Z^{\mu_1} Z^{\mu_2}/2$ 
however, in the following illustrational calculation we will not make use 
of this fact)

\newpage

\ba \label{6.37a}
\pi_\omega(Z^{k_1}_{\pm,M})\pi_\omega(Z^{k_2 k_3}_{\pm,M})
&=& \frac{1}{2}
\sum_{m_1=1}^M \sum_{1\le m_2\le m_3 \le M} 
\pi_\omega(Y^{k_1}_\pm(I_{m_1}))
\times\nonumber\\&&\times
[ 
\pi_\omega(Y^{k_2}_\pm(I_{m_2}))\pi_\omega(Y^{k_3}_\pm(I_{m_3})) 
+\pi_\omega(Y^{k_3}_\pm(I_{m_3}))\pi_\omega(Y^{k_2}_\pm(I_{m_2}))
\nonumber\\&&
+\pi_\omega(Y^{k_3}_\pm(I_{m_2}))\pi_\omega(Y^{k_2}_\pm(I_{m_3})) 
+\pi_\omega(Y^{k_2}_\pm(I_{m_3}))\pi_\omega(Y^{k_3}_\pm(I_{m_2}))
] 
\nonumber\\
&=& \frac{1}{2}
\sum_{1\le m_1\le m_2\le m_3 \le M} 
\pi_\omega(Y^{k_1}_\pm(I_{m_1}))
\times\nonumber\\&&\times
[ 
\pi_\omega(Y^{k_2}_\pm(I_{m_2}))\pi_\omega(Y^{k_3}_\pm(I_{m_3})) 
+\pi_\omega(Y^{k_3}_\pm(I_{m_3}))\pi_\omega(Y^{k_2}_\pm(I_{m_2}))
\nonumber\\&&
+\pi_\omega(Y^{k_3}_\pm(I_{m_2}))\pi_\omega(Y^{k_2}_\pm(I_{m_3})) 
+\pi_\omega(Y^{k_2}_\pm(I_{m_3}))\pi_\omega(Y^{k_3}_\pm(I_{m_2}))
] 
\nonumber\\
&+& \frac{1}{2}
\sum_{1\le m_2<m_1\le m_3 \le M} 
\pi_\omega(Y^{k_1}_\pm(I_{m_1}))
\times\nonumber\\&&\times
[ 
\pi_\omega(Y^{k_2}_\pm(I_{m_2}))\pi_\omega(Y^{k_3}_\pm(I_{m_3})) 
+\pi_\omega(Y^{k_3}_\pm(I_{m_3}))\pi_\omega(Y^{k_2}_\pm(I_{m_2}))
\nonumber\\&&
+\pi_\omega(Y^{k_3}_\pm(I_{m_2}))\pi_\omega(Y^{k_2}_\pm(I_{m_3})) 
+\pi_\omega(Y^{k_2}_\pm(I_{m_3}))\pi_\omega(Y^{k_3}_\pm(I_{m_2}))
] 
\nonumber\\
&+& \frac{1}{2}
\sum_{1\le m_2\le m_3<m_1 \le M} 
\pi_\omega(Y^{k_1}_\pm(I_{m_1}))
\times\nonumber\\&&\times
[ 
\pi_\omega(Y^{k_2}_\pm(I_{m_2}))\pi_\omega(Y^{k_3}_\pm(I_{m_3})) 
+\pi_\omega(Y^{k_3}_\pm(I_{m_3}))\pi_\omega(Y^{k_2}_\pm(I_{m_2}))
\nonumber\\&&
+\pi_\omega(Y^{k_3}_\pm(I_{m_2}))\pi_\omega(Y^{k_2}_\pm(I_{m_3})) 
+\pi_\omega(Y^{k_2}_\pm(I_{m_3}))\pi_\omega(Y^{k_3}_\pm(I_{m_2}))
] 
\nonumber\\
&=& \frac{1}{3!}
\sum_{1\le m_1\le m_2\le m_3 \le M} 
[3 \pi_\omega(Y^{k_1}_\pm(I_{m_1}))]
\times\nonumber\\&&\times
[ 
\pi_\omega(Y^{k_2}_\pm(I_{m_2}))\pi_\omega(Y^{k_3}_\pm(I_{m_3})) 
+\pi_\omega(Y^{k_3}_\pm(I_{m_3}))\pi_\omega(Y^{k_2}_\pm(I_{m_2}))
\nonumber\\&&
+\pi_\omega(Y^{k_3}_\pm(I_{m_2}))\pi_\omega(Y^{k_2}_\pm(I_{m_3})) 
+\pi_\omega(Y^{k_2}_\pm(I_{m_3}))\pi_\omega(Y^{k_3}_\pm(I_{m_2}))
] 
\nonumber\\
&+& \frac{1}{3!}
\sum_{1\le m_2 \le m_1\le m_3 \le M} 
[3\pi_\omega(Y^{k_1}_\pm(I_{m_1}))]
\times\nonumber\\&&\times
[ 
\pi_\omega(Y^{k_2}_\pm(I_{m_2}))\pi_\omega(Y^{k_3}_\pm(I_{m_3})) 
+\pi_\omega(Y^{k_3}_\pm(I_{m_3}))\pi_\omega(Y^{k_2}_\pm(I_{m_2}))
\nonumber\\&&
+\pi_\omega(Y^{k_3}_\pm(I_{m_2}))\pi_\omega(Y^{k_2}_\pm(I_{m_3})) 
+\pi_\omega(Y^{k_2}_\pm(I_{m_3}))\pi_\omega(Y^{k_3}_\pm(I_{m_2}))
] 
\nonumber\\
&+& \frac{1}{3!}
\sum_{1\le m_2\le m_3 \le m_1 \le M} 
[3\pi_\omega(Y^{k_1}_\pm(I_{m_1}))]
\times\nonumber\\&&\times
[ 
\pi_\omega(Y^{k_2}_\pm(I_{m_2}))\pi_\omega(Y^{k_3}_\pm(I_{m_3})) 
+\pi_\omega(Y^{k_3}_\pm(I_{m_3}))\pi_\omega(Y^{k_2}_\pm(I_{m_2}))
\nonumber\\&&
+\pi_\omega(Y^{k_3}_\pm(I_{m_2}))\pi_\omega(Y^{k_2}_\pm(I_{m_3})) 
+\pi_\omega(Y^{k_2}_\pm(I_{m_3}))\pi_\omega(Y^{k_3}_\pm(I_{m_2}))
] 
\nonumber\\
&-& \frac{1}{2}
\sum_{1\le m_2 \le m_3 \le M} 
\pi_\omega(Y^{k_1}_\pm(I_{m_2}))
\times\nonumber\\&&\times
[ 
\pi_\omega(Y^{k_2}_\pm(I_{m_2}))\pi_\omega(Y^{k_3}_\pm(I_{m_3})) 
+\pi_\omega(Y^{k_3}_\pm(I_{m_3}))\pi_\omega(Y^{k_2}_\pm(I_{m_2}))
\nonumber\\&&
+\pi_\omega(Y^{k_3}_\pm(I_{m_2}))\pi_\omega(Y^{k_2}_\pm(I_{m_3})) 
+\pi_\omega(Y^{k_2}_\pm(I_{m_3}))\pi_\omega(Y^{k_3}_\pm(I_{m_2}))
] 
\nonumber\\
&-& \frac{1}{2}
\sum_{1\le m_2\le m_3 \le M} 
\pi_\omega(Y^{k_1}_\pm(I_{m_3}))
\times\nonumber\\&&\times
[ 
\pi_\omega(Y^{k_2}_\pm(I_{m_2}))\pi_\omega(Y^{k_3}_\pm(I_{m_3})) 
+\pi_\omega(Y^{k_3}_\pm(I_{m_3}))\pi_\omega(Y^{k_2}_\pm(I_{m_2}))
\nonumber\\&&
+\pi_\omega(Y^{k_3}_\pm(I_{m_2}))\pi_\omega(Y^{k_2}_\pm(I_{m_3})) 
+\pi_\omega(Y^{k_2}_\pm(I_{m_3}))\pi_\omega(Y^{k_3}_\pm(I_{m_2}))
] 
\ea
The first three terms in the last equality of (\ref{6.37}) combine, up to 
commutators, to the expected expression 
$\pi_\omega(Z^{k_1 k_2 k_3}_\pm)+\pi_\omega(Z^{k_2 k_1 k_3}_\pm)$, see
section \ref{s3}, while the two remaining terms converge in the 
semiclassical limit to path ordered integrals of the form 
\be \label{6.38}
\frac{1}{M}\int_{x_1\le x_2} d^2x Y^{k_1}_\pm(x_2) Y^{k_2}_\pm(x_2)
Y^{k_3}_\pm(x_3)
\ee
and thus vanishes in the large $M$ limit, see below. For the general 
relations we get similar correction terms whose number depends only on $n$
and which are therefore suppressed compared to the correct leading term
as $M\to \infty$ and $\ell_s\to 0$.\\
\\
{\bf Commutation Relations:}\\
\\
Next we consider commutators. We have 
\ba \label{6.39}
&& [\pi_\omega(Z^{k_1..k_n}_\pm),
\pi_\omega(Z^{k'_1..k'_{n'}}_\pm)]
=C_n\cdot S_n\cdot \sum_{1\le m_1\le..\le m_n\le M}\;
C_{n'}\cdot S_{n'}\cdot \sum_{1\le m'_1\le..\le m'_{n'}\le M}
\times\\
&& \times
{[} \pi_\omega(Y^{k_1}_\pm(I_{m_1})\;..\;
\pi_\omega(Y^{k_n}_\pm(I_{m_n})),
\pi_\omega(Y^{k'_1}_\pm(I_{m'_1}))\;..\;
\pi_\omega(Y^{k'_{n'}}_\pm(I_{m'_{n'}}))]
\nonumber\\
&=& \mp 2i 
\sum_{l=1}^n\;\sum_{l'=1}^{n'}\; \sin(\ell_s^2[k_l\cdot k'_{l'}]/2)
\times\nonumber\\&&\times
C_n\cdot S_n\cdot \sum_{m_1,..,m_n=1}^M\; 
\theta(m_2-m_1)..\theta(m_n-m_{n-1})\;
C_{n'}\cdot S_{n'}\cdot \sum_{m'_1,..,m'_{n'}=1}^M
\;\theta(m'_2-m'_1)..\theta(m'_{n'}-m'_{n'-1})\;
\times\nonumber\\ && \times
{[}\delta_{m_l,m'_{l'}+1}-\delta_{m'_{l'},m_l+1}]
\times\nonumber\\
&& \times
\pi_\omega(Y^{k_1}_\pm(I_{m_1}))\;..\;
\pi_\omega(Y^{k_{l-1}}_\pm(I_{m_{l-1}}))
\pi_\omega(Y^{k'_1}_\pm(I_{m'_1}))\;..\;
\pi_\omega(Y^{k'_{l'-1}}_\pm(I_{m'_{l'-1}}))
\pi_\omega(a((I_{m_l},k_{m_l}),(I_{m'_{l'}},k'_{m'_{l'}})))
\nonumber\times\\
&& \times
\pi_\omega(Y^{k'_{l'+1}}_\pm(I_{m'_{l'+1}}))\;..\;
\pi_\omega(Y^{k'_{n'}}_\pm(I_{m'_{n'}}))
\pi_\omega(Y^{k_{l+1}}_\pm(I_{m_{l+1}}))\;..\;
\pi_\omega(Y^{k_n}_\pm(I_{m_n}))
\nonumber\\
&=& \mp 2i 
\sum_{l=1}^n\;\sum_{l'=1}^{n'}\; \sin(\ell_s^2[k_l\cdot k'_{l'}]/2)
\times\nonumber\\ 
&& \times
C_n\cdot S_n\cdot \sum_{m_1,..,\hat{m}_l,..,m_n=1}^M\; 
\;\theta(m_2-m_1)\;..\;\theta(m_{l-1}-m_{l-2})\;
\theta(m_{l+2}-m_{l+1})\;..\theta(m_{n}-m_{n-1})\;
\times\nonumber\\
&&\times
C_{n'}\cdot S_{n'}\cdot \sum_{m'_1,..,\hat{m}'_{l'},..,m'_{n'}=1}^M
\;\theta(m'_2-m'_1)\;..\;\theta(m'_{l'-1}-m'_{l'-2})\;
\theta(m'_{l'+2}-m'_{l'+1})\;..\theta(m'_{n'}-m'_{n'-1})\;
\times\nonumber\\
&& \times
\pi_\omega(Y^{k_1}_\pm(I_{m_1}))\;..\;
\pi_\omega(Y^{k_{l-1}}_\pm(I_{m_{l-1}}))
\pi_\omega(Y^{k'_1}_\pm(I_{m'_1}))\;..\;
\pi_\omega(Y^{k'_{l'-1}}_\pm(I_{m'_{l'-1}}))
\times\nonumber\\
&&\times 
\{
\sum_{m_l,m'_{l'}=1}^M \;
{[}(\delta_{m_l,m'_{l'}+1}-\delta_{m_l,m'_{l'}})
-(\delta_{m'_{l'},m_l+1}-\delta_{m_l,m'_{l'}})]\;
\times\nonumber\\&&\times
\theta(m_l-m_{l-1})\theta(m_{l+1}-m_l)
\theta(m'_{l'}-m'_{l'-1})\theta(m'_{l'+1}-m'_{l'})\;
\pi_\omega(a((I_{m_l},k_{m_l}),(I_{m'_{l'}},k'_{m'_{l'}}))
\}
\nonumber\times\\
&& \times
\pi_\omega(Y^{k'_{l'+1}}_\pm(I_{m'_{l'+1}}))\;..\;
\pi_\omega(Y^{k'_{n'}}_\pm(I_{m'_{n'}}))
\pi_\omega(Y^{k_{l+1}}_\pm(I_{m_{l+1}}))\;..\;
\pi_\omega(Y^{k_n}_\pm(I_{m_n}))
\nonumber
\ea
where $C_n$ and $S_n$ respectively enforce cyclic summation of the
$k_1..k_n$ and symmetric projection of the $((m_1,k_1),..,(m_n,k_n))$
respectively. Here $\theta$ is the Heavyside step function and a hat above 
the argument denotes its omission.

Consider the curly bracket in the last equality of (\ref{6.39}). We see 
that the square bracket that it contains is precisely the discretization 
of the distribution $(\partial_y-\partial_x)\delta(x,y)$ of the 
corresponding continuum calculation where the derivatives and 
$\delta-$distributions respectively are replaced by differences and 
Kronecker $\delta$ functions respectively. In the continuum calculation we 
would now perform an integration by parts, in the discrete calculation we 
perform a partial resummation. We have 
\ba \label{6.40}
&& \sum_{m,m'=1}^M\;[
(\delta_{m,m'+1}-\delta_{m,m'})
-(\delta_{m+1,m'}-\delta_{m,m'})] f(m,m')
\nonumber\\
&=& 
\sum_{m=1}^M[-f(m,M)\delta_{m,M}+\sum_{m'=1}^{M-1} 
(\delta_{m,m'+1}-\delta_{m,m'}) f(m,m')
\nonumber\\
&&-\sum_{m'=1}^M[-f(M,m')\delta_{M,m'}+\sum_{m=1}^{M-1} 
(\delta_{m+1,m'}-\delta_{m,m'}) f(m,m')
\nonumber\\
&=& 
\sum_{m=1}^M[\sum_{m'=2}^M \delta_{m,m'} f(m,m'-1)
-\sum_{m'=1}^{M-1}\delta_{m,m'} f(m,m')] 
\nonumber\\
&&-\sum_{m'=1}^M[\sum_{m=2}^M \delta_{m,m'} f(m-1,m')
\sum_{m=1}^{M-1} \delta_{m,m'}) f(m,m')]
\nonumber\\
&=& 
\sum_{m=1}^M[\delta_{m,M} f(m,M)+
\sum_{m'=2}^M \delta_{m,m'}(f(m,m'-1)-f(m,m'))] 
\nonumber\\&&
-\sum_{m'=1}^M[\delta_{M,m'} f(M,m')+ 
\sum_{m=2}^M \delta_{m,m'} (f(m-1,m')-f(m,m'))]
\nonumber\\
&=& 
\sum_{m,m'=1}^M \delta_{m,m'} \{[f(m,m'-1)-f(m,m')]-[f(m-1,m')-f(m,m')]\}
\ea
for any function defined on $\{1,..,M\}^2$ where we used the convention
$f(m,m')=0$ if one of $m,m'$ equals $0,M+1$.

Now for products of (possibly non commutative) functions defined on 
discrete values the difference 
replacing the derivative gives a discrete version of the Leibniz rule
\be \label{6.41}
[fg](m-1)-[fg](m)=[f(m-1)-f(m)]g(m-1)+f(m)[g(m-1)-g(m)]
\ee
Combining (\ref{6.40}) and (\ref{6.41}) we may write the curly bracket in 
the last equality of (\ref{6.39}) as 
\ba \label{6.42}
&&\sum_{m_l,m'_{l'}=1}^M \;
[(\delta{m_l,m'_{l'}+1}-\delta{m_l,m'_{l'}})
-(\delta{m'_{l'},m_l+1}-\delta{m_l,m'_{l'}}]\;
\times\\&&\times
\theta(m_l-m_{l-1})\theta(m_{l+1}-m_l)
\theta(m'_{l'}-m'_{l'-1})\theta(m'_{l'+1}-m'_{l'})
\;
\pi_\omega(a((I_{m_l},k_{m_l}),(I_{m'_{l'}},k'_{m'_{l'}}))
\nonumber\\
&=&
\sum_{m_l,m'_{l'}=1}^M \;\delta_{m_l,m'_{l'}}\;
\{
\theta(m_l-m_{l-1})\theta(m_{l+1}-m_l)\;
\times\nonumber\\&&\times
\{
[\theta(.-m'_{l'-1})\theta(m'_{l'+1}-.)
\pi_\omega(a((I_{m_l},k_{m_l}),(I_{.},k'_{.}))](m'_{l'}-1)
\nonumber\\&&
-
[\theta(.-m'_{l'-1})\theta(m'_{l'+1}-.)
\pi_\omega(a((I_{m_l},k_{m_l}),(I_{.},k'_{.}))](m'_{l'})
\}
\nonumber\\
&&
-\theta(m'_{l'}-m'_{l'-1})\theta(m'_{l'+1}-m'_{l'})\;
\times\nonumber\\&&\times
\{
[\theta(.-m_{l-1})\theta(m_{l+1}-.)
\pi_\omega(a((I_{.},k_{.}),(I_{m'_{l'}},k'_{m'_{l'}}))](m_l-1)
\nonumber\\&&
-
[\theta(.-m_{l-1})\theta(m_{l+1}-.)
\pi_\omega(a((I_{.},k_{.}),(I_{m'_{l'}},k'_{m'_{l'}}))](m_l)
\}
\}
\nonumber\\
&=&
\sum_{m_l,m'_{l'}=1}^M \;\delta_{m_l,m'_{l'}}\;
\{
\theta(m_l-m_{l-1})\theta(m_{l+1}-m_l)\;
\times\nonumber\\&&\times
\{
[(\theta(.-m'_{l'-1})\theta(m'_{l'+1}-.))(m'_{l'}-1)
-(\theta(.-m'_{l'-1})\theta(m'_{l'+1}-.))(m'_{l'})]
\pi_\omega(a((I_{m_l},k_{m_l}),(I_{.},k'_{.}))](m'_{l'}-1)
\nonumber\\&&
+
(\theta(.-m'_{l'-1})\theta(m'_{l'+1}-.))(m'_{l'})
[\pi_\omega(a((I_{m_l},k_{m_l}),(I_{.},k'_{.}))(m'_{l'}-1)
-\pi_\omega(a((I_{m_l},k_{m_l}),(I_{.},k'_{.}))(m'_{l'})]
\}
\nonumber\\
&&
-\theta(m'_{l'}-m'_{l'-1})\theta(m'_{l'+1}-m'_{l'})\;
\times\nonumber\\&&\times
\{
[(\theta(.-m_{l-1})\theta(m_{l+1}-.))(m_l-1)
-(\theta(.-m_{l-1})\theta(m_{l+1}-.))(m_l)]
\pi_\omega(a((I_{.},k_{.}),(I_{m'_{l'}},k'_{m'_{l'}}))](m_l-1)
\nonumber\\&&
+
(\theta(.-m_{l-1})\theta(m_{l+1}-.))(m_l)
[\pi_\omega(a((I_{.},k_{.}),(I_{m'_{l'}},k'_{m'_{l'}}))(m_l-1)
-\pi_\omega(a((I_{.},k_{.}),(I_{m'_{l'}},k'_{m'_{l'}}))(m_l)]
\}
\}
\nonumber
\ea
Now 
\ba \label{6.43}
&&
(\theta(.-m_{l-1})\theta(m_{l+1}-.))(m_l-1)
-(\theta(.-m_{l-1})\theta(m_{l+1}-.))(m_l)
\nonumber\\
&=&
[\theta(m_l-1-m_{l-1})-\theta(m_l-m_{l-1})]\theta(m_{l+1}-(m_l-1))
+
\theta(m_l-m_{l-1})[\theta(m_{l+1}-(m_l-1))-\theta(m_{l+1}-m_l)]
\nonumber\\
&=&
-\delta_{m_l,m_{l-1}}\theta(m_{l+1}-(m_l-1))
+\delta_{m_l,m_{l+1}+1}\theta(m_l-m_{l-1})
\ea
and similarly for the primed quantities. This allows us to carry out the 
sum over $m_l,m'_{l'}$ in the first and third term of (\ref{6.42}) which 
simplifies to 
\ba \label{6.44}
&&
\sum_{m_l,m'_{l'}=1}^M \;\delta_{m_l,m'_{l'}}\;
\{
\theta(m_l-m_{l-1})\theta(m_{l+1}-m_l)\;
\times\\&&\times
\{
[-\delta_{m'_{l'},m'_{l'-1}}\theta(m'_{l'+1}-(m'_{l'}-1))
+\delta_{m'_{l'},m'_{l'+1}+1}\theta(m'_{l'}-m'_{l'-1})]
\pi_\omega(a((I_{m_l},k_{m_l}),(I_{.},k'_{.}))](m'_{l'}-1)
\nonumber\\&&
+
(\theta(.-m'_{l'-1})\theta(m'_{l'+1}-.))(m'_{l'})
{[}\pi_\omega(a((I_{m_l},k_{m_l}),(I_{.},k'_{.}))(m'_{l'}-1)
-\pi_\omega(a((I_{m_l},k_{m_l}),(I_{.},k'_{.}))(m'_{l'})]
\}
\nonumber\\
&&
-\theta(m'_{l'}-m'_{l'-1})\theta(m'_{l'+1}-m'_{l'})\;
\times\nonumber\\&&\times
\{
{[}-\delta_{m_l,m_{l-1}}\theta(m_{l+1}-(m_l-1))
+\delta_{m_l,m_{l+1}+1}\theta(m_l-m_{l-1})]
\pi_\omega(a((I_{.},k_{.}),(I_{m'_{l'}},k'_{m'_{l'}}))](m_l-1)
\nonumber\\&&
+
(\theta(.-m_{l-1})\theta(m_{l+1}-.))(m_l)
{[}\pi_\omega(a((I_{.},k_{.}),(I_{m'_{l'}},k'_{m'_{l'}}))(m_l-1)
-\pi_\omega(a((I_{.},k_{.}),(I_{m'_{l'}},k'_{m'_{l'}}))(m_l)]
\}
\}
\nonumber
\ea
Now when comparing with the continuum calculation, (\ref{6.44}) is 
almost exactly 
the discrete counterpart of the result that one gets when integrating the 
derivatives of the $\delta-$distributions, coming from the Poisson 
brackets, by parts. The derivatives then hit the $\theta-$functions which 
results in a second $\delta-$distribution since $\theta'(x)=\delta(x)$. 
The only difference is that 
in the classical theory the operator $\pi_\omega(a(.,.,.,.))$ is replaced 
by the constant $1$ so that the second and fourth term in (\ref{6.44}) are 
missing. However, as we will see below, in a semiclassical state the 
difference between (in the sense of expectation values) the operator and  
the constant is of order $1/M$ and hence is suppressed semiclassically.
Thus, dropping the extra terms, up to quantum corrections which are the 
result from reordering terms, the above discrete calculation precisely 
reproduces the classical continuum calculation. In particular, after
carrying out the sum over the Kronecker $\delta$'s we obtain from 
(\ref{6.44}), dropping boundary terms of order $1/M$   
\ba \label{6.45}
&& \theta(m'_{l'+1}-m'_{l'-1}+1)
[\theta(m'_{l'+1}-m_{l-1}+1)\theta(m_{l+1}-m'_{l'+1}-1)
-\theta(m'_{l'-1}-m_{l-1})\theta(m_{l+1}-m'_{l'-1})]
\nonumber\\&&
-
\theta(m_{l+1}-m_{l-1}+1)
[\theta(m_{l+1}-m'_{l'-1}+1)\theta(m'_{l'+1}-m_{l+1}-1)
-\theta(m_{l-1}-m'_{l'-1})\theta(m'_{l'+1}-m_{l-1})]
\nonumber\\
&&
\ea
The effect is thus that the path ordering of the primed and unprimed 
labels gets intermingled. One now inserts a unity
\be \label{6.46}
1=\sum_{\pi\in S_{n+n'-2}} 
\theta(p_{\pi(2)}-p_{pi(1)}) .. \theta(p_{\pi(n+n'-2)}-p_{\pi(n+n'-3)}) 
\ee
where 
$(p_1,..,p_{n+n'-2})=(m_1,..,\hat{m}_l,..,m_n,m'_1,..,\hat{m}'_{l'},..,
m'_{n'})$ and notices that due to the partly still existing projection 
on the orderings $m_1\le .. \le m_n$ and $m'_1\le .. \le m'_{n'}$ 
only the reshuffle sums displayed in section \ref{s3} survive. The 
remaining calculation is thus identical to the continuum for which we 
refer the reader to \cite{6,16,7,15}.

\subsection{Classical Limit of the {\bf Quantum Pohlmeyer Charges}}
\label{s6.7}

Thus, we have verified that ${\cal H}_\omega$ carries a representation 
of the {\bf Quantum Pohlmeyer Algebra} with precise quantum corrections 
provided we show that these corrections are subleading in the 
semiclassical limit of the theory. We will now show that this is 
actually the case.
To do this we use the background 
independent semiclassical techniques developed in \cite{26}:\\
We choose a graph $\gamma$ with large $M=|\gamma|$ 
and use the same parameter $L$ with the dimension of 
length that we used in (\ref{6.32a}).
Then we consider the set 
${\cal S}_{\gamma,L}$ of momentum network labels $s$ such that 
$\gamma(s)=\gamma$ and such that $n^I_\mu(s):=k^I_\mu(s) L$ is an
integer
for every $\mu=0,..,D-1$ and 
every $I\in \gamma$. Given a point 
$m_0:=(\pi^0_\mu(x),X^\mu_0(x))_{x\in S^1}$ in the classical phase space 
${\cal M}$ we construct the following 
quantities for $s\in {\cal S}_\gamma$  
\be \label{6.47}
W_\pm(s,m_0):=\exp(i\sum_{I\in\gamma} k^I_\mu(s)\int_I\;
dx\;[\pi^\mu_0(x)\pm X^{\mu\prime}_0(x)])
\ee
We can rewrite 
(\ref{6.47}) in the form 
\be \label{6.48}
W_\pm(s,m_0)=\prod_{\mu=0}^{D-1}\;\prod_{I\in\gamma} 
[W_\pm(\mu,I,m_0)]^{n^I_\mu},\;\;
W_\pm(\mu,I,m_0)=
\exp(\int_I\;dx\;
[\pi^\mu_0(x)+iX^{\mu\prime}(x)]/L)
\ee
where $k^I_\mu(s)=n^I_\mu/L$, $n^I_\mu\in \Zl$.
Next we choose any $DM$ real numbers $r_\mu^I $ such that 
$r_\mu^I-r_\mu^J\not\in\Ql$ for $I\cap J\subset \partial I$ and 
$\mu=0,..,D-1$ and  
denote the corresponding momentum network labels with momenta
$k^I_{\mu 0}:=r_\mu^L /L$ by $s_0$. 

We now define a semiclassical state by  
\be \label{6.50}
\psi^\pm_{\gamma,L,m_0}:=\sum_{s\in {\cal S}_{\gamma,L}}\;
e^{-t\lambda(s)/2}\; \overline{W_\pm(s)}\; 
\pi_\omega(W_\pm(s+s_0))\Omega_\omega
\ee
where 
\be \label{6.51}
t:=(\frac{\ell_s}{L})^2,\;\;
\lambda(s)=\sum_{\mu=0}^{D-1}\sum_{I\in \gamma}\; (n^I_\mu)^2
\ee
For the motivation to consider precisely those states see \cite{26} or our 
companion paper \cite{11}. The states (\ref{6.50}) are normalizable due to 
the damping factor as we will see but not normalized. The parameter $t$ is 
called the ``classicality parameter'' for reasons that will become 
obvious in a moment. Notice that our choices imply that every state in the 
infinite sum in (\ref{6.51}) really has $\gamma$ as the underlying graph.

When we apply $\pi_\omega(Z_\pm)$ to (\ref{6.51}) we obtain a specific 
linear combination of operators of the form $\pi_\omega(W_\pm(s))$ with
$s$ in the ``lattice'' ${\cal S}_{\gamma,L}$. 
Again, due to the translation by $s_0$ in (\ref{6.50}) none of these 
operators maps us out of ${\cal H}^\pm_{\omega,\gamma}$ so that the 
projections in
(\ref{6.32}) act trivially. Let us compute the action 
of these operators on our semiclassical states. We find 
\ba \label{6.52}
\pi_\omega(W_\pm(s))\psi^\pm_{\gamma,L,m_0}
&=&
\sum_{s'\in {\cal S}_{\gamma,L}}\;e^{-\frac{t}{2}\lambda(s')}
\overline{W_\pm(s',m_0)}\; e^{\mp i \ell_s^2\alpha(s,s_0+s')/2}\;
\pi_\omega(W_\pm(s+s'+s_0))\Omega_\omega
\\
&=&
\sum_{s'\in {\cal S}_{\gamma,L}}\;e^{-\frac{t}{2}\lambda(s'-s)}
\overline{W_\pm(s'-s,m_0)}\; e^{\mp i \ell_s^2\alpha(s,s_0+s'-s)/2}\;
\pi_\omega(W_\pm(s'+s_0))\Omega_\omega
\nonumber\\
&=& W_\pm(s,m_0)\; e^{\mp i \ell_s^2\alpha(s,s_0)/2}\;
\sum_{s'\in {\cal S}_{\gamma,L}}\;e^{-\frac{t}{2}\lambda(s'-s)}
\overline{W_\pm(s',m_0)}\; e^{\mp i \ell_s^2\alpha(s,s')/2}\;
\pi_\omega(W_\pm(s'+s_0))\Omega_\omega
\nonumber
\ea
where we have used translation invariance of the lattice and antisymmetry 
as well as bilinearity of the function 
\be \label{6.53}
\alpha(s,s'):=\sum_{I\in \gamma(s),\;I'\in \gamma(s')}\;
[k^I(s)\cdot k^{I'}(s')]\; \alpha(I,I')
\ee
We thus find for the expectation value 
\ba \label{6.54}
&&\frac{<\psi^\pm_{\gamma,L,m_0},
\pi_\omega(W_\pm(s))\psi^\pm_{\gamma,L,m_0}>}{||\psi^\pm_{\gamma,L,m_0}||^2}
\nonumber\\
&=&W_\pm(s,m_0)\; e^{\mp i \ell_s^2\alpha(s,s_0)/2}\;
\frac{
\sum_{s'\in {\cal S}_{\gamma,L}}\;
e^{-\frac{t}{2}[\lambda(s'-s)+\lambda(s')]}\; 
e^{\mp i \ell_s^2\alpha(s,s')/2}
}
{
\sum_{s'\in {\cal S}_{\gamma,L}}\; e^{-t \lambda(s')}
}
\ea
In order to estimate this expression, let us write
\be \label{6.55}
\mp i \ell_s^2\alpha(s,s')/2
=\pm i\frac{t}{2}\sum_{I,\mu} n^I_\mu(s')[\sum_J n^{J\mu}(s)\alpha(I,J)]
=:\pm i\frac{t}{2}\sum_{I,\mu} n^I_\mu(s') c^\mu_I(s)
\ee
so that (\ref{6.54}) factorizes
\ba \label{6.56}
&&\frac{<\psi^\pm_{\gamma,L,m_0},
\pi_\omega(W_\pm(s))\psi^\pm_{\gamma,L,m_0}>}{||\psi^\pm_{\gamma,L,m_0}||^2}
\nonumber\\
&=&W_\pm(s,m_0)\; e^{\mp i \ell_s^2\alpha(s,s_0)/2}\;
e^{-\frac{t}{2}\lambda(s)}\;
\prod_{\mu,I} 
\frac{
\sum_{l\in \Zl} e^{-t(l^2-l [n^I_\mu(s)\pm i c^\mu_I(s)/2])}
}
{
\sum_{l\in \Zl} e^{-tl^2}
}
\ea
We are interested in the limit of small $t$ and large $M$ of this 
expression. In order to estimate it, the presentation (\ref{6.56})
is not very useful because the series in both numerator and denominator 
converge only slowly. Hence we apply the Poisson summation formula
\cite{26} and transform (\ref{6.56}) into
\ba \label{6.57}
&&\frac{<\psi^\pm_{\gamma,L,m_0},
\pi_\omega(W_\pm(s))\psi^\pm_{\gamma,L,m_0}>}{||\psi^\pm_{\gamma,L,m_0}||^2}
\\
&=&W_\pm(s,m_0)\; e^{\mp i \ell_s^2\alpha(s,s_0)/2}\;
e^{-\frac{t}{2}\lambda(s)}\;
\prod_{\mu,I} e^{\frac{(t[n^I_\mu(s)\pm i c^\mu_I(s)/2])^2}{4t}} 
\frac{
\sum_{l\in \Zl} \; e^{-\frac{\pi^2 l^2}{t}} \;
e^{-2\pi l[n^I_\mu(s)\pm i c^\mu_I(s)/2]}
}
{
\sum_{l\in \Zl} \; e^{-\frac{\pi^2 l^2}{t}} 
}
\nonumber\\
&=&
W_\pm(s,m_0)\; e^{\mp i \ell_s^2\alpha(s,s_0)/2}\;
e^{-\frac{t}{2}\lambda(s)}\;
e^{\frac{t}{4}\sum_{\mu,I} [n^I_\mu(s)\pm i c^\mu_I(s)/2]^2}\;
\prod_{\mu,I} 
\frac{1+
2\sum_{l=1}^\infty \; e^{-\frac{\pi^2 l^2}{t}} \;
\cosh(2\pi l[n^I_\mu(s)\pm i c^\mu_I(s)/2])
}
{1+
2\sum_{l\in \Zl} \; e^{-\frac{\pi^2 l^2}{t}} 
}
\nonumber\\
&=&
W_\pm(s,m_0)\; e^{\mp i \ell_s^2\alpha(s,s_0)/2}\;
e^{-\frac{t}{4}\sum_{\mu,I} [(n^I_\mu(s))^2+(c^\mu_I(s)/2)^2]}\;
\prod_{\mu,I} 
\frac{1+
2\sum_{l=1}^\infty \; e^{-\frac{\pi^2 l^2}{t}} \;
\cosh(2\pi l[n^I_\mu(s)\pm i c^\mu_I(s)/2])
}
{1+
2\sum_{l\in \Zl} \; e^{-\frac{\pi^2 l^2}{t}} 
}
\nonumber
\ea
where in the last step we have used $\sum_{\mu,I} n^I_\mu(s) c^\mu_I(s)=0$.

Notice that for the applications that we have in mind we have 
\be \label{6.58}
n_\mu^I(s)=\sum_{l=1}^n \delta_{I,I_{m_l}} \delta_\mu^{\mu_l}
\ee
and 
\be \label{6.59}
c^\mu_I(s)=\sum_J n^{\mu I}(s) \alpha(I,J)=
=\sum_{l=1}^n \eta^{\mu\mu_l} \alpha(I,I_{m_l})
=\sum_{l=1}^n \eta^{\mu\mu_l} 
{[}\delta_{I,I_{m_l+1}}-\delta_{I,I_{m_l-1}}]
\ee
It follows that from the $MD$ numbers $n_\mu^I(s)$ and $c^\mu_I(s)$  
respectively only $n$ respectively $2n$ are non vanishing. Thus
\be \label{6.60}
\sum_{\mu,I} [(n^I_\mu(s))^2+(c^\mu_I(s)/2)^2]=\frac{3}{2} n
\ee
which is actually independent of the specific configuration of 
$\mu_1..\mu_n$ and $I_{m_1},..,I_{m_n}$.\\ Next 
\be \label{6.61}
\ell_s^2\alpha(s,s_0)
=t\sum_{I,J} n^I_\mu(s) r^{\mu J}(s_0) \alpha(I,J)
=t\sum_{l=1}^n [r^{\mu_l I_{m_l+1}}-r^{\mu_l I_{m_l+1}}]
\ee
We can make (\ref{6.61}) vanish identically, for example, by choosing 
$M=|\gamma|$ to be an even number and by choosing 
$r^{I_m}_\mu=\sqrt{2}$ for $m$ even and 
$r^{I_m}_\mu=\sqrt{3}$ for $m$ odd. This meets all our requirements on the 
real numbers $r^I_\mu$ that we have imposed.\\
Finally, consider the product over the $MD$ pairs $(I,\mu)$ appearing in 
(\ref{6.57}). Precisely for the $n$ pairs $(I_{m_l},\mu_l),\;l=1,..,n$
and for the $2n$ pairs $(I_{m_l\pm 1},\mu_l),\;l=1,..,n$ the factor
is different from unity. Thus (\ref{6.57}) simplifies to 
\ba \label{6.62}
&&\frac{<\psi^\pm_{\gamma,L,m_0},
\pi_\omega(W_\pm(s))\psi^\pm_{\gamma,L,m_0}>}{||\psi^\pm_{\gamma,L,m_0}||^2}
\nonumber\\
&=& W_\pm(s,m_0)\; e^{-\frac{3nt}{8}} \;
[\frac{
1+2\sum_{l=1}^\infty \; e^{-\frac{\pi^2 l^2}{t}} \;\cosh(2\pi l)
}
{1+
2\sum_{l\in \Zl} \; e^{-\frac{\pi^2 l^2}{t}} 
}]^n \;
[\frac{
1+2\sum_{l=1}^\infty \; e^{-\frac{\pi^2 l^2}{t}} \;\cos(\pi l)
}
{1+
2\sum_{l\in \Zl} \; e^{-\frac{\pi^2 l^2}{t}} 
}]^{2n}
\ea
It is easy to see that the constants 
\ba \label{6.63}
c_1(t) &:=& 2\sum_{l\in \Zl} \; e^{-\frac{\pi^2 l^2}{t}} 
\nonumber\\
c_2(t) &:=& 2\sum_{l\in \Zl} \; e^{-\frac{\pi^2 l^2}{t}} \cosh(2\pi l) 
\nonumber\\
c_3(t) &:=& 2\sum_{l\in \Zl} \; e^{-\frac{\pi^2 l^2}{t}} \cos(\pi l) 
\ea
vanish faster than any power of $t^{-1}$ as $t\to 0$. Hence we may 
finish our computation by displaying the compact formula
\be \label{6.64}
\frac{<\psi^\pm_{\gamma,L,m_0},
\pi_\omega(W_\pm(s))\psi^\pm_{\gamma,L,m_0}>}{||\psi^\pm_{\gamma,L,m_0}||^2}
=W_\pm(s,m_0)\; e^{-\frac{3nt}{8}} \;
(\frac{[1+c_2(t)][1+c_3(t)]^2}{[1+c_1(t)]^3})^n
\ee
Obviously, the expectation value of $\pi_\omega(W_\pm(s))$ 
in the states $\psi^\pm_{\gamma,L,m_0}$, divided by the 
the classical value at the phase space point $m_0$ 
(a complex number of modulus one), differs from unity by a constant of
order $nt$. We could even get exact agreement by a finite 
``renormalization'' of 
the operator $\pi_\omega(W_\pm(s))$ by multiplying it by inverse of the 
$t-$dependent factor in (\ref{6.64}) which depends on $s$ only through 
the invariant quantity $n=|\gamma(s)|$ (counting non -- empty intervals 
only).

Let us now come to the expectation value of the $\pi_\omega(Z_\pm)$.
On the states $\psi^\pm_{\gamma,L,m_0}$ these operators reduce to 
\ba \label{6.65}
\pi_\omega(Z^{\mu_1..\mu_n}_\pm) &=&
(\frac{L}{2i})^n C_n\cdot 
\frac{1}{n!}\sum_{\pi\in S_n} \sum_{\sigma_1,..,\sigma_n=\pm 1}
\sum_{1\le m_1\le .. \le m_n\le M}
\nonumber \times\\
&& \times
\exp(\mp i t/2\sum_{j=1}^{n-1} \sum_{l=j+1}^n 
\sigma_{\pi(j)}\sigma_{\pi(l)} [n^{\pi(j)}\cdot n^{\pi(l)}]
[\delta_{m_{\pi(l)},m_{\pi(j)}+1}
-\delta_{m_{\pi(l)},m_{\pi(j)}-1}])
\nonumber\times\\
&& \times
\pi_\omega(W_\pm((I_{m_1},\sigma_1 \frac{1}{L} \delta^{(1)})\cup .. \cup
(I_{m_n},\sigma_n \frac{1}{L} \delta^{(n)}))
\ea
where $\delta^{(l)}_\mu=\delta^{\mu_l}_\mu$. Notice that the phase in the 
second 
line of (\ref{6.65}) is non-trivial only on configurations $I_{m_1},..,
I_{m_n}$ which would have zero mesure in the continuum and even in that 
case it differs from unity only by a term of order $nt$. Thus, up to 
$nt/M$
corrections we can ignore that phase (alternatively we could avoid the 
phase 
right from the beginning by redifining our operators as to perform the 
sum over $m_{l+1}\ge m_l+2,\;l=1,..,n-1,\;m_n\le M,\;m_1\ge 1$).
Then taking the expectation value of (\ref{6.65}) using (\ref{6.64})
gives up to $O(nt,\frac{1}{M})$ corrections 
\be \label{6.66}
\pi_\omega(Z^{\mu_1..\mu_n}_\pm) =
L^n 
C_n\cdot 
\sum_{1\le m_1\le .. \le m_n\le M}
\prod_{l=1}^n\;\;\sin(\frac{1}{L} Y^{\mu_l}_\pm(I_{m_l},m_0))
\ee
It is clear that for sufficiently large $M$ (\ref{6.66}) 
is a very good approximation to $Z^{\mu_1..\mu_n}_\pm(m_0)$. In fact the 
limit $\lim_{M\to\infty}$ reproduces the exact integral.

This kind of calculations can be used to confirm that the terms that 
we claimed to be subleading in the commutation relations for the 
{\bf Quantum Pohlmeyer Charges} are indeed neglible in the semiclassical 
limit. They can also be used to show that the relative fluctuations 
(absolute fluctuation divided by the square of the expectation value) of 
the quantum invariants are of order $O(nt/M)$.
We will not display these tedious but straightforward calculations 
here but refer the interested reader to \cite{26} for similar calculations 
performed in LQG.

\subsection{Physical Hilbert Space}
\label{s6.8}

What we have found so far is a representation $(\pi_\omega,{\cal 
H}_\omega)$ for the algebra of quantum invariants or Dirac 
Observables corresponding to
the classical $Z^{(n)}_\pm$. These operators commute with the unitary 
representations $U^\pm_\omega(\varphi)$ of the two copies of 
Diff$(S^1)$ generated by the two Virasoro constraints. The 
$\pi_\omega(Z^{k_1..k_n}_\pm)$ transform covariantly under Poincar\'e 
transformations. What is left to do is to find the physical Hilbert space.
In the present situation one can define the physical Hilbert space in two 
equivalent ways. The first one corresponds to gauge fixing, the second one 
to group averaging as defined in section \ref{s5}.\\
\\
{\bf Gauge Fixing:}\\
\\
Given any number $M$ we fix a graph $\gamma_M$ once and for all with
$|\gamma_M|=M$. We then consider the gauge fixed Hilbert space 
${\cal H}^\pm_{gf}$ defined by 
the completion of the finite linear span of states 
$\pi_\omega(W_\pm(s))\Omega_\omega$ with $\gamma(s)=\gamma_M$ whenever
$|\gamma(s)|=M$. By definition, the $\pi_\omega(Z^{(n)}_\pm)$ preserve
${\cal H}_{gf}$. Consider any other choice $M\mapsto \gamma'_M$. For each 
$M=0,1,..$ there is an element $\varphi\in $Diff$_\pm(S^1)$ such that 
$\varphi_M(\gamma_M)=\gamma'_M$. It follows that if 
$\gamma(s'_{kl})=\gamma'_{M_k}$ for all $l=1,..,L_k$ 
then
\ba \label{6.67}
&& 
<[\sum_{k=1}^K \sum_{l=1}^{L_k} z_{kl} \pi_\omega(W_\pm(s'_{kl}))]
\Omega_\omega, \pi_\omega(Z_\pm^{(n)})
[\sum_{k=1}^K \sum_{l=1}^{L_k} \tilde{z}_{kl} 
\pi_\omega(W_\pm(s'_{kl}))]\Omega_\omega>_\omega
\nonumber\\
&=&
\sum_{k=1}^K <[\sum_{l=1}^{L_k} z_{kl} \pi_\omega(W_\pm(s'_{kl}))]
\Omega_\omega, \pi_\omega(Z_\pm^{(n)})
[\sum_{l=1}^{L_k} \tilde{z}_{kl} 
\pi_\omega(W_\pm(s'_{kl}))]\Omega_\omega>_\omega
\nonumber\\
&=&
\sum_{k=1}^K <[\sum_{l=1}^{L_k} z_{kl} \pi_\omega(W_\pm(s_{kl}))]
\Omega_\omega, 
U^\pm_\omega(\varphi_{M_k})^{-1}\pi_\omega(Z_\pm^{(n)})
U^\pm_\omega(\varphi_{M_k})
[\sum_{l=1}^{L_k} \tilde{z}_{kl} 
\pi_\omega(W_\pm(s_{kl}))]\Omega_\omega>_\omega
\nonumber\\
&=&
<[\sum_{k=1}^K \sum_{l=1}^{L_k} z_{kl} \pi_\omega(W_\pm(s_{kl}))]
\Omega_\omega, \pi_\omega(Z_\pm^{(n)})
[\sum_{k=1}^K \sum_{l=1}^{L_k} \tilde{z}_{kl} 
\pi_\omega(W_\pm(s_{kl}))]\Omega_\omega>_\omega
\ea
so that expectation values of the invariants coincide. Here we have made 
use of the orthogonality relations of the states defined over different 
graphs. In fact, the two gauge fixed 
representations are unitarily equivalent because both are equivalent 
to direct sums
of Hilbert spaces ${\cal H}^{\pm}_{\omega,\gamma_M}$ and 
(${\cal H}^{\pm}_{\omega,\gamma'_M}$) respectively which are preserved by 
all charges 
and the unitary operator that maps between the Hilbert spaces is 
the one that maps ${\cal H}^\pm_{\omega,\gamma_M}$
to ${\cal H}^\pm_{\omega,\gamma'_M}$.\\
\\
{\bf Group Averaging:}\\
\\
Since on the $\pi_\omega(W_\pm(s))\Omega_\omega$ the gauge group acts by 
diffeomorphisms, we can directly copy the analysis from \cite{24}.
We will just summarize the main results.

To each 
momentum network label $s$ we assign a class $[s]$ defined by the orbit
of $s$, that is, $[s]:=\{\varphi(s);\;\varphi\in \mbox{Diff}(S^1)\}$. To 
each class $[s]$ we assign a distribution on the space $\Phi_{Kin}$,
consisting of the finite linear combinations of states of the form 
$\pi_\omega(W_\pm(s))\Omega_\omega$, defined by
\be \label{6.68}
\rho^\pm_{[s]}(\pi_\omega(W_\pm(s'))\Omega_\omega):=\chi_{[s]}(s')=
\delta_{[s],[s']}
\ee
where $\chi_.$ denotes the characteristic function. These states are 
the images of the anti -- linear rigging map
\be \label{6.69}
\rho^\pm_{[s]}:=\rho(\pi_\omega(W_\pm(s))\Omega_\omega)
\ee
and formally we have 
\be \label{6.69a}
\rho^\pm_{[s]}=
\sum_{s'\in 
[s]}\;<\pi_\omega(W_\pm(s'))\Omega_\omega,.>_\omega
\ee
which explains the word ``group averaging''.

The distributions $\rho_{[s]}$ belong to the dual $\Phi_{Kin}^\ast$ of 
$\Phi_{Kin}$ on which one defines duals of operators $O$ by
\be \label{6.70}
[O'\rho^\pm_{[s]}](f):=\rho_{[s]}(O^\dagger f)
\ee
where $O^\dagger$ is the adjoint of $O$ in 
${\cal H}_{Kin}={\cal H}_\omega$ and $f\in \Phi_{Kin}$. It follows that 
\be \label{6.71}
[U^\pm_\omega(\varphi)]'\rho^\pm_{[s]}=\rho^\pm_{[s]}
\ee
is invariant, hence they solve the Virasoro constraints exactly. One 
defines the physical Hilbert space 
${\cal H}^\pm_{Phys}$ as the completion of the finite linear span 
of the $\rho^\pm_{[s]}$ under the inner product
\be \label{6.72}
<\rho^\pm_{[s]},\rho_{[s']}>_{Phys}
:=\rho_{[s']}(\pi_\omega(W_\pm(s))\Omega_\omega)
=\delta_{[s],[s']}
\ee

The action of the charges $\pi_\omega(Z^{(n)}_\pm)$ is again by duality
\be \label{6.73}
[(\pi_\omega(Z^{(n)}_\pm))'\rho^\pm_{[s]}](f)
\rho^\pm_{[s]}]((\pi_\omega(Z^{(n)}_\pm)f)
\ee
where we have used symmetry of the operators. It is not difficult to see 
that 
\be \label{6.74}
(\pi_\omega(Z^{(n)}_\pm))'\rho(\pi_\omega(W_\pm(s))\Omega_\omega)
=\rho((\pi_\omega(Z^{(n)}_\pm) \pi_\omega(W_\pm(s))\Omega_\omega)
\ee
so that the rigging map $\rho$ commutes with the invariants due to 
diffeomorphism invariance. It follows from the general properties of a 
rigging map \cite{19a} 
that the dual operators $(\pi_\omega(Z^{(n)}_\pm))'$ are symmetric as well
on ${\cal H}^\pm_{Phys}$. 

Finally, it is clear that dual representation of the charges 
is unitarily equivalent to the gauge fixed representation above by simply 
identifying ${\cal H}^\pm_{Phys}$ with ${\cal H}^\pm_{gf}$.

\subsection{Gravitons}
\label{s6.9}

It is easy to check that in our notation
the graviton states in usual string theory in the 
lightcone gauge are 
given by the symmetric, transverse and traceless components of
\be \label{6.75}
|a,b;p>:=[\int_{S^1}\;dx\; e^{ix}\;\hat{Y}^a_-]\;[\int_{S^1}\;dx\; 
e^{-ix}\;\hat{Y}^b_+]\;|p>
\ee
with $a,b=1,..D-1$ given by transversal indices and $|p>$ is the usual 
string theory vacuum (tachyon with momentum $p$). See e.g. \cite{11} 
for a 
derivation. Due to the mode functions $e^{\pm ix}$ appearing in 
(\ref{6.75}), graviton states are not gauge invariant states in the sense 
of section \ref{s6.8}. We can, however, describe them in our gauge fixed 
Hilbert space as the massless states ($p\cdot p=0$)
\be \label{6.76}
\Omega^{ab}_{\omega_p}:=\pi_{\omega_p}(W^{aM}_-)
\pi_{\omega_p}(W^{bM}_+)\Omega_{\omega_p}
\ee
in the limit $M\to\infty$ where ($S_M$ denotes symmetric projection)
\ba \label{6.77}
\pi_\omega(W^{aM}_\pm) &=&(\frac{L}{2i})^M\; S_M\cdot\; \prod_{m=1}^M 
\{[\pi_\omega(W_\pm(s^a_{c,m}))-\pi_\omega(W_\pm(s^a_{c,m}))^{-1}]
\mp 
i[\pi_\omega(W_\pm(s^a_{s,m})-\pi_\omega(W_\pm(s^a_{s,m}))^{-1}]\}
\nonumber\\
s^a_{c,m} &=& (\frac{\delta^a_\mu}{L} 
\cos(\mp\frac{2\pi i [m+1/2]}{M}),[\frac{m-1}{M},\frac{m}{M}]),\;\;
s^a_{s,m}=(\frac{\delta^a_\mu}{L} 
\sin(\mp\frac{2\pi i [m+1/2]}{M}),[\frac{m-1}{M},\frac{m}{M}])
\ea
Since, however, the operators (\ref{6.77}) are not gauge invariant, it is 
not clear what their meaning is in the light of the invariant description 
of this paper. Clearly, more work is needed in order to obtain a 
meaningful notion of graviton creation operators in terms of the invariant 
charges $\pi_\omega(Z^{(n)}_\pm)$. Within LQG this has only recently been 
understood in the linearized sector \cite{28} but an understanding in the 
full theory is still lacking. We will come back to this question in the 
companion paper \cite{11}. 

\section{Conclusions, Open Questions and Outlook}
\label{s7}

In this paper we have combined ideas from Loop Quantum Gravity, 
Algebraic Quantum Field Theory and Pohlmeyer's Theory of the 
invariant charges in order to construct quantum field 
theories for the closed, bosonic string in flat Minkowskian target space
which differ significantly from usual
string theory. Let us list once more the main differences:
\begin{itemize}
\item[1.] {\it Target Space Dimension}\\
There is no sign, neither from a ghost free spectrum requirement 
(covariant
quantization of usual string theory) nor from a Lorentz invariance 
requirement (lightcone quantization of usual string theory), of a critical
dimension. Our construction works in any dimension, especially $D=4$.
\item[2.] {\it Ghosts}\\
We always work with honest Hilbert spaces, mathematically ill -- defined 
objects such
as negative norm states are strictly avoided. Hence there are no ghosts 
to get rid of.
\item[3.] {\it Weyl Invariance}\\
We never introduce a worldsheet metric because we are working directly 
with the more geometrical Nambu -- Goto string rather than the Polyakov 
string. Thus, there is no artificial Weyl invariance introduced which is 
to be factored out later. 
\item[4.] {\it Conformal Invariance}\\
We never have to fix the (Weyl and) worldsheet diffeomorphism invariance 
by going to a conformal worldsheet gauge. Our formulation is manifestly 
worldsheet diffeomorphism invariant. Hence, there is never a residual 
gauge freedom corresponding to the conformal diffeomorphism group of the
flat worldsheet metric to be taken care of. For the same reason, conformal 
field theory does not play any role whatsoever in our approach.
\item[5.] {\it Virasoro Anomalies and Central Charge}\\
Following the tradition of algebraic QFT, we have separated the 
quantum algebra of string theory from its representation theory. On a 
properly chosen Weyl algebra of kinematical operators we have the 
local gauge symmetry group of the string corresponding to worldsheet 
diffeomorphisms and the global Poincar\'e symmetry acting by 
automorphisms. Then by standard operator theoretical constructions one 
obtains automatically an anomaly -- free, moreover unitary representation 
of both symmetry groups on an important subclass of cyclic 
representations, provided they exist. Hence, there are no anomalies 
in our formulation, the central charge vanishes. This is a direct 
consequence of carrying out a true Dirac quantization of the constraints
in contrast to standard string theory where only half of the constraints
are imposed strongly. We will come back to that point in our companion 
paper \cite{11}
\item[6.] {\it Existence of Representations}\\
We did not (yet) carry out a full analysis of the representation 
theory of the quantum string. However, we found at least one 
representation which fulfills all requirements.
\item[7.] {\it Invariants}\\
To the best of our knowledge, in standard string theory the problem of 
defining the classical Dirac observables as operators on the Hilbert space 
has not been addressed so far. The closest construction that we are aware 
of are the DDF operators in the lightcone gauge of the Virasoro 
constraints \cite{1}. In a fully worldsheet background independent and 
diffeomorphism invariant formulation 
that one is forced to from an LQG perspective, dealing with the Dirac
observables is mandatory. Fortunately, the invariant charges have been 
constructed already by Pohlmeyer and his collaborators. The example 
representation that we have constructed actually supports a 
specific quantum deformation of the classical charge algebra. The 
corresponding operators define invariant n -- point functions which are 
finite without UV -- divergences.
\item[8.] {\it Tachyon}\\
We saw that we can construct representations with arbitrary, non -- 
negative mass spectrum , so there is no tachyon in the spectrum. 
Usually the tachyon is (besides
the phenomenological need for fermionic matter) one of the motivations
for considering the superstring. Our example shows that this depends on 
the representation and is not always necessary. 
\end{itemize}
~~~~~~~~~~\\
In the jargon of standard string theory, one could summarize this by 
saying that {\it the LQG -- String presents a 
new and consistent solution to quantizing string theory}. Actually, there 
is not {\it the} LQG -- String, presumably there exist infinitely many 
solutions to the representation problem (which are consistent by 
definition).\\ 
\\
Of course, we do not claim that the particular representation we found is 
necessarily 
of any physical significance. In fact it cannot be since we have not 
included (yet) any fermionic degrees of freedom. Also, besides not having 
carried out a full analysis of the representation theory,
our analysis is incomplete in many respects as 
for instance we have not yet developed the S -- Matrix theory for 
the LQG -- String (however, sinze the {\bf Pohlmeyer Charges} are nothing 
else than invariant n -- point functions, this is presumably not very 
difficult). What is also missing, so far, is a comparison with the objects 
of usual string theory because it is hard to translate gauge dependent 
notions such as graviton states into our invariant language, see section 
\ref{s6.9}. Finally, there are four immediate open questions:\\
i)\\
First of all the {\bf Pohlmeyer Charges}
together with the the boost generators reconstruct the string embedding
$X^\mu$ 
(up to gauge transformations) completely only up to scalar multiples of 
the momentum $p^\mu$. It seems to be hard to define an invariant which 
captures this one parameter degree of freedom unless one includes string 
scattering \cite{6,15}.\\ 
ii)\\
Secondly, an 
interesting open question is whether one can find a supersymmetric (or
at least fermionic) extension of our Weyl algebra and if curved target 
spaces can be treated the same way. What is needed is an analogue of the 
$Y_\pm$ with the same simple commutation relations and the same 
simple behaviour under gauge transformations. If that would be possible
and if appropriately generalized {\bf Pohlmeyer Charges} could be 
found, then one 
could repeat the analysis of this paper because the structure of the
constraint algebra of the $V_\pm$ remains the same even for the 
supersymmetric extension and for curved target spaces.\\
iii)\\
Thirdly one might wonder whether an approach based on invariants as 
carried out in this paper is not
possible also for higher p -- brane theories such as the 
(super)membrane \cite{10} which is a candidate for M -- Theory.\\
iv)\\
Lastly one may wonder which other GNS representations 
one gets by constructing the folium of 
$\omega$. Of course, the folium should be based on $G-$invariant positive 
trace class operators, see section \ref{s4}. Thus one would try to define 
those from bounded operators constructed from the 
$\pi_\omega(Z^{(n)}_\pm)$. Notice that trace class operators are in 
particular compact, thus they must have discrete spectrum with all non 
vanishing eigenvalues of finite multiplicity. Since ${\cal H}_\omega$ is 
not separable, such an operator would have to have uncountably infinite 
multiplicity for the eigenvalue zero.\\ 
\\ 
Let us conclude by stressing once more that
the claim of this paper is certainly not to have found a 
full solution of string theory.
Rather, we wanted to point out two things:\\
First of all, that canonical and algebraic methods can be 
fruitfully combined in order to analyze the string. Secondly, that 
the specific Fock representation that one always uses in string theory 
is by far not the end of the story: The invariant representation theory of 
the quantum string, as we have defined it here, is presumably very rich 
and we encourage string theorists to study the string from the algebraic 
perspective and to systematically analyze all its representations.
This might lead to a natural resolution of major current puzzles 
in string theory, such as the cosmological constant puzzle \cite{27}
(120 orders of magnitude too large),
the tachyon condensation puzzle \cite{9} (unstable bosonic string 
vacua), the vacuum degeneracy puzzle
\cite{29} (huge moduli space of vacua upon compactification), the 
phenomeology puzzle \cite{30} (so far the standard model has not been 
found among all possible string vacua of the five superstring theories 
currently defined, even when including D -- branes) and finally the 
puzzle of proving perturbative finiteness beyond two loops \cite{31}. 
See the beautiful review \cite{32} for a status report on these issues.
Namely, it might
be that there are much simpler representations of the string, especially
in lower dimensions and possibly without supersymmetry, which avoid or 
simplify all or some these problems.

While this would be attractive, the existence of new, phenomenologically
sensible representations would demonstrate that $D=10,11,26$ dimensions,
supersymmetry and the matter content of the world are tied to a specific 
representation of string theory and hence would not be a prediction 
in this sense. We believe, however, that the potential discovery of new, 
physically interesting representations for string theory, in the sense of 
this paper, is a fascinating research 
project which could lead to major progress on the afore mentioned 
puzzles.\\
\\
\\
{\large Acknowledgements}\\
\\
We are grateful to Detlev Buchholz and Klaus Fredenhagen for 
repeatedly encouraging to look at the string from the algebraic point of 
view and to Jan Ambjorn and Hermann Nicolai for suggesting to apply Loop 
Quantum Gravity methods to string 
theory. We also would like to thank Dorothea Bahns, Gerrit 
Handrich, Catherine Meusburger and Karl -- Henning Rehren for very
fruitful discussions about Pohlmeyer's programme for string theory. 
Special thanks to Dorothea Bahns for a copy of her diploma thesis.

This research project was supported in part by 
funds from NSERC of Canada to the Perimeter Institute for Theoretical 
Physics.


\begin{thebibliography}{99}

\parskip -5pt


\bibitem{1} M.B. Green, J. Schwarz, E. Witten, ``Superstring Theory", vol.
1, 2, Cambridge University Press, Cambridge, 1987\\
J. Polchinski, ``String Theory", vol.1 : ``An Introduction to the Bosonic
String", vol. 2 : ``Superstring Theory and Beyond", Cambridge University
Press, Cambridge, 1998

\bibitem{2} 
C. Rovelli, ``Quantum Gravity'', Cambridge University Press,
Cambridge, 2004, at press\\
T. Thiemann, ``Modern Canonical Quantum General
Relativity'', Cambridge University Press, Cambridge, 2004, at press

\bibitem{2a}
A. Ashtekar, ``Quantum Geometry and Gravity: Recent Advances'',
gr-qc/0112038\\
L. Smolin, ``Quantum Gravity with a Positive Cosmological Constant'',
hep-th/0209079\\
C. Rovelli, ``Loop Quantum Gravity", Living Rev. Rel. {\bf 1} (1998) 1,
gr-qc/9710008\\
T. Thiemann,``Lectures on Loop Quantum Gravity'', Lecture Notes in
Physics, {\bf 631} (2003) 41 -- 135, gr-qc/0210094

\bibitem{3} R. Haag, ``Local Quantum Physics'', 2nd ed., Springer Verlag,
Berlin, 1996

\bibitem{4} D. Buchholz, ``Algebraic Quantum Field Theory: A Status 
Report'', Plenary talk given at 13th International Congress in 
Mathematical Physics (ICMP 2000), London, England, 17-22 Jul 2000, 
math-ph/0011044

\bibitem{5} G. B. Folland, ``Harmonic Analysis in Phase Space'', Ann. 
Math. Studies, no. 122, Princeton University Press, Princeton N.J., 1989

\bibitem{6} 
K. Pohlmeyer, ``A Group Theoretical Approach to the Quantization of the
Free Relativistic Closed String", Phys. Lett. {\bf B119} (1982) 100;
``The Invariant Charges of the Nambu-Goto Theory in WKB
Approximation", Commun. Math. Phys. {\bf 105} (1986) 629;
``The Algebra formed by the Charges of the Nambu-Goto Theory: Casimir
Elements", Commun. Math. Phys. {\bf 114} (1988) 351;
``Uncovering the Detailed Structure of the Algebra Formed by the Invariant
Charges of Closed Bosonic Strings Moving in (1+2)-Dimensional Minkowski
Space", Commun. Math. Phys. {\bf 163} (1994) 629-644;
``The Invariant Charges of the Nambu-Goto Theory: Non-Additive
Composition Laws", Mod. Phys. Lett. {\bf A10} (1995) 295-308;
``The Nambu-Goto Theory
of Closed Bosonic Strings Moving in (1+3)-Dimensional Minkowski
Space: The Quantum Algebra of Observables", Annalen Phys. {\bf 8}
(1999) 19-50, [hep-th/9805057]

\bibitem{16} K. Pohlmeyer, K.H. Rehren,
``Algebraic Properties of the Invariant Charges of the Nambu-Goto Theory",
Commun.Math.Phys. {\bf 105} (1986) 593;
``The Algebra formed by the Charges of the Nambu-Goto Theory:
Identification of a Maximal Abelean Subalgebra"
Commun. Math. Phys. {\bf 114} (118) 55;
``The Algebra formed by the Charges of the Nambu-Goto Theory: Their
Geometric Origin and Their Completeness", Commun. Math. Phys.
{\bf 114} (1988) 177

\bibitem{7}
K. Pohlmeyer, M. Trunk, ``The Invariant Charges of the Nambu-Goto Theory:
Quantization of Non-Additive Composition Laws", hep-th/0206061\\
G. Handrich, C. Nowak, ``The Nambu-Goto Theory
of Closed Bosonic Strings Moving in (1+3)-Dimensional Minkowski
Space: The Construction of the Quantum Algebra of Observables up to
Degree Five", Annalen
Phys. {\bf 8} (1999) 51-54, [hep-th/9807231]\\
G. Handrich, ``Lorentz Covariance of the Quantum Algebra of Observables:
Nambu-Goto Strings in 3+1 Dimensions",
Int. J. Mod. Phys. {\bf A17} (2002) 2331-2349\\
G. Handrich, C. Paufler, J.B. Tausk, M. Walter,
``The Representation of the Algebra of Observables of the Closed Bosonic 
String
in 1+3 Dimensions: Calculation to Order $\hbar^7$", math-ph/0210024 

\bibitem{17}
C. Meusburger, K.H. Rehren, ``Algebraic Quantization of the Closed Bosonic
String", math-ph/0202041

\bibitem{8} 
R. F. Streater, A. S. Wightman, ``PCT, Spin and Statistics, and
all that", Benjamin, New York, 1964

\bibitem{10} R. Helling, H. Nicolai, ``Supermebranes and Matrix Theory",
hep-th/9809103

\bibitem{11} T. Thiemann, ``The LQG -- String: Loop Quantum Gravity 
Quantization of String Theory II. Curved Target Space''

\bibitem{11a} A. Starodubtsev, ``String Theory in a Vertex Operator 
Representation: A Simple Model for Testing Loop Quantum Gravity'',
gr-qc/0201089

\bibitem{11b} J. Magueijo, L. Smolin, ``String Theories with Deformed 
Energy Momentum Relations, and a Possible Non -- Tachyonic Bosonic 
String'', hep-th/0401087

\bibitem{12} A. Hanson, T. Regge, C. Teitelboim, ``Constrained Hamiltonian
Systems", Accdemia Nazionale dei Lincei, Roma, 1976

\bibitem{13} M. Henneaux, C. Teitelboim, ``Quantization of Gauge Systems"
Princeton University Press, Princeton, 1992

\bibitem{14} V. I. Arnol'd, ``Dynamical Systems III. Mathematical Aspects
of Classical and Celestial Mechanics", Springer Verlag, Berlin, 1993

\bibitem{14a} A.M. Perelomov, ``Integrable systems of classical mechanics 
and Lie algebras'', Birkhauser Verlag, Basel, 1990

\bibitem{15}
D. Bahns, ``Die Invariantenalgebra des Nambu -- Goto -- Strings in 
Erzeugungs -- und Vernichtungsoperatoren'', Diploma Thesis (in German),
Albert -- Ludwigs -- Universit\"at Freiburg, November 1999 

\bibitem{18}
O. Bratteli, D. W. Robinson, ``Operator Algebras and Quantum Statistical
Mechanics", vol. 1,2, Springer Verlag, Berlin, 1997

\bibitem{19}
R. Brunetti, K. Fredenhagen, M. Kohler, ``The Microlocal Spectrum 
Condition and Wick Polynomials of Free Fields on Curved Spacetimes'', 
Commun. Math. Phys. {\bf 180} 633-652,1996, [gr-qc/9510056]\\
R. Brunetti, K. Fredenhagen, ``Microlocal Analysis and Interacting Quantum 
Field Theories: Renormalization on Physical Backgrounds'',
Commun. Math. Phys. {\bf 208}, 623-661,2000, [math-ph/9903028]\\
R. Brunetti, K. Fredenhagen,, R. Verch, ``The Generally Covariant Locality
Principle: A New Paradigm for Local Quantum Field Theory'',
Commun. Math. Phys. {\bf 237}, 31-68,2003, [math-ph/0112041]\\
R. Verch, ``A Spin Statistics Theorem for Quantum Fields on Curved 
Spacetime Manifolds in a Generally Covariant Framework'', 
Commun. Math. Phys. {\bf 223}, 261-288, 2001, [math-ph/0102035]\\
S. Hollands, R. M. Wald, ``Local Wick Polynomials and Time Ordered 
Products of Quantum Fields in Curved Spacetime'', Commun. 
Math. Phys. {\bf 223} 289-326,2001, [gr-qc/0103074];
``Existence of Local Covariant Time Ordered Products of Quantum Fields 
in Curved Spacetime'', Commun. Math. Phys. {\bf 231} 309-345, 2002,
[gr-qc/0111108];
``On the Renormalization Group in Curved Spacetime'',
Commun.Math.Phys. {\bf 237}, 123-160, 2003, [gr-qc/0209029]\\
S. Hollands, ``A General PCT Theorem for the Operator Product Expansion
in Curved Spacetime'', Commun. Math. Phys. {\bf 244} 209-244, 2004,
[gr-qc/0212028]\\
W. Junker, E. Schrohe, ``Adiabatic Vacuum States on General Spacetime 
Manifolds: Definition, Construction and Physical Properties'' 
Annales Poincare Phys.Theor. {\bf 3}, 1113-1182, 2002 [math-ph/0109010]

\bibitem{19a} D. Giulini, D. Marolf , ``A Uniqueness Theorem for 
Constraint
Quantization"
Class. Quant. Grav. 16:2489-2505,1999, [gr-qc/9902045];
``On the Generality of Refined Algebraic
Quantization", Class. Quant. Grav. 16:2479-2488,1999, [gr-qc/9812024]

\bibitem{9a} T. Thiemann, ``
The Phoenix Project: Master Constraint Programme for Loop Quantum 
Gravity'',
gr-qc/030580

\bibitem{13a} B. Dittrich, T. Thiemann, ``Testing the 
Master Constraint Programme for Loop Quantum Gravity'', in preparation

\bibitem{24}
A. Ashtekar, J. Lewandowski, D. Marolf, J. Mour\~ao, T.
Thiemann, ``Quantization for diffeomorphism invariant theories
of connections with local degrees of freedom", Journ. Math. Phys.
{\bf 36} (1995) 6456-6493, [gr-qc/9504018]

\bibitem{21}
H. Sahlmann, ``When do Measures on the Space of Connections
Support the Triad Operators of Loop Quantum Gravity?'', gr-qc/0207112;
``Some Comments on the Representation Theory of the Algebra Underlying
Loop Quantum Gravity'', gr-qc/0207111\\
A. Okolow, J. Lewandowski, ``Diffeomorphism Covariant
Representations of the Holonomy Flux Algebra'', gr-qc/0302059\\
H. Sahlmann, T. Thiemann, ``On the Superselection Theory of
the Weyl Algebra for Diffeomorphism Invariant Quantum Gauge Theories'',
gr-qc/0302090;
``Irreducibility of the Ashtekar-Isham-Lewandowski Representation'',
gr-qc/0303074

\bibitem{20}
J. Glimm, A. Jaffe, ``Quantum Physics", Springer-Verlag, New York, 1987

\bibitem{23} W. Thirring, ``Lehrbuch der Mathematischen Physik", vol. 3,
Springer Verlag, Berlin, 1978

\bibitem{24aa} 
A. Ashtekar, C.J. Isham, ``Representations of the Holonomy
Algebras of Gravity and Non-Abelean Gauge Theories",
Class. Quantum Grav. {\bf 9} (1992) 1433, [hep-th/9202053]\\
A. Ashtekar, J. Lewandowski, ``Representation
theory of analytic Holonomy $C^\star$ algebras", in ``Knots and
Quantum Gravity", J. Baez (ed.), Oxford University Press, Oxford 1994

\bibitem{24ab} T. Thiemann, ``Kinematical Hilbert Spaces for Fermionic and 
Higgs Quantum Field Theories'', Class. Quant. Grav. {\bf 
15}, 1487-1512, 1998, [gr-qc/9705021]

\bibitem{24ac}
A. Ashtekar, J. Lewandowski, H. Sahlmann, ``Polymer and Fock 
Representations for a Scalar Field'', Class.Quant.Grav. {\bf 20}
L11-1,2003, gr-qc/0211012

\bibitem{24a} Y. Yamasaki, ``Measures on Infinite Dimensional Spaces",
World Scientific, Singapore, 1985

\bibitem{24b} G.W. Mackey, ``Unitary Group Representations in Physics,
Probability Theory and Number Theory'', 
Benjamin-cummings Publ.Comp., Reading, 1978  

\bibitem{25}
M. Reed, B. Simon, ``Methods of Modern Mathematical Physics",
vol. 2, Academic Press, New York, 1978

\bibitem{26} T. Thiemann, ``Quantum Spin Dynamics (QSD): VII.
Symplectic Structures and Continuum Lattice Formulations of
Gauge Field Theories", Class.Quant.Grav.18:3293-3338,2001,
[hep-th/0005232]; ``Gauge Field Theory Coherent States (GCS): I.
General Properties", Class.Quant.Grav.18:2025-2064,2001, [hep-th/0005233];
``Complexifier Coherent States for Canonical Quantum General Relativity",
gr-qc/0206037\\
T. Thiemann, O. Winkler, ``Gauge Field Theory Coherent States
(GCS): II. Peakedness Properties", Class.Quant.Grav.18:2561-2636,2001,
[hep-th/0005237]; ``III. Ehrenfest Theorems",
Class. Quantum Grav. {\bf 18} (2001) 4629-4681, [hep-th/0005234];
``IV. Infinite Tensor Product and Thermodynamic Limit",
Class. Quantum Grav. {\bf 18} (2001) 4997-5033, [hep-th/0005235]\\
H. Sahlmann, T. Thiemann, O. Winkler, ``Coherent States for
Canonical Quantum General Relativity and the Infinite Tensor Product
Extension", Nucl.Phys.B606:401-440,2001
[gr-qc/0102038]\\
H. Sahlmann, T. Thiemann, ``Towards the QFT on
Curved Spacetime Limit of QGR. 1. A General Scheme'', [gr-qc/0207030];
``2. A Concrete Implementation", [gr-qc/0207031]

\bibitem{28}
M. Varadarajan, ``Fock representations from U(1) Holonomy Algebras",
Phys. Rev. {\bf D61}, 104001, 2000, [gr-qc/0001050]\\
M. Varadarajan, ``Photons from Quantized Electric Flux Representations",
Phys.Rev. {\bf D64} (2001) 104003, [gr-qc/0104051]\\
M. Varadarajan, ``Gravitons from a Loop Representation of Linearized
Gravity", Phys. Rev. {\bf D66} (2002) 024017, [gr-qc/0204067]
%

\bibitem{27} E. Witten, ``The Cosmological Constant Problem from the 
Viewpoint of String Theory'', hep-ph/0002297

\bibitem{9} A. Sen, B. Zwiebach, ``Tachyon Condensation in String Field 
Theory'', JHEP {\bf 0003} 002, 2000; [hep-th/9912249]

\bibitem{29}  W. Lerche, ``Recent Developments in String Theory´´,
hep-th/9710246

\bibitem{30} 
M. Haack, B. Kors, D. L\"ust, ``Recent Developments in String Theory:
From Perturbative Dualities to M Theory", hep-th/9904033

\bibitem{31} E. D'Hoker, D.H. Phong, ``Lectures on Two Loop 
Superstrings'', hep-th/0211111 

\bibitem{32} L. Smolin, ``How far are we from a quantum theory of 
gravity'', hep-th/0303185









\end{thebibliography}
\end{document}